\documentclass[aps,twocolumn,pra,tightenlines,floatfix,showkeys,showpacs]{revtex4-1}

\usepackage[dvips]{graphicx}
\usepackage[english]{babel}
\usepackage{amsmath}
\usepackage{amssymb}
\usepackage{times}
\usepackage{color}

\newcommand{\xik}{\xi_{\bf k}}

\newcommand{\ek}{\epsilon_{\mathbf{k}}}

\newcommand{\Ek}{E_{\mathbf{k}}}

\newcommand{\phik}{\varphi_{\mathbf{k}}}

\newcommand{\mb}[1]{{\mathbf{#1}}}
\newcommand{\sumk}{\sum_{\mathbf{k}}}

\newcommand{\sumq}{\sum_{\mathbf{q}}}

\newcommand{\Omegaq}{\Omega_{\mathbf{q}}}

\newcommand{\uk}{u_{\mathbf{k}}}
\newcommand{\vk}{v_{\mathbf{k}}}

\renewcommand{\d}{\mathrm{d}}

\begin{document}

\title{Pseudogap phenomena in ultracold atomic Fermi gases}

\author{Qijin Chen}
\email[Corresponding author: ]{qchen@zju.edu.cn}
\affiliation{Department of Physics and Zhejiang Institute of Modern
  Physics, Zhejiang University, Hangzhou, Zhejiang 310027, China}
%\author{Jibiao Wang}
%\author{Yi Yu}
%\email[Corresponding authors: ]{qchen@zju.edu.cn}

\author{Jibiao Wang}
\affiliation{Department of Physics and Zhejiang Institute of Modern
  Physics, Zhejiang University, Hangzhou, Zhejiang 310027, China}

\date{\today}

\begin{abstract}
  The pairing and superfluid phenomena in a two-component ultracold
  atomic Fermi gas is an analogue of Cooper pairing and
  superconductivity in an electron system, in particular, the high
  $T_c$ superconductors. Owing to the various tunable parameters that
  have been made accessible experimentally in recent years, atomic
  Fermi gases can be explored as a prototype or quantum simulator of
  superconductors.  It is hoped that, utilizing such an analogy, the
  study of atomic Fermi gases may shed light to the mysteries of high
  $T_c$ superconductivity. One obstacle to the ultimate understanding
  of high $T_c$ superconductivity, from day one of its discovery, is
  the anomalous yet widespread pseudogap phenomena, for which a
  consensus is yet to be reached within the physics community, after
  over 27 years of intensive research efforts.  In this article, we
  shall review the progress in the study of pseudogap phenomena in
  atomic Fermi gases in terms of both theoretical understanding and
  experimental observations. We show that there is strong, unambiguous
  evidence for the existence of a pseudogap in strongly interacting
  Fermi gases. In this context, we shall present a pairing fluctuation
  theory of the pseudogap physics and show that it is indeed a strong
  candidate theory for high $T_c$ superconductivity.

\end{abstract}



\maketitle
\tableofcontents

\section{Introduction}
\label{sec:1}

Study of atomic Fermi gases, especially the pairing and superfluid
phenomena, has become a major field in physics research over the last
decade \cite{ourreview,BlochRMP}. Intrinsically a many-body system,
atomic Fermi gases have attracted physicists from both condensed
matter and atomic, molecular and optics (AMO) communities, as well as
from other communities, e.g., nuclear and particle physics and
astrophysics.  Even superstring theorists have now found it a play
ground for the ingenious idea of the AdS/CFT correspondence \cite{Maldacena98,Witten98,MaldacenaPhysRep,Zaanen}. This is
primarily due to the fact that many tunable parameters have been made
accessible experimentally for atomic Fermi gases, including
temperature, pairing interaction strength, pairing symmetry,
population imbalance, mass imbalance, geometric aspect ratio of the
trap, optical lattices, and dimensionality, etc., as well as extra
degrees of freedom such as spin-orbit coupling and synthetic gauge
fields, which make atomic gases a suitable system for quantum
simulation and quantum engineering of existing and previously unknown
systems, and have thus provided a great opportunity for studying many
exotic quantum phenomena.

In terms of superfluidity, atomic Fermi gases can be thought of as the
charge neutral counterpart of superconductors, which have been an
important subject in contemporary condensed matter physics. In
particular, high $T_c$ superconductivity has been a great challenge
since its discovery a quarter century ago. With tunable interactions,
it is strongly hoped that one may learn about the notoriously
difficult problem of high $T_c$ superconductivity via studying atomic
Fermi gases.

At the heart of the high $T_c$ problem is the widespread anomalous
normal state gap \cite{Timusk} in the single particle excitation
spectrum, which has been referred to as the \emph{pseudogap}, and has
emerged since the discovery of high $T_c$ superconductors. It is
essential to understand the pseudogap phenomena in order to reach a
consensus on the mechanism of high $T_c$ superconductivity. Due to the
analogy between superfluidity and superconductivity, it is expected
that study of the pairing and superfluid phenomena in ultracold Fermi
gases may eventually shed light on the pseudogap physics and thus the
mechanism of high $T_c$ superconductivity.

The first and most widely explored parameter in ultracold atomic Fermi
gases is the pairing interaction strength. Using an $s$-wave Feshbach
resonance, one can tune the effective pairing strength from the weak
coupling limit of Bardeen-Cooper-Schrieffer (BCS) superfluidity
\cite{Schrieffer} all the way through the strong coupling limit of
Bose-Einstein condensation (BEC) \cite{Bose,Einstein,PSB03,PS02}. In
this way, the theoretical idea of BCS-BEC crossover, which was first
proposed by Eagles \cite{Eagles} and Leggett \cite{Leggett} at zero
temperature $T$ and then extended to finite $T$ by Nozieres and
Schmitt-Rink \cite{NSR} and many others
\cite{TDLee1,TDLee2,SadeMelo,Randeriareview,Janko,Maly1,Maly2,Chen2,Chen1,Kosztin1,MicnasRMP,Micnas1,Micnas95,Ranninger,Drechsler,Haussmann,Haussmann2,Tchern,Gorbar,Gusynin,Marini},
can be realized and studied systematically in experiment.

There have been a few milestones in experimental studies of the
superfluidity and BCS-BEC crossover of ultracold Fermi
gases. Degenerate Fermi gases was achieved a few years \cite{Jin}
after the experimental realization of BEC of dilute gases of bosonic
alkali atoms \cite{CornellBEC,HuletBEC1,HuletBEC2,KetterleBEC}, such as
$^{23}$Na, $^{87}$Rb, and $^7$Li. BEC of diatomic molecules on the BEC
side of a Feshbach resonance was first reported in 2003 in Fermi gases
of $^{40}$K and of $^6$Li \cite{Jin3,Grimm,Ketterle2}. Superfluidity
in a Fermi gas in the entire BCS-BEC crossover was achieved and
reported in 2004 \cite{Jin4,Grimm2,Grimm4,Ketterle3}. A continuous
thermodynamic superfluid transition was not observed until late 2004
\cite{ThermoScience-full}. A smoking gun of superfluidity came from
the Ketterle group in 2005 which reported observation of vortex
lattices, a macroscopic manifestation of quantum phenomena, from the
BCS through BEC regimes \cite{KetterleV}. Population (or spin)
imbalance has been the second experimental parameter which has been
explored in ultracold Fermi gases since 2006 \cite{ZSSK06,Rice1}. It
is expected to lead to new phases such as phase separation and the
exotic Fulde-Ferrell-Larkin-Ovchinnikov (FFLO) states
\cite{FF,LO_ru,LO}. Further parameters which have been gradually explored
experimentally include geometric aspect ratio and dimensionality, mass
imbalance, pairing symmetry such as $p$-wave, synthetic gauge fields
and spin-orbit coupling, long range interactions as in dipolar
molecules and magnetic atoms, as well as periodic potential, i.e.,
optical lattices.

\begin{figure*}[t]
\label{fig:DOS}
%\centerline{\includegraphics[width=5.4in, clip]{DOSSf_5-2.eps}}
\centerline{\includegraphics[width=5.4in, clip]{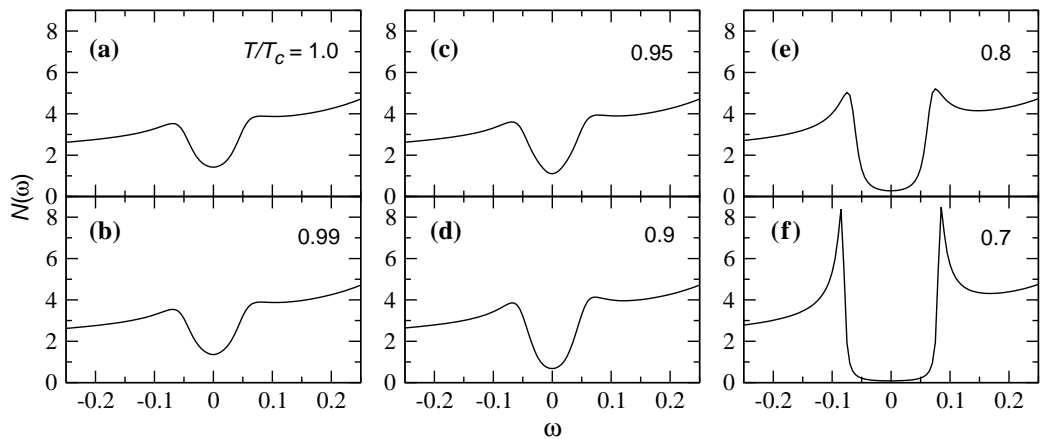}}
\caption{Typical evolution of the density of states in the presence of
  a pseudogap for an $s$-wave superconductor as a function of
  temperature, calculated for an quasi-2D superconductor on a square
  lattice at quarter filling. Panels (a)-(f) correspond to various
  temperatures (as labeled) decreasing from $T=T_c$. The DOS drops
  rapidly as the system enters the superconducting state below
  $T_c$. At $T/T_c\sim 0.7$, as shown in (f), the DOS is close to a
  true gap as that of strict BCS theory. The frequency $\omega$ is in
  units of the 2D half band width. Taken from Ref.~\cite{ChenPhD}. }
\end{figure*}

There have been a few reviews on the subject of atomic Fermi
gases. References \cite{ourreview} and \cite{ReviewLTP-Full} are the
earliest reviews on this subject, emphasizing the similarity between
Fermi gases and high $T_c$ superconductivity as well as BCS-BEC crossover
physics. Reference \cite{GiorginiRMP} reviewed the progress on the
physics of degenerate Fermi gases from the theoretical
perspective. Strong correlation effects in terms of many-body physics
were only quickly mentioned as ``other theoretical approaches''. The
review by Chin \textit{et al} \cite{ChinRMP} focuses more on Feshbach
resonances, with a very brief touch on the experiments on BCS-BEC
crossover. A few papers in the Varenna proceedings \cite{VarennaProc},
as well as Ref.~\cite{BlochRMP}, also gave an review on the
experimental and theoretical progress on atomic Fermi gases, without
much emphasis on the pseudogap physics.  It is the purpose of the
current paper to give a more or less systematic review on the study of
the pseudogap physics in cold atomic Fermi gases.

The rest of this paper is arranged as follows. In
Sec.~\ref{sec:Concept}, we shall first introduce the concept of
pseudogap in the context of high $T_c$ superconductivity, and then
provide examples of the pseudogap phenomena above and below $T_c$, and
finally give an overview of the theoretical debate on the nature of
the pseudogap. In Sec. III, we shall start by a summary of various
theories of pairing fluctuations in the context of BCS-BEC crossover,
and then present a particular pairing fluctuation theory for the
pseudogap phenomena for a homogeneous system and later extend to Fermi
gases in a trap. We shall end this section by presenting theoretical
results on the thermodynamics and superfluid density. In Sec. IV, we
shall show key results from the present pairing fluctuation theory on
the pseudogap phenomena in both the 3D homogeneous case (Subsec. IVA)
and in a trap (Subsec. IVC). In Subsec. IVB, we shall also give a
summary of the applications of the present theory to high $T_c$
superconductors with a $d$-wave pairing symmetry.  In Sec. V, we will
present a series of experiments which show strong evidence or support
of a pseudogap in unitary Fermi gases. While we focus mainly on
population balanced two component Fermi gases, we shall show one case
of population imbalanced Fermi gas experiment. In Sec. VI, we shall
discuss the effect of particle-hole fluctuations and propose further
experiments on pseudogap physics. Finally, we will conclude in
Sec. VII.

\section{What is a pseudogap? -- Pseudogap phenomena in high $T_c$ superconductors}

\label{sec:Concept}

\subsection{What is a pseudogap?}

We begin by introducing the concept of pseudogap, which has emerged
since day one of high $T_c$ superconductivity. In BCS theory, when the
superconducting order parameter $\Delta$ becomes nonzero below the
transition temperature $T_c$, a gap opens up at the Fermi level in the
single particle excitation spectrum. The density of states (DOS)
becomes zero within the gap. This gap originates purely from the order
parameter and therefore vanishes at and above $T_c$.  Soon after the
high $T_c$ superconductivity was discovered in cuprates, an excitation
gap was observed already above $T_c$, below a higher temperature $T^*$
(which is referred to as the pseudogap \emph{crossover}
temperature). Without phase coherence, such a gap does not lead to
complete depletion of the DOS within the gap, but rather the DOS was
only partially depleted. As $T$ approaches $T_c$ from above, the DOS
drops quickly to zero at the Fermi level once phase coherence sets in
as the system enters the superconducting state. In contrast to the
true gap below $T_c$, \emph{the gap observed experimentally above
  $T_c$ has been referred to as a pseudogap}. Whether the pseudogap
persists below $T_c$ has been under debate.

The typical behavior of the DOS near the Fermi level for a
pseudogapped superconductor is shown in Fig.~\ref{fig:DOS} for various
temperatures from $T_c$ to slightly below $T_c$. The curves are
calculated theoretically for an $s$-wave superconductor on a quasi-two
dimensional (2D) lattice. From Fig.~\ref{fig:DOS}(a), one can see
clearly a partial depletion of the DOS around the Fermi level
($\omega=0$). As $T$ lowers into the broken symmetry state, phase
coherence sets in, and the DOS drops rapidly. At $T=0.7T_c$, the
depletion within the gap becomes almost complete so that the DOS looks
like one in a strict BCS mean-field theory, with two sharp coherent
peaks at the gap edges.

\subsection{Pseudogap in the normal state above $T_c$}

Above $T_c$, the pseudogap manifests itself in various physical
quantities, including the $dI/dV$ characteristics in tunneling
spectroscopy \cite{Renner,Renner1998,Krasnov2000,Fischer2}, specific
heat \cite{Loram,Williams1998}, dc resistivity
\cite{Walker1995,Graf1995,Yan1995}, nuclear magnetic resonance (NMR)
\cite{Williams1996,Williams1997,Magishi1996,Goto1997,Bobroff1997,Ishida1998},
infrared and ac conductivity \cite{Puchkov,Basov,Basov2}, neutron
scattering \cite{Tranquada,MookH,Aeppli3}, Raman scattering
\cite{Ruani1997,Chen1997,Nemetschek1997,Quilty1998}, Nernst effect
\cite{Nernst,Ong2,Ong3,Tan}, spin susceptibility, etc., as a function
of temperature. The most direct probe, of course, is the
angle-resolved photoemission spectroscopy (ARPES)
\cite{arpesanl1,arpesstanford,ANLPRL}, which probes directly the
spectral function $A(\mathbf{k}, \omega)$. Along the Fermi surface,
the quasiparticle coherence peak position in the measured spectral
function reveals directly the gap parameter. A review on various
experiments on the pseudogap phenomena can be found in
Ref.~\cite{Timusk}. Here we only show a couple of examples to
illustrate the pseudogap phenomena.

\begin{figure}
%  \centerline{\hspace*{0.1in}\includegraphics[width=2.4in]{Fig2a.eps}}
%\vskip 1.5ex
% \centerline{\includegraphics[width=2.8in]{Fig2b.eps}} 
 \centerline{\includegraphics[]{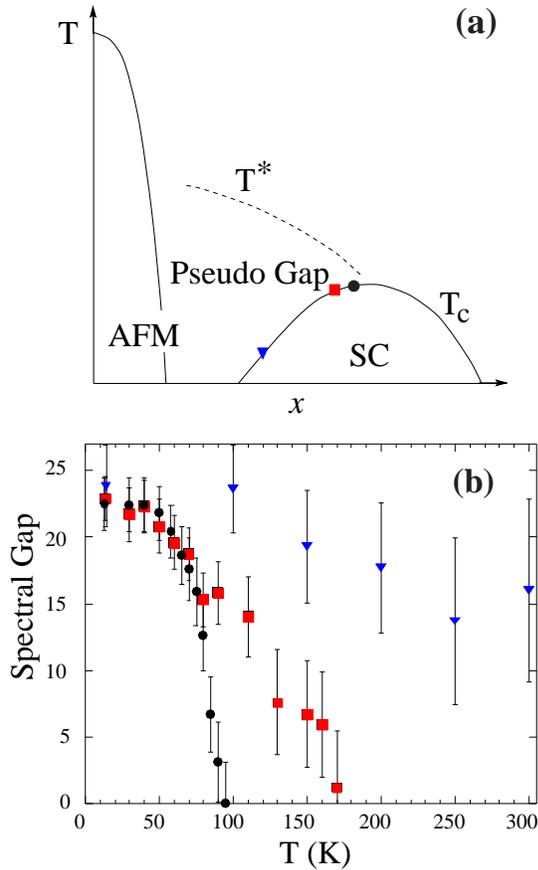}} 
 \caption[Schematic phase diagram for the cuprates and evidence from
 ARPES measurements for pseudogap in underdoped samples from ARPES.]
 {(a) Schematic phase diagram for the cuprate superconductors (The
   horizontal axis is the hole doping concentration), and (b) ARPES
   measurement of the temperature dependence of the excitation gap at
   ($\pi$, 0) in a near-optimal $T_c=87$~K sample ({\large
     $\bullet$}), and underdoped 83~K ({\color{red}\scriptsize
     $\blacksquare$}) and 10~K ({\color{blue}$\blacktriangledown$})
   samples. The gap values were determined via leading edge shift from
   the Fermi level. The units for the gap are meV. Panel (b) is taken
   from Ref.~\cite{arpesanl1}.}
\label{fig:PhaseDiagram_ARPES} 
\end{figure}

High $T_c$ superconductors of the cuprates, such as
YBa$_2$Cu$_3$O$_{7-\delta}$ (YBCO), Bi$_2$Sr$_2$CaCu$_2$O$_{8+\delta}$
(Bi2212) and La$_{2-x}$Sr$_x$CuO$_4$ (LSCO), have a layered structure,
with charge carriers moving in the copper-oxide planes. The electron
transport along the $c$-axis (i.e.,the direction perpendicular to the
planes) is largely incoherent. This makes the cuprates quasi-2D
materials. While the parent compounds are insulating antiferromagnets,
superconductivity occurs at low $T$ upon hole doping
\cite{holedopingfootnote}. Within the superconducting ($ab$-)planes,
it is now known that the order parameter $\Delta$ of the cuprate
superconductors has a $d_{x^-y^2}$ symmetry, such that $\Delta =
\Delta_0(\cos k_x - \cos k_y)/2$, where we have set the in-plane
lattice constant $a$ to unity. Thus the gap has a maximum in the
anti-nodal directions near $(\pi, 0)$, whereas it closes in the nodal
directions from $\Gamma$ to $(\pi,\pi)$ in the Brillouin zone
(BZ). The order parameter changes sign across the nodal points along
the Fermi surface.

Shown in Fig.~\ref{fig:PhaseDiagram_ARPES}(a) is a schematic phase
diagram for the cuprate superconductors. The transition temperature
$T_c$ reaches a maximum around doping concentration $x = 0.155$. There
is a temperature range between $T_c$ and $T^*$ in the underdoped
regime where a finite pseudogap exists. Shown in
Fig.~\ref{fig:PhaseDiagram_ARPES}(b) are the ARPES measurements of the
excitation gap near ($\pi$, 0) for Bi2212 at different doping
concentrations. At optimal doping ($T_c=87$~K sample), the gap closes
roughly at $T_c$, similar to that predicted in BCS theory. However,
for the underdoped samples, it is clear that the gap persists at very
high $T$. This is the most direct measurement, and hence evidence of
the existence, of a pseudogap above $T_c$.

\begin{figure}[t]
%  \centerline{\includegraphics[width=3in]{Renner_dIdV.eps}}
  \centerline{\includegraphics[width=3in]{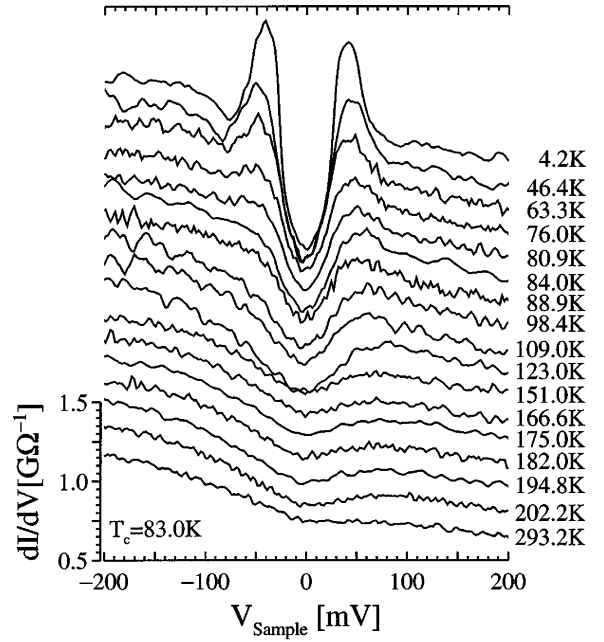}}
  \caption{Typical tunneling spectra for an underdoped cuprate
    superconductor as a function of temperature. Shown data were
    measured on an underdoped Bi2212 sample of $T_c=83$~K. The
    horizontal axis is the bias. The conductance scale corresponds to
    the 293 K spectrum, the other spectra are offset vertically for
    clarity.  Taken from Ref.~\cite{Renner1998}.}
\label{fig:Renner_dIdV} 
\end{figure}

In Fig.~\ref{fig:Renner_dIdV}, we present, as an example, typical
normal-insulator-superconductor (SIN) tunneling spectra measured for
an underdoped cuprate superconductor as a function of temperature,
using scanning tunneling microscopy (STM). Here the $dI/dV$
characteristics can be regarded as the DOS, but broadened by thermal
effects. In sharp contrast with a BCS mean-field true gap, the
pseudogap does not close even at $T^*$. But rather, the coherence
peaks broaden and the DOS fills in with increasing $T$. At $T^*$, the
sign of DOS depletion disappears and so does the pseudogap.  The way
the pseudogap disappears at high $T$ is a clear distinction from that
of a true gap in a weak coupling BCS superconductor, which shrinks in
magnitude to zero at $T_c$.

\begin{figure}
%  \centerline{\includegraphics[width=2.2in]{Kanigel_PRL_2008.Fig4.eps}}
  \centerline{\includegraphics[width=2.2in]{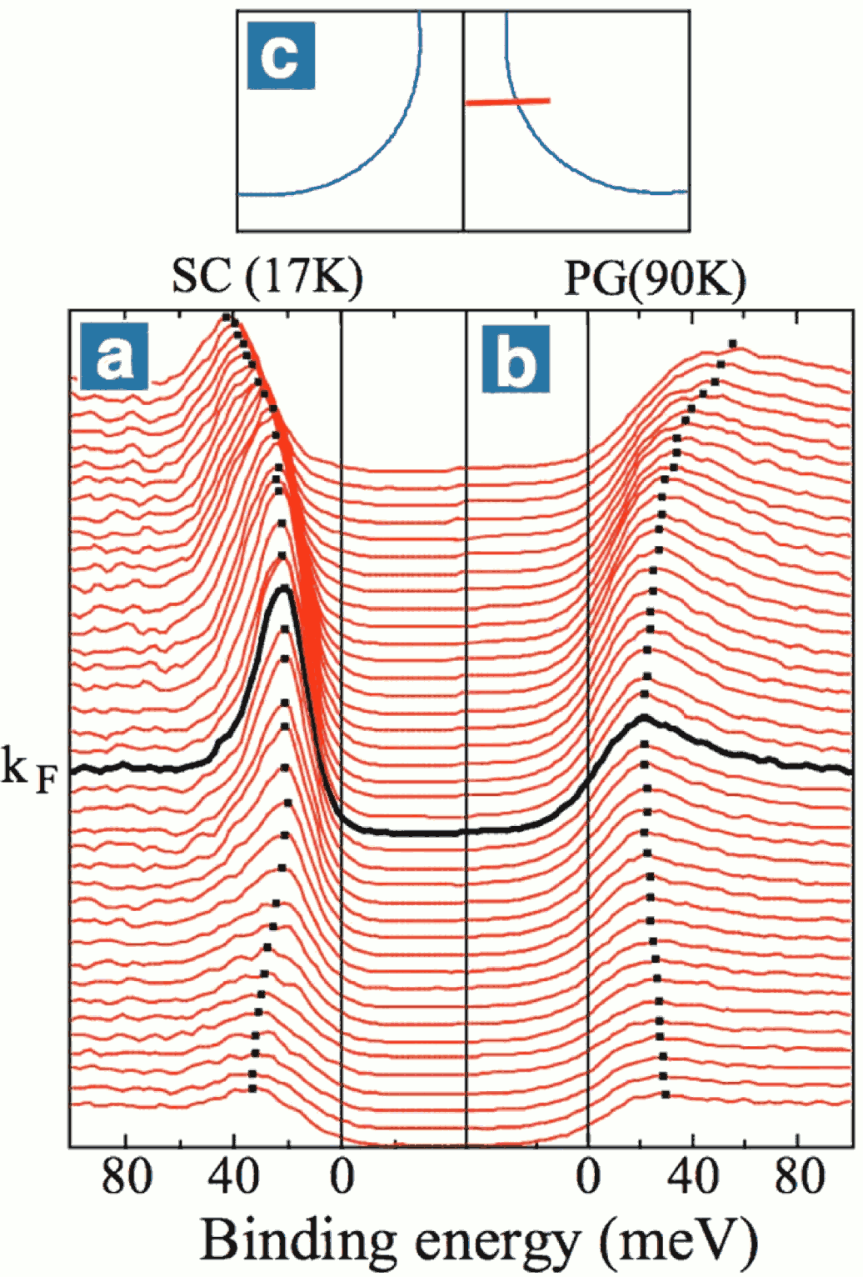}}
  \caption{Comparison of EDCs between (a) the superconducting state
    ($T = 17$~K) and (b) the pseudogap phase ($T = 90$~K) for a Bi2212
    film with $T_c = 80$~K for the cut in the zone shown in (c). The
    thick curves in (a) and (b) correspond to where the cut goes
    through the Fermi surface. Here SC and PG denote superconducting
    and pseudogap state, respectively. Taken from Ref.~\cite{ANLPRL}.}
\label{fig:ARPES_EDC} 
\end{figure}

Similar behavior can be found in the ARPES energy distribution curves
(EDC) for underdoped samples as well
\cite{arpesanl1,arpesstanford}. Usually, an ARPES EDC curve consists
of a quasiparticle coherence peak on top of an incoherent
background. At $T \ll T_c$, the coherence peak is sharp and
pronounced. Once the temperature rises above $T_c$, for an underdoped
sample, it becomes broadened quickly, and the spectral weight under
the peak decreases with $T$ rapidly, until it merges with the large
incoherent background. At the same time, the peak location almost does
not move with $T$. This can be seen in Fig.~\ref{fig:ARPES_EDC}. The
EDC curves in Figs.~\ref{fig:ARPES_EDC}(a) and (b) were taken along
the cut in the Brillouin zone shown Fig. \ref{fig:ARPES_EDC}(c), which
goes across the Fermi level. Panels (a) and (b) correspond to low
$T\ll T_c$ and above $T_c$ cases, respectively, with curves of the
same $\mathbf{k}$ lined up together. As the cut goes through the Fermi
surface, the coherence peak reaches the minimum quasiparticle
excitation energy, as determined by the excitation gap
$\Delta_{\mathbf{k}}$. It is obvious that the coherence peak in the
pseudogap case is much broader and less pronounced than its
superconducting counterpart.

\subsection{Pseudogap in the superfluid state below $T_c$}
\label{subsec_below}

Figures \ref{fig:PhaseDiagram_ARPES}--\ref{fig:ARPES_EDC} and most
experimental measurements show clear evidence of the existence of a
pseudogap at and above $T_c$. It is a natural question to ask how the
pseudogap above $T_c$ and the superconducting gap below $T_c$ connect
to each other at $T_c$. There have been intensive debates on this
issue over the years. The answer to this important question depends on
the interpretation of the pseudogap in different theories. Despite the
differences from one theory to another, one can think of two
possibilities in general. One possibility is that the pseudogap
becomes the superconducting gap instantly once the system enters the
superconducting state across $T_c$. The other is that the order
parameter or the superconducting gap increases gradually from zero at
$T_c$. For the former, one would see a first order phase transition
and a jump in the order parameter and superfluid density across
$T_c$. For the latter possibility, the pseudogap necessarily persists
into the superfluid state, in order to keep the total excitation gap
smooth across $T_c$ as observed in ARPES data and other experiments.
Given these rather obvious differences between these two possibilities,
superfluid density ($n_s/m$) or in-plane London penetration depth
($\lambda$) measurements seem to have unambiguously ruled out the
former possibility. Indeed, superfluid density $n_s/m \propto \lambda
^{-2}$ vanishes continuously as $T$ approaches $T_c$ from below in
bulk cuprate superconductors.

At a more concrete level, compatible with the first possibility may be
a school of microscopic theories which consider the pseudogap above
$T_c$ as a signature of a competing hidden order, such as the
$d$-density wave (DDW) order \cite{Laughlin}, the staggered orbital
current \cite{LeePhysicaC2000,LeeWenPRB2001,LeePRL2003}, loop current
order \cite{VarmaPRB1997,VarmaPRB2006}, etc. Given the hidden order
assumption, a natural prediction would be that the hidden ordered
phase gives way completely to the superconducting order across $T_c$,
as in most other phase transitions. However, if this is true, not only
a first order transition is necessary, but also it would take a
miracle for the total excitation gap to remain so smooth across $T_c$
as observed experimentally. Then to pass the experimental test, these
hidden order theories may also need to associate themselves with the
second possibility, namely the hidden order parameter (and thus the
pseudogap) survives the superconducting phase transition and coexists
with the superconducting order below $T_c$. Together they contribute
to the total excitation gap.

Among the second possibility, there are different scenarios which give
rise to different interpretations and different temperature
dependencies of the pseudogap. These differences are associated with
the origins of the pseudogap in these theories or conjectures. In
order to fit the specific heat data for underdoped superconductors,
Loram and coworkers \cite{Loram,Loram98,LoramPhysicaC} contemplated
that the pseudogap below $T_c$ may take its value at (and above) $T_c$
such that it is relatively temperature independent. As a consequence,
(the magnitude of) the order parameter ($\Delta_{sc}$) is much smaller
than the total excitation gap ($\Delta$) at all $T < T_c$ for an
underdoped cuprate superconductor.  We note that this was a rather
simple recipe \emph{without any theoretical justification}. Among
microscopic theories, while the competing hidden order theories may be
associated with the second possibility, a most natural school of
theories in this category would be the \emph{precursor
  superconductivity}, in which the pseudogap is a precursor to the
superconductivity and originates from the same pairing as that causes
the superconducting order at low $T$. We will elaborate further on
this in the next subsection.

\begin{figure}[tb]
%  \centerline{\includegraphics[width=3in]{RennerFig3a3.eps}}
  \centerline{\includegraphics[width=3in]{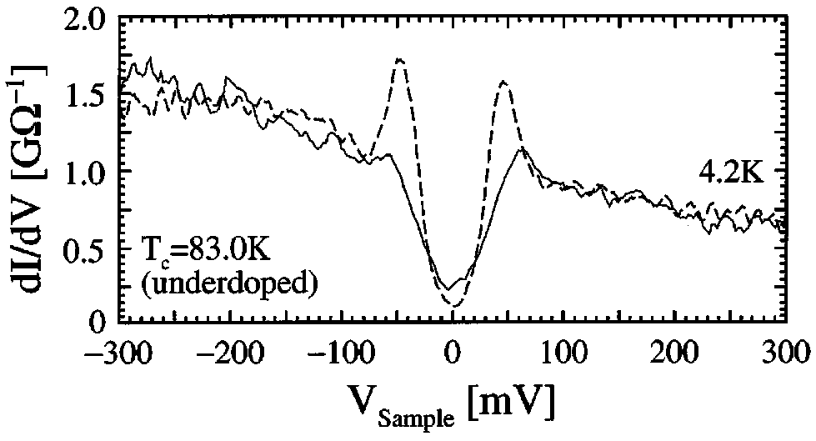}}
  \caption{STM measurements of the $dI/dV$ characteristics of an
    underdoped Bi2212 sample inside (solid) and outside (dashed) a
    vortex core at low $T=4.2$~K. Here $T_c = 83$~K. $dI/dV$ is
    proportional to the density of states. Taken from
    Ref.~\cite{Renner}.}
\label{fig:Renner_Vortex} 
\end{figure}

To probe the pseudogap below $T_c$, if it does exist, the best way is
arguably to suppress the order parameter. Luckily, this can be
achieved inside a vortex core. Figure \ref{fig:Renner_Vortex} shows
STM measurements of the $dI/dV$ characteristics of an underdoped
Bi2212 sample inside and outside a vortex core at very low
$T$. Outside the vortex core, the order parameter is large at low $T$,
and there are sharp peaks at the gap edges (dashed curve).  At the
center of the vortex core, the superconducting order parameter is
suppressed to zero (solid curve). Nevertheless, it is clear that the
$dI/dV$ curve shows a strong depletion of the DOS within the
peaks. The peak locations are roughly the same as those outside the
core. Of course, such a depletion is absent above the pseudogap
crossover temperature $T^*$. Therefore, this plot serves as evidence
of pseudogap below $T_c$ in an underdoped cuprate. Indeed, this can be
naturally explained within a pairing fluctuation theory \cite{Chen2};
In addition to noncondensed pairs, the magnetic field inside the
vortex core causes the originally condensed pairs lose phase coherence
and thus contribute to the pseudogap rather than the order parameter
\cite{Chen4}. On the other hand, as one may notice, in order for the
competing hidden order theories to explain the survival of the
total gap at the center of the vortex core at very low $T$, one would
have to assume that the superconducting order is converted into hidden
order parameter by the magnetic field. This, however, is very
unnatural.

\subsection{Theoretical debate about the nature of the pseudogap}

The pseudogap phenomena is widespread in high $T_c$ superconductivity
experiments. However, a consensus of its origin is yet to be
reached. There have been many different theories attempting to explain
the nature of the pseudogap. Most of these theories only provide
qualitative pictures, incapable of quantitative calculations.

Early models include the resonating valence bond (RVB) theory of
Anderson \cite{RVB,Vanilla} and the closely related spin-charge
separation idea \cite{SpinChargeSeparation,Spin-ChargeSeparation}. In
these theories, the pseudogap originates from the spin gap of the
antiferromagnetic spin pairing (i.e., the spinon pairing). Despite the
well established phenomena of spin-charge separation in 1D, so far
there has been no experimental support for spin-charge
separation in 2D, not to mention 3D.
 
About the same time, Uemura and coworkers \cite{UemuraPRL89,Uemura}
noticed possible connections between the cuprate superconductivity and
BEC via the well-known Uemura plot of $T_c$ versus superfluid density
$n_s/m^*$. This has been used to suggest that the cuprates have to do
with BCS-BEC crossover.  In an attempt to explain the pseudogap
phenomena, Lee and coauthors \cite{TDLee1,TDLee2} proposed a
boson-fermion model. The pseudogap phenomena was then explored using
the BCS-BEC crossover idea in 3D continuum
\cite{SadeMelo,Randeriareview,Janko,Maly1,Maly2} or the
negative-$U$ Hubbard model on a lattice
\cite{MicnasRMP,Micnas1,Micnas95,Ranninger}, assuming an $s$-wave
pairing symmetry. These theories belong to the school of precursor
superconductivity, in which the pseudogap above $T_c$ and the
superconducting gap below $T_c$ originate from the same pairing
interaction, and thus the pseudogap in the normal state is a precursor
to superconductivity below $T_c$. A theory of the broken symmetry
state and the presence of the pseudogap below $T_c$, especially for
$d$-wave pairing, was not available until the work of
Refs. \cite{Chen2,Chen1,Kosztin1}.

It is worth mentioning that in a very recent work \cite{Norman2014},
Mishra and coworkers showed that a pseudogap which is not
associated with pairing would suppress $T_c$ to zero. Therefore, they
concluded that the pseudogap observed in the cuprates must be due to
pairing.

Compatible with but distinct from the precursor superconductivity
school are theories based on phase fluctuations such as that of Emery
and Kivelson \cite{Emery} and the QED3 theory of Tesanovic and
coworkers \cite{TesanovicQED3}. The
former addresses mainly spin-wave type of phase fluctuations whereas
the latter has an emphasis on vortex fluctuations. In both theories,
the pseudogap originates from a pairing field without phase coherence;
the pairing field emerges at $T^*$ but the phase coherence does not
lock in until a lower temperature $T_c$. The strong Nernst signals
observed above $T_c$ in underdoped cuprates \cite{Nernst,Ong2,Ong3}
may be regarded as a support for the latter theory. On the other hand,
it should be noted that the Nernst effect data can be explained
within a pairing fluctuation theory as well \cite{Ussishkin,Tan}.

Both the spin gap scenarios as in RVB and spin-charge separation and
charge gap scenarios as in precursor superconductivity and phase
fluctuation pictures have to do with pairing in the particle-particle
channel. A big departure from this common feature are the competing
hidden order ideas, mentioned in Subsec.~\ref{subsec_below}, which
take the pseudogap as a hidden order parameter. For example, the DDW
order is associated with the particle-hole channel. The staggered
current and loop current order are not related to pairing,
either. They rely on the underlying quasi-2D lattice structure of the
cuprates.

The RVB and spin-charge separation ideas can be traced back to the
fact that the parent compounds of the cuprate superconductors are
insulating antiferromagnets in the Mott state, with an underlying
quasi-2D, layered lattice structure. The DDW, staggered current and
loop current ideas have also to do with the underlying lattice
structures, which are apparently not pertinent to the atomic Fermi
gases in a big single trap. Deeper than but closely related to the
pseudogap phenomena is the mechanism of superconductivity in the
cuprates, namely, what provides the glue for the electrons to pair up.

Luckily, for atomic Fermi gases, the underlying pairing interaction is
known and can be precisely manipulated experimentally. While one may
continue to debate on the origin of the pseudogap phenomena in the
cuprates, as far as the atomic Fermi gases are concerned, this fact
does make the pairing fluctuation theory the most natural candidate
for the theory of the superfluidity and pairing.

\section{Pairing fluctuation theory for the pseudogap}

\subsection{Various pairing fluctuation theories for BCS-BEC crossover}

Pairing fluctuation theories belong to the school of precursor
superfluidity. There are different pairing fluctuation
theories. Nevertheless, common to these theories are strong pairing
fluctuations or pairing correlations already above $T_c$, which
necessarily cause deviation of the system behavior from those
described by the BCS mean-field theory. The first thing that has been
looked into is the superfluid transition temperature $T_c$. Not all of
these theories contain a pseudogap in their single particle excitation
spectrum, nor are they all self-consistent.
As the pairing strength varies, a pairing fluctuation theory is often
used to address the BCS-BEC crossover problem, and thus is often
referred to as a BCS-BEC crossover theory as well.  Note that from
this section on, we will use the term ``superfluidity'' in place of
``superconductivity'', in order to be appropriate for both
superconductors and charge neutral superfluids.

The very first work on finite temperature BCS-BEC crossover, by
Nozi\'eres and Schmitt-Rink (NSR) \cite{NSR} in 1985, can be regarded
as the earliest pairing fluctuation theory. However, in the NSR
theory, only bare Green's functions are involved so that the pairing
fluctuations induced self energy does not feed back into the $T_c$
equation. As a consequence, pseudogap does not appear in the NSR
theory. Although one may find features of pseudogap via further
calculation of the spectral function with the self-energy included,
this procedure certainly breaks self-consistency. Indeed, not
including the self energy in the $T_c$ equation itself introduces
inconsistency. For example, the $T_c$ equation is inconsistent with
the condition 
\begin{equation}
\frac{\partial \Omega_S}{\partial\Delta} = 0 
\label{eq:Gapeq}
\end{equation}
as $T_c$ is approached from below, where $\Omega_S$ is the
thermodynamic potential in the superfluid state.
Sa de Melo \textit{et al.}  \cite{SadeMelo} obtained identical
equations as NSR, using a Saddle point approximation plus Gaussian
fluctuations. There have been many studies in the literature using a
similar approximation \cite{Randeriareview}. Milstein \textit{et al}
\cite{Milstein} used a similar treatment but within a
two-channel model. Indeed, it has turned out that the saddle point
approximation with Gaussian fluctuations and the NSR approximation are
equivalent.  With a narrow Feshbach resonance in a two-channel model,
Ohashi and Griffin \cite{Griffin} caculated $T_c$ using the NSR
approximation.
Strinati and coworkers \cite{Strinati6,Strinati4} also followed the
NSR calculation found the same $T_c$ and number equations as
NSR. There are other pairing fluctuation theories on BCS-BEC crossover
based on the NSR approximation. Noticeably, rather than fixing the
inconsistency in the $T_c$ equation, Hu and Drummond \cite{Drummond3}
proposed to add an extra term in the number equation. This necessarily
leads to two unphysical results: (i) This extra term does not exist
above $T_c$, so that it will give rise to a different $T_c$ and
chemical potential $\mu$, depending on whether $T_c$ is approached
from above or below. (ii) In a trap, a uniform global chemical
potential requires that the density jumps across the edge of the
superfluid core. In fact, should the $T_c$ equation is fixed so that
Eq.~(\ref{eq:Gapeq}) is satisfied, this extra term would vanish
automatically.  More systematic and detailed comparison between the
NSR-based theories and the pairing fluctuation theory which we will
present soon below can be found in Ref.~\cite{Levin_AnnPhys}.

Using a $GG$ scheme for the $T$-matrix approximation, which is
sometimes referred to as a conserving approximation, Haussmann
\cite{Haussmann2} and Tchernyshyov \cite{Tchern} \textit{et al}
developed a different kind of pairing fluctuation theory, which leads
to a substantially lower $T_c$ than others, especially in the BCS
through unitary regimes. This is primarily because the $GG$ scheme
double counts certain self energy diagrams.  This theory is rather
similar to the FLEX approximation of Scalapino and coworkers
\cite{FLEX1,FLEX2} for the cuprates. Recently, Haussmann \textit{et al.}
\cite{Zwerger} improved upon the NSR theory but found unphysical
non-monotonic first-order-like behavior in the temperature dependence
of entropy, $S(T)$.

It should be emphasized that none of these above mentioned theories
contained pseudogap self energy in the $T_c$ equation. All NSR based
theories essentially inherit the inconsistency of NSR treatment as
well.

A pairing fluctuation theory which does contain a pseudogap was
developed by the Levin group. Levin and coworkers
\cite{Janko,Maly1,Maly2} did intensive numerical study and found
that a pseudogap opens up as $T$ approaches $T_c$ from above once the
pairing correlation self energy is fed back into the $T_c$
equation. Chen, Kosztin and coworkers \cite{Chen2,Kosztin1} extended
this work into a systematic theory for the superfluid state, and
applied it to $d$-wave cuprate superconductors. With proper inclusion
of low dimensionality and lattice effects \cite{Chen1}, Chen
\textit{et al.} \cite{Chen2} found an excellent (semi-quantitative)
agreement of their computed cuprate phase diagram with that observed
experimentally. This theory also gives a very natural explanation of
the anomalous quasi-universal behavior of the superfluid density as a
function of $T$ for different hole doping concentrations.  In contrast
to other theories mentioned above, pseudogap is a natural unavoidable
consequence of strong pairing correlations in this theory.

Now with experimental evidence of a pseudogap in atomic Fermi gases,
more people are finding in their theories evidence of a pseudogap
\cite{OhashiPRA82,Bulgac2009_rep,PieriPRA84}. Various quantum Monte Carlo simulations are also
finding a pseudogap at unitarity above $T_c$.

\subsection{Pairing fluctuation theory for the pseudogap}

In this subsection, we will present a particular pairing fluctuation
theory, in which the pairing correlation self energy is fed back into
the $T_c$ and gap equation in a self-consistent fashion. 

As in all pairing fluctuation theories, the key difference between
this theory and the BCS mean field theory is that it includes finite
center-of-mass momentum pairing. It is the finite momentum pairing
that will give rise to a self energy beyond the strict BCS mean-field
treatment.

What makes this theory unique is that finite momentum pairs and single
particles are treated on an equal footing. As a consequence, these
finite momentum pairs will cause a single particle excitation gap
without phase coherence. In fact, the physical picture here is very
intuitive. When strong pairing correlations are present, to excite a
single fermion above $T_c$, one necessarily has to pay extra energy in
order to break the pairing. This extra energy is associated with the
pseudogap. In the BCS limit, this extra energy is negligible. However,
in the BEC limit, stable two-body bound pairs will form at high $T$ so
that one has to pay at least the binding energy to break the pairs. In
the unitary or crossover regime, the pairs are meta-stable with a zero
two-body binding energy so that the pseudogap is most
pronounced. Needless to say, very much like the superfluid order
parameter, the pseudogap is a many-body effect. While the pseudogap
persists deep into the BEC regime, where a Fermi surface no longer
exists, the big two-body binding energy may obscure the pseudogap
effects. While the low energy excitations are Bogoliubov
quasiparticles and finite momentum pairs for the BCS and BEC limits,
respectively, a mix of both types necessarily takes place in the
crossover regime. This is a requirement of the \emph{smoothness} of
the crossover.

The derivation of this theory \cite{ChenPhD} follows the early work of
Kadanoff and Martin \cite{Kadanoff}. Through the equation of motion
approach with a truncation of the infinite series of equations at the
three-particle level, and decomposing the three particle Green's
function $G_3$ into a sum of products of single particle Green's
function $G$ and two particle Green's function $G_2$, we rigorously
derived our self-consistent set of equations, with reasonable
simplifications. While our equations may be conveniently cast
diagrammatically into a $T$-matrix approximation, we emphasize that
this theory is \emph{not} a diagrammatic approach. For example, the
pair susceptibility $\chi$ consists of a mix of bare Green's function
$G_0$ and full Green's function $G$. This mix is \emph{not an ad hoc
  diagrammatic choice}, but rather a natural consequence of the
equation of motion approach.  A main and nice feature of this theory
is that it naturally recovers the BCS-Leggett result at zero $T$ and
in the BCS limit. In addition, throughout the superfluid phase, our
Green's function and the equations take the BCS form, except for the
extra pseudogap contribution in the quasiparticle dispersion. Finally,
the pseudogap (squared) is directly proportional to the density of
finite momentum pairs, so that it provides a good measure of the
contributions of finite momentum pairing fluctuations.

Instead of giving a full derivation of the theory, which can be found
elsewhere \cite{ChenPhD}, here we only give a summary and present the
key equations so that we can focus on the physical picture. In
addition, here we only consider the one-channel model, which is
appropriate for high $T_c$ superconductors as well as atomic Fermi
gases with a wide Feshbach resonance. A two-channel version of this theory
can be found in Refs.~\cite{JS2,ourreview}.

\begin{figure}
%\centerline{\includegraphics[clip]{Tmatrix_SelfE.eps}}
\centerline{\includegraphics[clip]{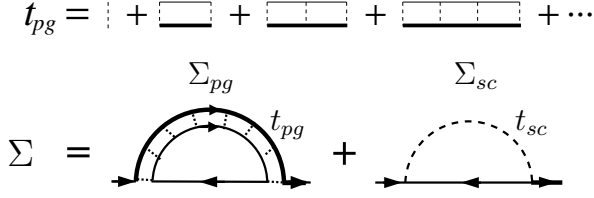}}
\caption{Feynman diagrams of the particle-particle scattering
  $T$-matrix $t_{pg}$ and the single particle self energy
  $\Sigma$. The self energy $\Sigma$ contains two contributions, from
  the condensate and finite momentum pairs, respectiveily.}
\label{fig:SelfE_T}
\end{figure}

It is known that superfluidity concerns primarily the
particle-particle channel. The main processes are summarized in the
Feynman diagrams shown in Fig.~\ref{fig:SelfE_T}. Here the finite
momentum $T$-matrix $t_{pg}$ may be regarded as (the central part of)
a two particle propagator, and the dashed line represents
non-propagating, zero-momentum pairs in the condensate.  The
self-energy $\Sigma$ of the single fermions comes from scattering with
condensed and non-condensed pairs. Alternatively, a fermion may decay
into a pair and a hole, which then recombine at a later point in
spacetime, as shown in the second line in the figure. From the first
line, it is not hard to conclude that the $T$-matrix can be regarded
as a renormalized pairing interaction. Indeed, summing up the ladder
diagrams, one obtain 
\begin{equation}
t_{pg}(Q) = \frac{U}{1+U\chi(Q)}, 
\end{equation}
with the same dimensionality as the interaction, where we have assumed
a separable pairing interaction
$V_\mathbf{k,k'}=U\varphi_\mathbf{k}\varphi_\mathbf{k'}$, with
$\varphi_\mathbf{k}=1$ for a short range contact potential in atomic
Fermi gases and $\varphi_\mathbf{k}=\cos k_x -\cos k_y$ for $d$-wave
cuprate superconductors \cite{separableinteraction}. Here the pair
susceptibility
\begin{equation}
\chi(Q) = \sum_K G_0(Q-K)G(K) \varphi_{\mathbf{k}-\mathbf{q}/2}^2\,.
\label{eq:chi}
\end{equation}
For clarity, a four-vector notation has been used, i.e.,
$K=(i\omega_l, \mathbf{k})$, $Q=(i\Omega_n, \mathbf{q})$, $\sum_K
\equiv T \sum_l\sum_\mathbf{k}$, etc., where $\Omega_n = 2n\pi T$ and
$\omega_l = (2l+1)\pi T$ are even and odd Matsubara frequencies,
respectively. Here and throughout we shall use the natural units
$\hbar = k_B=1$ and set the volume to unity.

From Fig.~\ref{fig:SelfE_T}, it is straightforward to write down the
self-energy $\Sigma$ and its superconducting component $\Sigma_{sc}$
and pseudogap component $\Sigma_{pg}$, as follows.
\begin{subequations}
\begin{eqnarray}
\Sigma(K) &=& \Sigma_{sc}(K) + \Sigma_{pg}(K) \,,\\
\Sigma_{sc}(K) &=& -\Delta_{sc}^2 G_0(-K)\varphi_\mathbf{k}^2
=\frac{\Delta_{sc}^2\varphi_\mathbf{k}^2}{i\omega_l + \xi_\mathbf{k}} \,,\\
\Sigma_{pg}(K) &=& \sum_Q t_{pg}(Q) G_0(Q-K)\varphi_\mathbf{k-q/2}^2
\,,
\label{eq:sigmapg}
\end{eqnarray}
\end{subequations}
where $\xi_\mathbf{k} = \epsilon_\mathbf{k}-\mu$ is the free fermion
dispersion measured with respect to the Fermi level.

At this point, an approximation is needed in order to simplify the
final result. Notice that pairing instability condition, i.e., the
Thouless criterion, $t_{pg}^{-1}(0) =0$, implies that the main
contribution in Eq.~(\ref{eq:sigmapg})  comes from the vicinity of
$Q=0$. This leads to a good mathematical simplification over the
complicated convolution,
\begin{equation}
\Sigma_{pg}(K) \approx \left[\sum_Q t_{pg}(Q)\right] G_0(-K)\varphi_\mathbf{k}^2
= -\Delta_{pg}^2 G_0(-K)\varphi_\mathbf{k}^2 \,,
\label{eq:pgsimplification}
\end{equation}
where we have defined the pseudogap $\Delta_{pg}$ via
\begin{equation}
\Delta_{pg}^2 = -\sum_Q t_{pg}(Q) \,.
\label{eq:PG_def}
\end{equation}

Now it is clear that, under approximation
Eq.~(\ref{eq:pgsimplification}), we have the total self energy in the
BCS form, 
\begin{equation}
\Sigma(K) = -\Delta^2 G_0(-K)\varphi_\mathbf{k}^2\,,
\label{eq:SelfE}
\end{equation}
where we have defined a total excitation gap $\Delta$ via
\begin{equation}
  \Delta^2 =  \Delta_{sc} ^2 + \Delta_{pg}^2 \,.
\end{equation}
Therefore, one immediately concludes that the full Green's function
$G(K)$ also takes the BCS form,
\begin{equation}
G(K) = \frac{u_k^2}{i\omega_l -\Ek} + \frac{v_k^2}{i\omega_l +\Ek} \,,
\label{eq:G}
\end{equation}
where $\Ek = \sqrt{\xik^2 + \Delta^2 \varphi_\mathbf{k}^2}$ is the
dispersion of the Bogoliubov quasiparticles, and $\uk^2, \vk^2 =
\frac{1}{2} (1\pm \xik/\Ek)$ are formally the usual BCS coherence
factors.

Upon substituting the expressions for $G_0$ and $G$ into the Thouless
criteria, $U^{-1} + \chi(0) =0$, one obtains immediately the gap (or
$T_c$) equation after carrying out the Matsubara summation,
\begin{equation}
1+ U\sumk \frac{1-2f(\Ek)}{2\Ek}\varphi_\mathbf{k}^2 = 0 \,,
\label{eq:gap} 
\end{equation}
where $f(x)$ is the Fermi distribution function. For a short range
contact potential with $\phik=1$, as in atomic Fermi gases, one may
conveniently regularize the ultraviolet divergence via the relation
\cite{Kokkelmans}
\begin{equation}
\frac{m}{4\pi a}= \frac{1}{U} + \sumk \frac{\phik^2}{2\ek} \,,
\label{eq:unitary}
\end{equation}
based on the Lippmann-Schwinger equation, so that the interaction
strength $U$ is replaced with (the inverse of) the low energy $s$-wave
scattering length $a$, which is a widely used experimental parameter
in the AMO community.  In this way, the gap equation becomes
\begin{equation}
\frac{m}{4\pi a} + \sumk \left[\frac{1-2f(\Ek)}{2\Ek} -
\frac{1}{2\ek} \right]\varphi_\mathbf{k}^2  =0 \,.
\label{eq:gap_amo}
\end{equation}
Note that Eq.~(\ref{eq:unitary}) defines a critical coupling $U_c$,
which is more familiar to the condensed matter community; $U_c$
corresponds to the threshold for two fermions to form a bound state in
vacuum, where the scattering length $a$ diverges, 
\begin{equation}
  U_c = -1{\Big/}\sumk  \dfrac{\phik^2}{2\ek}\,. 
\end{equation}
Obviously, $U_c$ depends on the ultraviolet cutoff momentum in
$\phik^2$. It is worth mentioning that in 2D and the contact limit in
3D, $U_c$ goes to zero for an $s$-wave pairing interaction
\cite{noteonatomicgases}.

Now it should be emphasized that it is the mixed form of the pair
susceptibility $\chi$ in Eq.~(\ref{eq:chi}) that gives rise to the BCS
form in the gap equation (\ref{eq:gap}). This is very satisfying since
it is known that BCS theory works well in the weak coupling
regime. Such a feature was already recognized in the early paper by
Kadanoff and Martin \cite{Kadanoff}. Throughout the entire BCS-BEC
crossover, this BCS form of gap equation reproduces the BCS-Leggett
ground state \cite{Leggett}. This is an important merit of the present
pairing fluctuation theory, since, while one may argue that the
BCS-Leggett ground state is not perfect in the BEC regime, it has
nonetheless been a basis for various theoretical works. It is apparent
that the gap equation in other competing $T$-matrix approximations
with a $G_0G_0$ or $GG$ in the pair susceptibility will deviate
substantially from the BCS form.

Given the full Green's function Eq.~(\ref{eq:G}), it is
straightforward to write the fermion number constraint,
\begin{equation}
  n = 2\sumk \left[ v_k^2 + \frac{\xik}{\Ek} f(\Ek)\right] \,,
\label{eq:number}
\end{equation}
which is the number equation.

Equations (\ref{eq:gap}), (\ref{eq:G}) and (\ref{eq:PG_def}) now forms
a closed set of self-consistent equations, which can be used to solve
for $T_c$ and $\mu$, and $\Delta_{pg}$ at $T_c$, or for
$\Delta$, $\mu$, and $\Delta_{pg}$ for $T<T_c$. To simplify and
facilitate the computation of Eq.~(\ref{eq:PG_def}), we Taylor-expand the
inverse $T$ matrix after analytic continuation, $i\Omega_n \rightarrow
\Omega + i 0^+$, as
\begin{equation}
t_{pg}^{-1}(\Omega,\mathbf{q}) = Z\left(\Omega -\frac{q^2}{2M^*} + \mu_{pair}\right) \,,
\label{eq:PG_exp}
\end{equation}
at the lowest order in $\Omega$ and $\mathbf{q}$. A more elaborate
treatment which includes higher order terms such as $\Omega^2$ as well
as the imaginary part can be found in Ref.~\cite{ChenPhD}. Use of such
higher order expansion is made in cases where it makes a substantial
(quantitative) difference \cite{Guo2009PRA,WangFFLO}. Here the inverse
residue $Z$, the effective pair mass $M^*$, and the effective pair
chemical potential $\mu_{pair}$ can be obtained in the process of the
expansion. One can immediately extract the pair dispersion
$\Omega_\mathbf{q} = q^2 /2M^*-\mu_{pair}$. It now follows that
\begin{equation}
Z\Delta_{pg}^2 \approx \sumq b(\Omega_\mathbf{q}) \,,
\label{eq:PG}
\end{equation}
where $b(x)$ is the usual Bose distribution function.  Evidently,
Eq.~(\ref{eq:PG}) suggests that $\Delta_{pg}^2$ represents the density
of finite momentum pairs (up to a nearly constant coefficient).

\begin{figure}
%\centerline{\includegraphics[width=3.4in,clip]{InvTk4g1T1q0.3_2.eps}}
\centerline{\includegraphics[width=3.4in,clip]{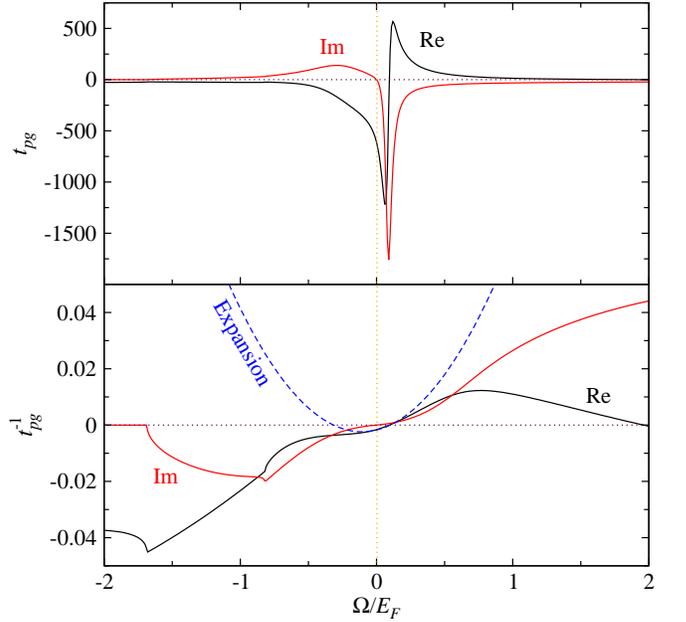}}
\caption{Typical behavior of the $T$-matrix and its inverse. Shown are
  calculated at unitarity $U=U_c$ at $T_c$, for a 3D continuum case
  with a finite range of interaction of the Lorentzian type
  $\varphi_\mathbf{k}^2 = [1+(k/k_0)^2]^{-1}$, with $k_0/k_F=4$ and
  pair momentum $q/k_F = 0.3$. Here Re and Im denote real and
  imaginary parts, respectively. The blue dashed curve is from the
  expansion of the inversion $T$-matrix, which coincides with the full
  Re~$t_{pg}^{-1}$ curve in the neighborhood of $\Omega =
  \Omega_\mathbf{q}$.}
\label{fig:Tmatrix}
\end{figure}

Typical behaviors of the $T$-matrix $t_{pg}(\Omega, \mathbf{q})$ and
its inverse are shown in Fig.~\ref{fig:Tmatrix}. The curves are
calculated for a 3D unitary Fermi gas at $T_c$, with a Lorentzian type
of pairing potential, $\varphi_\mathbf{k}^2 = [1+(k/k_0)^2]^{-1}$ at
$k_0/k_F=4$ and $q/k_F = 0.3$.  From the lower panel, one can see that
the Taylor expansion of the inverse $T$-matrix, $t_{pg}^{-1}(\Omega,
\mathbf{q})$, up to the order of $\Omega^2$, agrees with the real part
of the full curve very well near the dispersion relation $\Omega
\approx \Omega_\mathbf{q}$, where the imaginary part, $\mbox{Im\,}
t_{pg}^{-1}(\Omega, \mathbf{q})$, becomes very small. This leads to a
sharp resonance peak in $\mbox{Im\,} t_{pg}(\Omega, \mathbf{q})$ at
$\Omega \approx \Omega_\mathbf{q}$. This peak becomes sharper for
smaller $q $ and at lower $T$, as expected. When the order parameter
develops below $T_c$, for $q=0$, there is an extended range of
$\Omega$ at which $\mbox{Im\,} t_{pg}^{-1}(\Omega, 0)$
vanishes. This is an effect of a finite excitation gap.

\begin{figure}
%\centerline{\includegraphics[width=3.4in,clip]{InvTk4g1-T_2.eps}}
\centerline{\includegraphics[width=3.4in,clip]{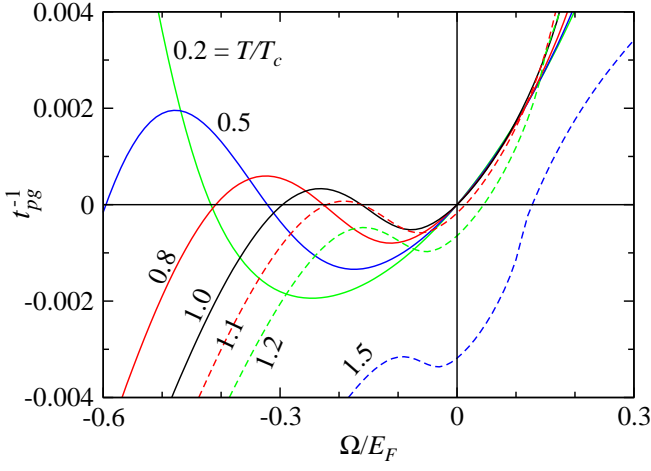}}
\caption{Typical behavior of the real part of the inverse $T$-matrix
  $t_{pg}^{-1}(\Omega, 0)$ near $\Omega = 0$ for different $T$, as
  labeled, below (solid) and above $T_c$ (dashed curves). Shown are
  results calculated at unitarity $U=U_c$, for a 3D continuum case
  with a finite range of interaction of the Lorentzian type
  $\varphi_\mathbf{k}^2 = [1+(k/k_0)^2]^{-1}$, with $k_0/k_F=4$. }
\label{fig:InvT}
\end{figure}

From the expansion Eq.~(\ref{eq:PG_exp}), it is easy to see that the
Thouless criterion requires
\begin{equation}
 \mu_{pair}=0,\qquad \text{for}\quad T\le T_c\,,
\label{eq:BECcondition}
\end{equation}
which is precisely the BEC condition of the (bosonic) fermion
pairs. Therefore, it is transparent that \emph{the present pairing
  fluctuation theory unifies BCS theory and Bose-Einstein
  condensation} using the BCS-BEC crossover picture; They are two
sides of the same coin. Such a unification has not been made so
obvious in other competing pairing fluctuation theories. Technically,
it is the Taylor expansion Eq.~(\ref{eq:PG_exp}) that has made this
unification transparent; a similar expansion has not been seen in
competing theories.

Indeed, as shown in Fig.~\ref{fig:InvT}, for different $T\leq T_c$
(solid curves), the real part Re~$t_{pg}^{-1}(\Omega, \mathbf{q=0})$
always goes through the origin. However, for $T>T_c$ (dashed curves),
this is no longer true. The nonzero intercept $t_{pg}^{-1}(0,
0)=Z\mu_{pair}$ determines the effective pair chemical potential above
$T_c$.

In fact, there are various situations where we need to know the
approximate value of the pseudogap above $T_c$. In such a case, we
need to extend the gap equation (\ref{eq:gap}) or (\ref{eq:gap_amo})
to situations above $T_c$, as %$t^{-1}_{pg}(0,0)= Z \mu_{pair}$, i.e.,
\begin{equation}
\frac{m}{4\pi a} + \sumk \left[\frac{1-2f(\Ek)}{2\Ek} -
\frac{1}{2\ek} \right]\varphi_\mathbf{k}^2  =Z\mu_{pair} \,.
\label{eq:gap_aboveTc}
\end{equation}
The pseudogap equation is still given by Eq.~(\ref{eq:PG}) but with a
nonzero $\mu_{pair}$, along with the number equation
(\ref{eq:number}). Since $\Delta_{sc}=0$ above $T_c$, it is clear that
the three unknowns are now ($\Delta_{pg}=\Delta$, $\mu$,
$\mu_{pair}$), as compared to ($\Delta_{pg}$, $\mu$, $\Delta_{sc}$)
below $T_c$.

It should be pointed out that as $T$ increases above $T_c$, the $T$
matrix $t_{pg}(Q)$ no longer diverges at $Q=0$. Therefore,
Eq.~(\ref{eq:pgsimplification}) is no longer a good approximation for
the pseudogap self energy $\Sigma_{pg}(K)$. In this sense, the use of
the extended gap equation (\ref{eq:gap_aboveTc}) should be restricted
to a temperature regime not far above $T_c$, where $-\mu_{pair}$ is
still very small.

Generalization of the above equations to population imbalanced as well
as mass imbalanced situations is straightforward, which can be found in
Refs.~\cite{Chien06,ChienPRL,Chen2007PRB,Guo2009PRA,Wang2013,WangFFLO}.

Finally, a few remarks are in order. The pseudogap self energy given
in Eq.~(\ref{eq:sigmapg}) formally contains all contributions at the
$T$-matrix level. However, by Eq.~(\ref{eq:pgsimplification}), the
pseudogap self energy is approximated by a BCS-like, off-diagonal,
coherent form. When the pseudogap $\Delta_{pg}$ vanishes, the
pseudogap self energy is gone. Therefore, the diagonal incoherent
contributions are dropped out.  The incoherent contributions, $\delta
\Sigma (K)$, is dominant in the weak coupling BCS limit, and becomes
less important in the intermediate through strong coupling BEC
regimes. It mainly causes a chemical potential shift, as well as a
slight fermion mass renormalization. Such contributions are usually
neglected in the study of superconductivity. Nevertheless, for atomic
Fermi gases, as the strong couplings regime becomes accessible, it is
known that these contributions have a substantial \emph{quantitative}
impact on the so-called beta factor at unitarity \cite{Thomas}, which
is defined as $1 + \beta = \mu(0) /E_F $, where $\mu(0)$ and $E_F$ are
the zero $T$ chemical potential at unitarity and the noninteracting
Fermi energy, respectively. Without the incoherent contributions, the
present theory produces the same prediction as the BCS mean-field
result, $\beta \approx -0.41$, whereas the experimental values and
quantum Monte Carlo (QMC) simulation results are found between -0.5
and -0.7 \cite{ThermoScience-full,Salomon3,Carlson3}. When the
incoherent contributions are included, theoretical calculations of
Perali \textit{et al} found $\beta \approx -0.545$, in better
agreement with experiment, as expected. Here our attention is focused
mainly on the moderate and strong coupling regimes, where the
pseudogap effect is strong so that the incoherent self energy
contribution is less important and only causes minor
\emph{quantitative} corrections.

Despite the simple BCS form of the self energy, Eq.~(\ref{eq:SelfE}),
our result does include the contributions of pairing fluctuations, as
in Eq.~(\ref{eq:sigmapg}). It is the simplification via
Eq.~(\ref{eq:pgsimplification}) that encapsulates the fluctuations
into a single parameter, $\Delta_{pg}$, via an integration of the
fluctuation spectrum, as given in Eq.~(\ref{eq:PG_def}).

\subsection{Extended to Fermi gases in a trap}

When placed in a 3D isotropic harmonic trap, with a trapping potential
$V(r) = \frac{1}{2}m\omega^2r^2$, one can resort to the local density
approximation (LDA), by imposing a local chemical potential $\mu(r) =
\mu - V(r)$ and  the total particle number constraint, 
\begin{equation}
N = \int \mathrm{d}^3 r\, n(r) \,,
\label{eq:TrapN}
\end{equation}
where $n(r)$ is the local number density, and $\mu\equiv\mu_0=\mu(0)$
is the chemical potential at the trap center, often referred to as the
global chemical potential. Note that here the trap potential does not
necessarily have to be isotropic; it may be anisotropic with a
variable aspect ratio, including the quasi-2D pancake or quasi-1D
cigar shapes as the limit of a large aspect ratio. It may also be an
optical lattice, which we shall not cover in this review.

With LDA, at any given location, the fermions are subject to
pairing. Below $T_c$, there exists a superfluid core in the center of
the trap. Outside the core, the fermions may or may not be paired,
depending on their concrete radial position and the strength of the
pairing interaction. When the pairing correlation is strong, one
expects to find a pseudogap in the outskirt of the superfluid core.
Inside the superfluid core, the fermions locally satisfy the gap
equation as well as the pseudogap equation, while the local chemical
potential $\mu(r)$ determines the local density $n(r)$. Outside the
superfluid core, the fermions are in the normal state, so that the
effective pair chemical potential $\mu_{pair}$ becomes nonzero. In
this case, we need to use the extended gap equation
(\ref{eq:gap_aboveTc}) in place of equation (\ref{eq:gap}).
As mentioned earlier, the use of the extended gap equation
(\ref{eq:gap_aboveTc}) should be restricted to a temperature regime
not far above $T_c$.  In the trapped case, this translates into a
narrow shell outside the superfluid core. Nevertheless, as the density
gets lower towards the trap edge, the gap becomes small and the error
introduced into the total number $N$ via the local $n(r)$ is
negligible. Thus in our actual numerical calculations, we apply
Eq.~(\ref{eq:gap_aboveTc}) for the entire shell of Fermi gases outside
the superfluid core, and switch to unpaired normal Fermi gas state
when the gap becomes tiny, e.g., when $\Delta < 10^{-5}$.

\subsection{Thermodynamics and superfluid density}

The pseudogap and finite momentum pair excitations necessarily affect
the thermodynamic behavior and transport properties, such as the
superfluid density.  Away from the BCS regime, both Bogoliubov
quasiparticles and finite momentum pairs are present at finite
$T$. They serve to destroy the superfluid density and contribute to
the entropy. Knowing the excitation spectra, it is straightforward to
write down the entropy $S$, as a sum of fermionic ($S_f$) and bosonic
($S_b$) contributions. In a trap, the total entropy involves an
integral over the trap, given by $S=\int \mathrm{d}^3r\, s(r)$ (and
similarly for $S_f$ and $S_b$), where
\begin{eqnarray}
s &=&s_f + s_{b} \nonumber\,, \\
s_f &=& -2 \sumk [ f_k \ln f_k + (1-f_k) \ln (1-f_k)], \nonumber \\
s_{b} &=& - \sum_{q\ne 0} [b_q \ln b_q -(1+b_q) \ln (1+b_q) ] \,.
\end{eqnarray}
Here $f_k \equiv f(\Ek)$, and $b_q \equiv b(\Omegaq-\mu_{pair})$.  The
fermion contribution coincides \emph{formally} with the standard BCS
result for noninteracting Bogoliubov quasiparticles [although here
$\Delta(T_c) \neq 0$].  And the bosonic contribution is given by the
expression for non-directly-interacting bosonic pairs with dispersion
$\Omegaq$, with an effective mass $M^*$ which is not necessarily equal
to $2m$. When the chemical potential $\mu$ becomes negative, the
entropy $S$ becomes dominantly bosonic, since the fermionic part $S_f$
becomes exponentially suppressed. 

One can also write down the energy of the Fermi gas, which consists of
a fermionic and bosonic part in a similar fashion. Thus, in a trap,
the local energy is given by
\begin{eqnarray}
E &=& \mu n(r) + E_f + E_b \,, 
\label{eq:E}
\\
E_f &=& \sum_K (i\omega_n +\ek - \mu(r)) G(K) \nonumber\\ 
&=& \sumk [2\Ek f_k -(\Ek-\ek+\mu(r))]
+ \Delta^2 \chi(0) \,, \nonumber\\
E_b &=& \sum_q (\Omegaq-\mu_{pair}) b_q \,,\nonumber
\end{eqnarray}
where the pair susceptibility $\chi(0)$ is given by Eq.~(\ref{eq:chi})
at $Q=0$. One obtains the total energy by integrating
Eqs.~(\ref{eq:E}) over the entire trap.

To end this subsection, we present the expression for the superfluid
density, which can be derived using the linear response theory with a
generalized Ward identity \cite{ChenPhD,Kosztin1,Kosztin2}. In a
homogeneous case, it is given by
\begin{eqnarray} 
  \left(\frac{n_s}{m}\right) &=&
  \frac{2}{d}\sum_{\mb{k}}\frac{\Delta_{sc}^2}{\Ek^2} \left [
    \frac{1-2f(\Ek)} {2\Ek}+f^\prime(\Ek)\right] \nonumber\\
  &&\times{} \left [
    \left(\vec{\nabla}_\mathbf{k}^{}\xik^{}\right)^2 \phik^2 -\frac{1}{4}
    (\vec{\nabla}^{}_\mathbf{k}\xik^2)\cdot
    (\vec{\nabla}^{}_\mathbf{k}\phik^2)\right] \nonumber\\
  &=& \frac{\Delta_{sc}^2}{\Delta^2} \left(\frac{n_s}{m}\right)^{MF}
  \,,   
\label{eq:Ns}
\end{eqnarray}
where $d=3$ is the dimensionality, $f'(x)=\d f(x)/\d x$, and we have
kept $(n_s/m)$ as a combination since on a lattice the mass $m$ is not
well defined, but the combination is. The \emph{key result} here is
the last line in Eq.~(\ref{eq:Ns}), where $(n_s/m)^{MF}$ is the BCS
mean-field expression for $(n_s/m)$, which necessarily persists into
the normal state when a pseudogap exists above $T_c$. It is the
prefactor $\Delta_{sc}^2/\Delta^2$ that guarantees that there be no
Meissner effects above $T_c$. Indeed, within the present pairing
fluctuation theory, the superfluid density vanishes continuously and
nicely as $T$ approaches $T_c$ from below, following the $T$
dependence of $\Delta_{sc}^2$ in the vicinity of $T_c$. A population
imbalanced version of Eq.~(\ref{eq:Ns}) can be found in
Ref.~\cite{Stability}.
In a trap, all one needs to do is to integrate the local superfluid
density $n_s(r)$ over the entire trap, $N_s=\int \mathrm{d}^3r\, n_s(r)$.

\section{Key results of the present pairing fluctuation theory}

In this section, we will present some key results related to the
pseudogap phenomena. We first present the results for a dilute
two-component 3D Fermi gas with a short range $s$-wave pairing
interaction, which serves to demonstrate the simple physical picture,
and will be a basis of comparison for other cases. Next we shall
present the main results for the cuprate superconductors, and then
quickly switch to results relevant for atomic Fermi gases, which is
the main subject of this Review.

\begin{figure}
%\centerline{\includegraphics[width=3.4in,clip]{3DTccontact-InvkFa.eps}}
\centerline{\includegraphics[width=3.4in,clip]{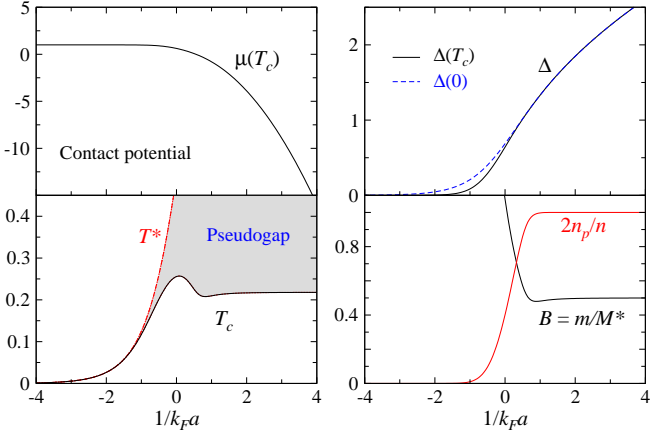}}
\caption{Calculated phase diagram and $T_c$, $T^*$,
  $\mu(T_c)$, $\Delta (0)$, $\Delta (T_c)$ and pair fraction $2n_p/n$
  and inverse pair mass $m/M^*$ (at $T_c$) of a 3D homogeneous Fermi
  gas with a contact potential as a function of $1/k_Fa$. Here $n_p$
  denotes the number density of pairs. }
\label{fig:3D}
\end{figure}

\subsection{Two-component homogeneous Fermi gases in the 3D continuum}

Figure \ref{fig:3D} summarizes the main results of the present theory
on the behavior of a 3D Fermi gas with a contact potential. Shown are
the phase diagram and related quantities, including $T_c$, $T^*$,
$\mu(T_c)$, $\Delta (0)$, $\Delta (T_c)$, as well as the pair fraction
$2n_p/n$ and the effective inverse pair mass $m/M^*$ at $T_c$. Here
the pair formation temperature $T^*$, as a crossover temperature, is
approximated by the mean field solution of $T_c$. While the $T_c$
curve is close to its mean-field counterpart in the weak coupling BCS
regime, a (shaded) pseudogap phase emerges in the intermediate
(crossover or unitary) through strong pairing BEC regimes. Along with
the BCS-BEC crossover, the fermionic chemical potential $\mu$
decreases from $E_F$ in the noninteracting limit, and approaches a
large negative given by $-E_b/2$ in the deep BEC regime, where $E_b$
is the two-body binding energy. At the same time, the excitation gaps
$\Delta(0)$ and $\Delta(T_c)$, at zero $T$ and $T_c$, respectively,
grow with $1/k_Fa$. While $\Delta(T_c)$ at $T_c$ roughly vanishes in
the BCS regime, it becomes nearly equal to $\Delta(0)$ in the BEC
regime. A scrutiny also reveals that the ratio $\Delta/|\mu|$
approaches 0 in the BEC limit, implying that in the deep BEC regime,
many-body effects are relatively unimportant so that pairing is
dominated by two-body physics. Indeed, the curve of $2n_p/n$ shows
that in the BEC regime, essentially all fermions form pairs. A
calculation of the pair size reveals that it shrinks in real space
with increasing pairing strength \cite{ChenPhD,ourreview}, leading to
a dilute Bose gas of tightly bound fermion pairs in the deep BEC
regime.

\begin{figure}
%\centerline{\includegraphics[width=3.05in,clip]{AtGapsg1-T.eps}}
\centerline{\includegraphics[width=3.05in,clip]{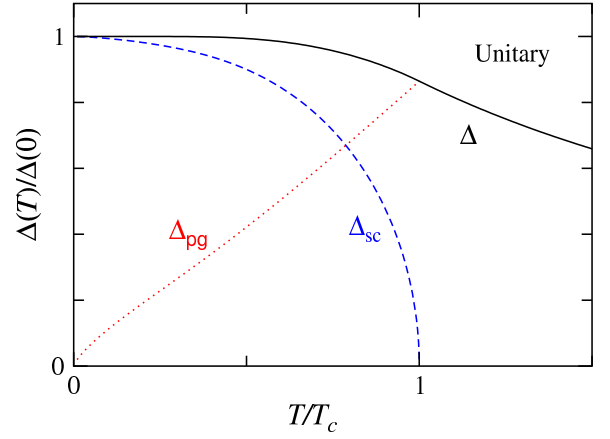}}
\caption{Normalized gaps as a function of reduced temperature $T/T_c$
  at unitarity. The gap at $T_c$ is comparable to the gap at
  $T=0$. The curves are calculated for a homogeneous 3D Fermi gas in
  continuum, with pairing symmetry $\phik^2 = 1/[1+(k/k_0)^2]$ at
  $k_0/k_F=4$.  }
\label{fig:Gaps}
\end{figure}

One feature that seems unique to the present theory is that the $T_c$
curve reaches a maximum near unitarity. At the same time, there is a
minimum where $\mu$ changes sign. In the BEC limit, $T_c$ approaches
its ideal BEC asymptote from below, as expected on physical grounds.
This nonmonotonic behavior of $T_c$ can be readily explained. Starting
from the intermediate pairing strength regime, the formation of pairs
quickly depletes the effective density of fermions, making the
effective fermionic density of state decrease, associated with a
shrinking Fermi surface. This leads to a decrease in $T_c$. On the
other hand, the bosonic part of the system emerges and grows, as given
by the increasing pair density $n_p$. Beyond the $\mu=0$ point, the
Fermi surface completely vanishes, and $n_p$ reaches its maximum value
$n/2$, so that $T_c$ is controlled by the BEC temperature, which
increases slowly with $m/M^*$.

It should be emphasized that, as shown by the $m/M^*$ curve, except in
the deep BEC regime, the effective pair mass differ significantly from
$2m$. This should be contrasted with NSR-based theories, which has
$M^* = 2m$ in all cases.

As one can see from Fig.~\ref{fig:3D}, in the pseudogap phase, the
pseudogap $\Delta(T_c)$ and the pair density $n_p$ grow hand in hand.

Figure \ref{fig:Gaps} illustrates the behavior of the gaps as a
function of temperature for a 3D homogeneous Fermi gas at
unitarity. The pseudogap at $T_c$ is close to the zero $T$ gap
$\Delta(0)$. For weaker coupling toward the BCS limit, the pseudogap
$\Delta_{pg}/\Delta(0)$ decreases and vanishes eventually. On the
contrary, with increasing pairing strength toward the BEC regime, the
ratio $\Delta(T_c)/\Delta(0)$ approaches unity so that the gap becomes
essentially temperature independent except at very high $T$. At low
$T$, following Eq.~(\ref{eq:PG}), the pseudogap scales as
$\Delta_{pg}(T) \propto T^{3/4}$.

\begin{figure}
%\centerline{\includegraphics[width=3.3in,clip]{Figure3-n_2.eps}}
\centerline{\includegraphics[width=3.3in,clip]{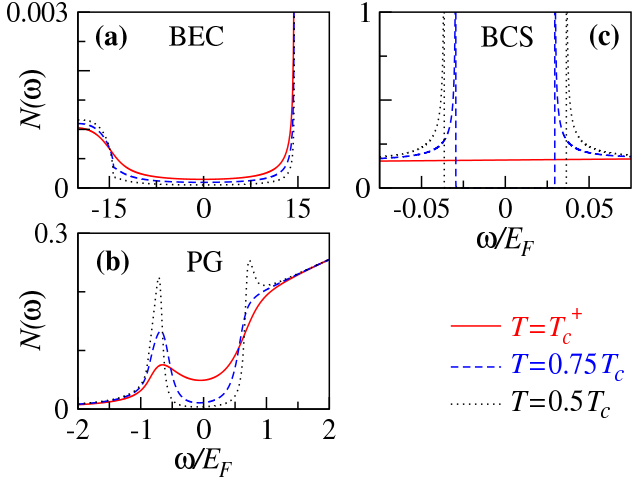}}
\caption{Fermionic density of states vs energy for the three regimes
  at three indicated temperatures from slightly above $T_c$ down to
  the superfluid phase. Note the big difference in the
  scales. Replotted from Ref.~\cite{JS2}.  }
\label{fig:DOS2}
\end{figure}

In Fig.~\ref{fig:DOS2}, we present the typical density of states
$N(\omega)=-2 \sum_{\bf k} {\rm Im}\, G(\omega+i0,{\bf k})$ for (a)
BEC, (b) BCS and (c) pseudogap regimes for a homogeneous 3D Fermi gas,
at different temperatures from slightly above $T_c$ down to $0.5T_c$,
half way into the superfluid phase. Note that to distinguish the
incoherent pair contributions from that of the condensate to the self
energy, we have used a more realistic form of the pseudogap self
energy,
\begin{equation}
\Sigma_{pg}(\omega, \mathbf{k}) \approx
\frac{\Delta^2_{pg}}{\omega+\xik + i\gamma}\,,
\end{equation}
where $\gamma$ is treated as a phenomenological parameter independent
of temperature \cite{footnoteongamma}. The very low but finite DOS for $\omega
< 0$ in the BEC regime is purely a consequence of particle-hole mixing
due to pairing. Except for the BCS regime, where the gap closes at
$T_c$, a pseudogap is already present at $T_c$ for both the BEC and
the pseudogap regimes.

From Fig.~\ref{fig:DOS2}, one can easily conclude that unless one has
a very high resolution in experiment, one can no longer use the
opening of a gap in the DOS as a signature of superfluid transition in
the presence of a pseudogap. Instead, it is a signature of pairing
which in general takes place before superfluid phase coherence sets
in. This is a very important effect of the pseudogap.

\begin{figure}
%\centerline{\includegraphics[width=3.3in,clip]{3DNsk04-T-g.eps}}
\centerline{\includegraphics[width=3.3in,clip]{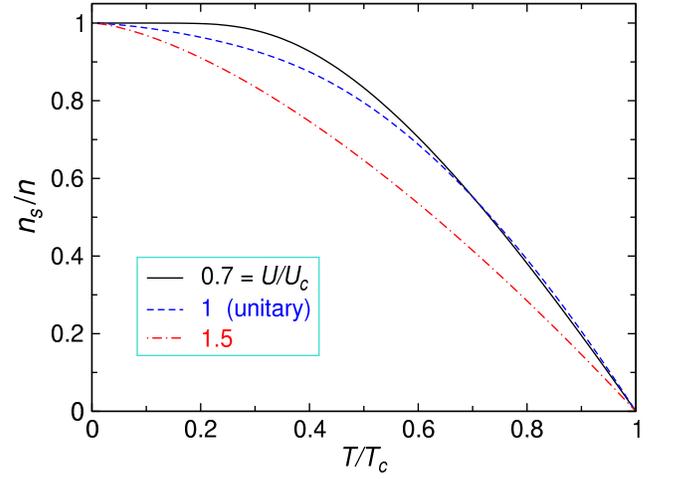}}
\caption{Normalized superfluid density $n_s/n$ of a 3D homogeneous
  Fermi gas as a function of $T/T_c$ for three different regimes. Here
  $U/U_c=1$ is equivalent to $1/k_Fa = 0$, the unitary limit. The
  pseudogap or finite momentum pairs contribute a $T^{3/2}$ power law
  to the low $T$ dependence, which becomes dominant as $U/U_c$
  increases. The calculation was done for an NSR type of potential,
  $\phik^2 = 1/[1+(k/k_0)^2] $ with $k_0/k_F = 4$.}
\label{fig:Ns}
\end{figure}

Shown in Fig.~\ref{fig:Ns} is the normalized superfluid density
$n_s/n$ in a 3D homogeneous Fermi gas as a function of the reduced
temperature $T/T_c$, for different pairing strengths $U/U_c = 0.7$,
1.0, and 1.5, corresponding to the BCS, unitary, and BEC regimes,
respectively. In comparison with the exponential $T$ dependence of the
BCS case (black solid curve), a clear deviation can be seen in the
unitary case (blue dashed curve) already. This is due to the bosonic
pair excitations, which obey the $T^{3/2}$ power law at low $T$. In
the BEC regime (red dot-dashed curve), the low $T$ behavior is
dominated by the $T^{3/2}$ power law of the bosonic
excitations. Indeed, in the BEC regime, fermionic quasiparticles are
essentially absent below $T_c$, due to the large negative chemical
potential $\mu$.

Figure \ref{fig:Ns} confirms that in our theory, due to the
generalized Ward identity \cite{Kosztin1}, Meissner effect is
necessarily absent above $T_c$. In this way, our superfluid density
vanishes nicely at $T_c$, unlike some competing scenarios (see, e.g.,
theories based on NSR) which predicts a first order jump or
nonmonotonic temperature dependence at $T_c$
\cite{Zwerger,Strinati8,Griffingroup2,Drummond3}.

\begin{figure}
%\centerline{\includegraphics[width=3.4in,clip]{Dwave_2.eps}}
\centerline{\includegraphics[width=3.4in,clip]{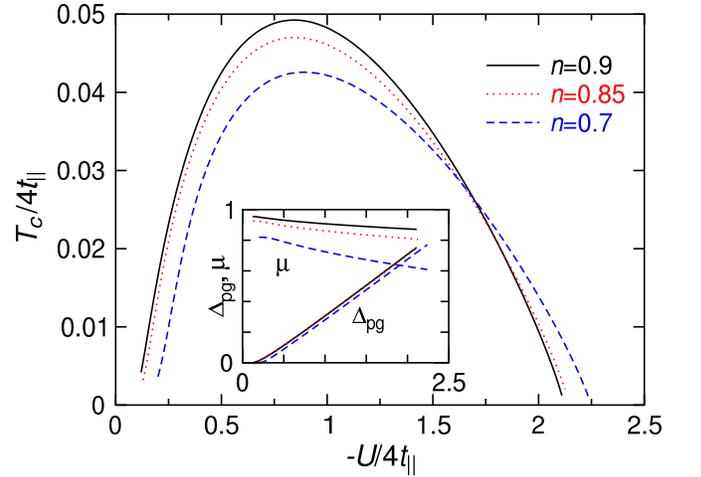}}
\caption{Superfluid transition temperature $T_c$ as a function of
  $-U/4t_\parallel$ for fermions on a quasi-2D square lattice, with a
  $d$-wave pairing symmetry, at density $n=0.9$ (black solid line),
  0.85 (red dotted), and 0.7 (blue dashed line). Shown in the
  inset are corresponding $\Delta(T_c)$ and $\mu$. The system is deep
  in the fermionic regime when $T_c$ vanishes, where the chemical
  potential $\mu$ is not far from its noninteracting value. Here
  $t_\perp/t_\parallel = 0.01$. Taken from Ref.~\cite{Chen1}.}
\label{fig:Dwave}
\end{figure}

\begin{figure*}
%\centerline{\includegraphics[height=3.42in,clip]{Cuprate_Phase_Fit.eps}
%\hskip 3ex
%\includegraphics[height=3.4in,clip]{RPP_71_062501_Fig2_600dpi_edited.eps}}
\centerline{\includegraphics[clip]{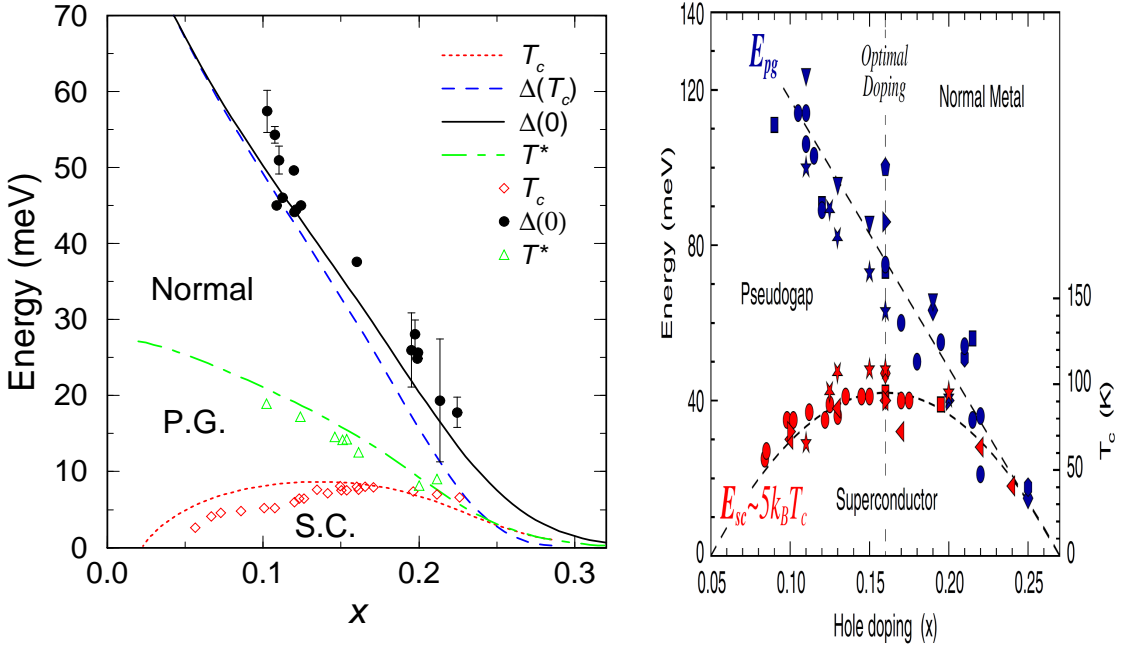}}
\caption[Cuprate phase diagram, and comparison with experiment.]
{ (Left) Cuprate phase diagram, taken from Ref.~\cite{ChenPhD},
  showing $\Delta (0)$, $T_c$, $\Delta (T_c)$, and $T^*$, calculated
  for $-U/4t_0=0.047$, and $t_\perp/t_\parallel =0.003$.  Shown as
  symbols are experimental data.  The normal, pseudogap, and
  superconducting phases are labeled with ``Normal'', ``P.G.'', and
  ``S.C.'', respectively.  (Right) Plot of a recent collection of
  experimentally measured pseudogap data (with $E_{pg} =
  2\Delta_{pg}$, blue symbols), taken from Ref.~\cite{Sawatzky}. The
  right axis shows the temperature scales (for $T_c$ and $T^*$). Note
  that the right panel has been horizontally squeezed so that it can
  be overlaid on top of the left panel in the range of $x= 0.05 \sim
  0.27$. These experimental pseudogap data in the right panel agree
  with the blue dashed curve for $\Delta_{pg} = \Delta(T_c)$ in the
  left panel very well. }
\label{fig:Cuprate_Phase}
\end{figure*}

\subsection{Application for the cuprates: quasi-2D superconductors on
  a lattice with a $d$-wave pairing symmetry}

When the pairing fluctuation theory is applied to a quasi-2D lattice,
it turns out that the lattice periodicity and the low dimensionality
bear important consequences. The periodic lattice imposes an upper
cut-off in the momentum space, and fermion pairs have to move via
virtual ionization. As a result, the superfluid transition temperature
$T_c$ scales as $t_\parallel^2/U$ at low densities, where
$t_\parallel$ is the in-plane nearest neighbor hopping integral, and
$U$ is the on-site attractive (pairing) interaction. At high
densities, calculations show that $T_c$ vanishes abruptly at an
intermediate pairing strength so that the BEC regime is not
accessible. For high $T_c$ superconductors, the $d$-wave pairing
symmetry further restricts the lower bound of the pair size to that of
a unit cell. Along with the non-local effect \cite{Kosztin_nonlocal}
of the $d$-wave pairing, the system is essentially always in the high
density regime so that $T_c$ vanishes abruptly at an intermediate
pairing strength. The high density strongly suppresses the motion of
the (finite size) pairs so that at certain point, the pairing strength
is so strong that the pairs become localized, with a diverging
effective mass $M^*\rightarrow \infty$.   More details
regarding the lattice, low dimensionality and $d$-wave effects may be
found in Ref.~\cite{Chen1}.

Shown in Fig.~\ref{fig:Dwave} are the $T_c$ curves for a $d$-wave
superconductor on a quasi-2D square lattice at relatively high
densities relevant to the cuprate superconductors. For all three
densities, $T_c/4t_\parallel$ shuts off around 2.2. The chemical
potential shown in the inset reveals that the system is still deep in
the fermionic regime when $T_c$ vanishes abruptly.  The $\Delta_{pg}$
curves show that the pseudogap effect is strong. More details
regarding the lattice, low dimensionality and $d$-wave effects may be
found in Ref.~\cite{Chen1}.

In Fig.~\ref{fig:Cuprate_Phase}, we present the theoretical cuprate
phase diagram calculated using this theory (lines) and compare with
experimental data (symbols) in the left panel. In our calculations, we
take $U$ to be doping independent, and incorporate the effect of the
Mott transition at half filling, by introducing a doping concentration
$x$ dependence into the in-plane hopping matrix elements $t_\parallel=
t_0 x$, as would be expected in the limit of strong on-site Coulomb
interactions in a Hubbard model \cite{Anderson87}. Therefore, except
the weak logarithmic dependence \cite{Chen1} of $T_c$ on the
anisotropy $t_\perp/t_\parallel$, there is only one free parameter,
i.e., $U/t_0$. Without further tweaking details such as next nearest
neighbor hopping $t'$, the agreement between theory and experiment in
terms of low $T$ gap $\Delta(0)$, $T^*$ and $T_c$ is remarkable. A
later collection of pseudogap at and above $T_c$ are shown in the
right panel, which is so scaled that a direct comparison can be
made by overlaying it on top of the left panel. Note that the
experimental $T_c$ data points in both panels fit the same empirical
formula $T_c = T_c^{\mbox{max}}[1 - 82.6(0.16 - x)^2 ]$ fairly well
\cite{JohnZ,Sawatzky}.

This remarkable (semi-)quantitative agreement between theory and
experiment really distinguishes the present theory from other rival
theories of high $T_c$ superconductivity. Despite the fact many
different high $T_c$ theories have been proposed, one finds it awkward
that it is hard to find a high $T_c$ theory that is capable of
quantitative computations.

\begin{figure}
%\centerline{\includegraphics[width=3.2in,clip]{TrapTcU0.89g35k050-kFa3.eps}}
\centerline{\includegraphics[width=3.2in,clip]{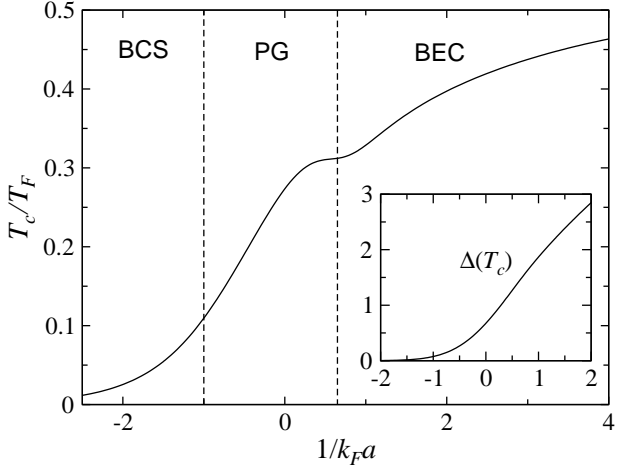}}
\caption{ Behavior of $T_c$ as a function of $1/k_Fa$ for a
  Fermi gas in a 3D isotropic harmonic trap with a short range
  potential. The plateau is clearly a residue of the maximum-minimum
  feature in the $T_c$ curve of the homogeneous Fermi gas. As $1/k_Fa
  \rightarrow +\infty$, $T_c$ approaches its BEC asymptote in a trap,
  $0.518 T_F$, where $T_F$ is the global Fermi temperature in the
  noninteracting limit. The inset shows the behavior of
  corresponding $\Delta(T_c)$. Here ``PG'' denotes pseudogap. }
\label{fig:3DTrap}
\end{figure}

\subsection{3D Fermi gas in an isotropic trap}

When the typical pair size or coherence length is far smaller than the
trap size (more precisely, the size of the Fermi gas cloud), LDA is a
good approximation. In Fig.~\ref{fig:3DTrap}, we present the solution
of $T_c$ under LDA as a function of $1/k_Fa$ for a Fermi gas in a 3D
isotropic harmonic trap with a contact potential. With no surprise, a
pseudogap is found to emerge as the pairing strength grows. The
behavior of the pseudogap at $T_c$ is shown in the inset. Near
unitarity, the plateau in the $T_c$ curve is clearly a residue of the
maximum-minimum feature in the $T_c$ curve of the homogeneous Fermi
gas. Meanwhile, due to the shrunk cloud size and thus increased
density at the trap center in the BEC regime, $T_c$ approaches a much
greater BEC asymptote in a trap, $0.518 T_F$, as $1/k_Fa \rightarrow
+\infty$, as compared to its homogeneous counterpart, $0.218
T_F$. Note that in a trap the global $T_F$ is defined by the Fermi
temperature in the noninteracting limit. While the Fermi gas locally
satisfies the gas equation as if it were in a homogeneous case, it is
easy to conclude that from the weak coupling BCS limit through the deep BEC
limit, the central density $n(0)$ is enhanced by a factor of
$(0.518/0.218)^{3/2}=3.66$ by the pairing interaction. (Here we have
made use of the relation $E_F\propto n^{2/3}$ for a homogeneous Fermi gas).

\begin{figure}
%\centerline{\includegraphics[width=3.2in,clip]{DensityGapProfilesT0.5-InvkFa.eps}}
\centerline{\includegraphics[width=3.2in,clip]{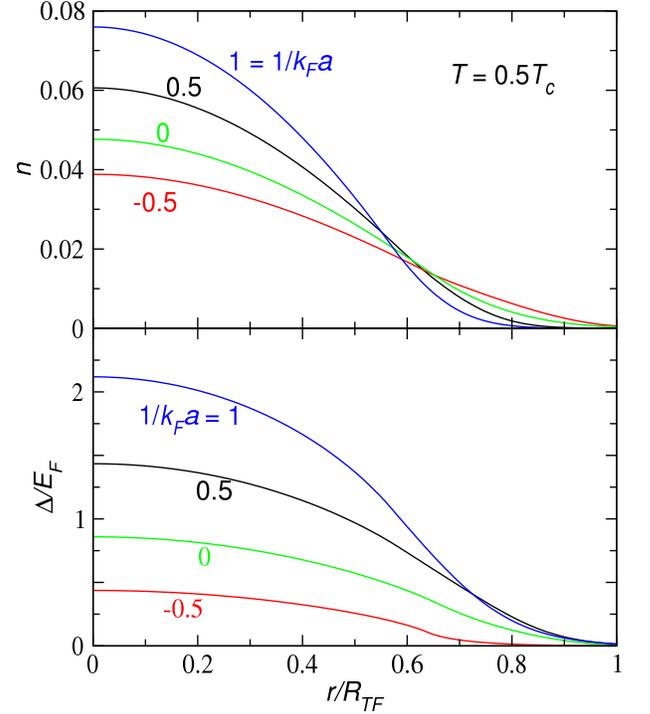}}
\caption{ (a) Density and (b) gap profile of a Fermi gas
  in a 3D harmonic trap for various pairing strengths from (near-)BCS
  to BEC, as labeled. All curves are calculated half way below their
  corresponding $T_c$. Here $R_{TF}$ is the Thomas-Fermi (TF) radius,
  given by the zero $T$ radius in the noninteracting limit.}
\label{fig:DensityGapProfile}
\end{figure}

Shown in Fig.~\ref{fig:DensityGapProfile} in the evolution of the
spatial density and gap profiles in the 3D harmonic trap as a function
of the pairing strength. For illustration purpose, it suffices to
focus in the near-BCS through near BEC regimes, without going to the
extreme BCS or BEC limits. Indeed, it is the crossover or unitary
regime, where the scattering length becomes large, that has been the
focus of most studies. In the weak coupling limit, the scattering
length is proportional to the interaction strength. For this reason,
the unitary regime has often been referred to (mainly by the AMO
community) as ``strongly interacting''.  As can be seen from the
figure, as the pairing strength increases, the Fermi gas cloud shrinks
toward the trap center (upper panel), where the density necessarily
increases as a result. At the same time, the spatial distribution of
the pairing gap (lower panel) also becomes more focused at the trap
center, despite its growing with the pairing strength.

To distinguish the present theory from competing theories, it is worth
mentioning that the density profiles are smooth spatially, and evolve
monotonically with temperature. This should be contrasted with the
theory of Strinati and coworkers \cite{Strinati4}, which predicts
non-monotonic radial dependence and non-monotonic temperature
dependence. 

\begin{figure}
%\centerline{\includegraphics[width=3.in,clip]{TrapU0.89g35k050nu1-dens-T4.eps}}
\centerline{\includegraphics[width=3.in,clip]{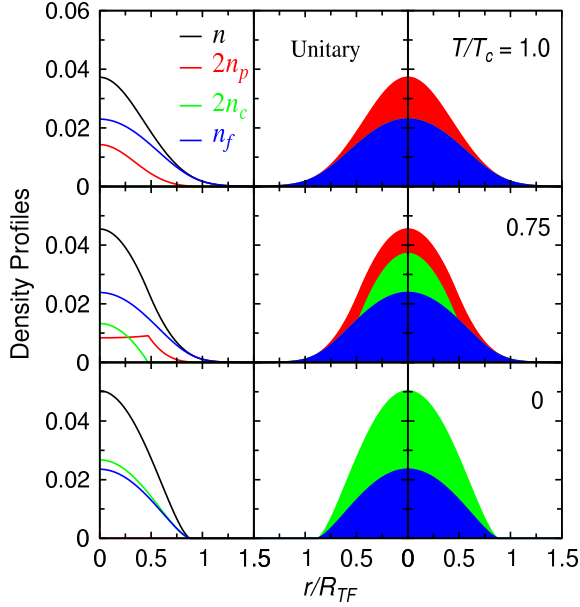}}
\caption{ Decomposition of the density profile $n(r)$ of
  a Fermi gas in a trap at unitarity for representative temperatures
  $T/T_c = 1$, 0.75, and 0, as labeled. At $T_c$, there are only
  finite momentum pairs (red) and fermions (blue). Below $T_c$, the
  condensate (green) develops, and the finite momentum pair
  contributions decreases. The pair density $n_p$ is nearly uniform
  inside the superfluid core. At $T=0$, finite momentum pairs
  disappear and all pairs are condensed. }
\label{fig:DensityDecomp}
\end{figure}

It should also mentioned that mean-field calculations predict as a
signature of superfluidity a kink at the edge of the superfluid core
in the density profile \cite{JasonHo,Chiofalo}.  Such a kink is absent
in our theory as well as experimental observations.  Indeed, it can be
easily shown that \cite{ChenPhD}
\begin{eqnarray}
n &=& 2Z\Delta^2 + 2\sumk f(\xik)  = 2 Z\Delta_{sc}^2 + 2 Z\Delta_{pg}^2
+ 2\sumk f(\xik)\nonumber\\ &\equiv& 2n_{c} + 2n_{p} + n_f \,,
\label{eq:n_decomp}
\end{eqnarray}
where $n_c \equiv Z\Delta^2_{sc}$ is the number density of condensed
Cooper pairs \cite{noteonpairdensity}, $n_f \equiv 2\sumk f(\ek -
\mu(r))$ is the density of fermions as though they were free. In
Fig.~\ref{fig:DensityDecomp}, we show the density profile $n(r)$
(black curve) and its component contributions from the condensate
$2n_c$ (green), finite momentum pairs $2n_p$ (red) and free fermions
$n_f$ (blue), for three representative temperatures $T/T_c = 1$, 0.75,
and 0. The right columns show the (de-)composition of the density. At
$T/T_c = 0.75$, the density profile is composed of all three
components. It is evident that the contribution of finite momentum
pairs (red area) is essential in eliminating the kink, which would exist
otherwise at the edge of the superfluid core (green area). It is worth
mentioning that finite momentum pair density $n_p$ (red curve) is
nearly flat inside the superfluid core.  This is because
$\mu_{pair}=0$ and the effective pair mass is nearly the same across
the core. Then $n_p$ starts to decrease gradually outside the core,
when $-\mu_{pair}$ acquires a finite value and grows with radius. At
$T_c$, the superfluid core disappears. On the other hand, at $T=0$,
the finite momentum pairs disappear; all pairs are condensed. In the
BCS mean-field theory, where $\Delta_{pg}=0$ and the fermion
propagator contains no self energy feedback, Eq.~(\ref{eq:n_decomp})
reduces to $n = 2n_c + n_f$. In fact, this is how the density was
decomposed in Refs.~\cite{JasonHo,Chiofalo}. Figure
\ref{fig:DensityDecomp} shows that without finite momentum pairs,
there would be an unphysical kink at the edge of the superfluid core.

\begin{figure}
%\centerline{\includegraphics[width=3.in,clip]{cond_fraction.eps}}
\centerline{\includegraphics[width=3.in,clip]{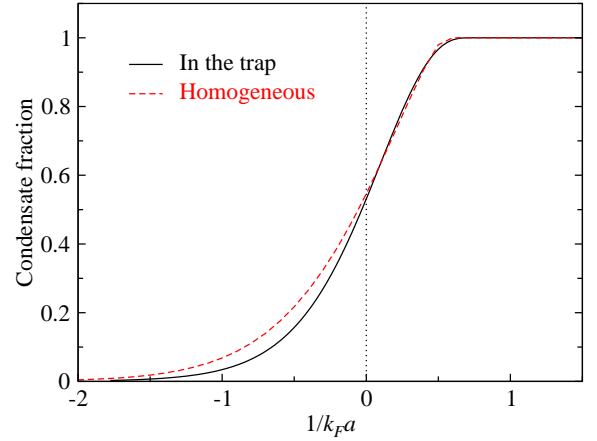}}
\caption{ Condensate fraction at $T=0$ as a function of
  $1/k_Fa$ in a trapped (solid line) and homogeneous (red dashed line)
  Fermi gas with a short-range potential, based on the decomposition
  given by Eq.~(\ref{eq:n_decomp}) and Fig.~\ref{fig:DensityDecomp}. }
\label{fig:condensatefraction}
\end{figure}

Interestingly, it is worth mentioning that under this decomposition,
the condensate fraction (green area) $2n_c/n$ is not 100\% at $T=0$ at
unitarity since $\mu > 0$, even if the superfluid density is. This
result is shown in Fig.~\ref{fig:condensatefraction}, where the
condensate fraction is plotted as a function of pairing strength for
the entire BCS-BEC crossover for a short range potential in both the
trapped (black solid) and homogeneous (red dashed line) cases.  As one
can imagine from Eq.~(\ref{eq:n_decomp}), the figure shows that the
condensate fraction does not rise to 100\% until the BEC regime is
reached, where $\mu$ changes sign and becomes negative. To understand
the small condensate fraction in the BCS regime, one notices that,
based on Eq.~(\ref{eq:n_decomp}), at zero $T$ this fraction covers the
rest part that is not accounted for by the $n_f$ term. Therefore,
\emph{it is a measure of the extent to which a fermion lives a life as
  a component of a Cooper pair rather than an individual fermion.} At
unitarity, the condensate fraction is about 0.55 in the homogeneous
case, and 0.53 in the trap, respectively. These numbers are close to
that from quantum Monte Carlo simulations \cite{Giorgini2005}, 0.57
for a total particle number $N=66$.

\begin{figure}
%\centerline{\includegraphics[width=3.4in,clip]{TrapS-kFa6PRL3.eps}}
\centerline{\includegraphics[width=3.4in,clip]{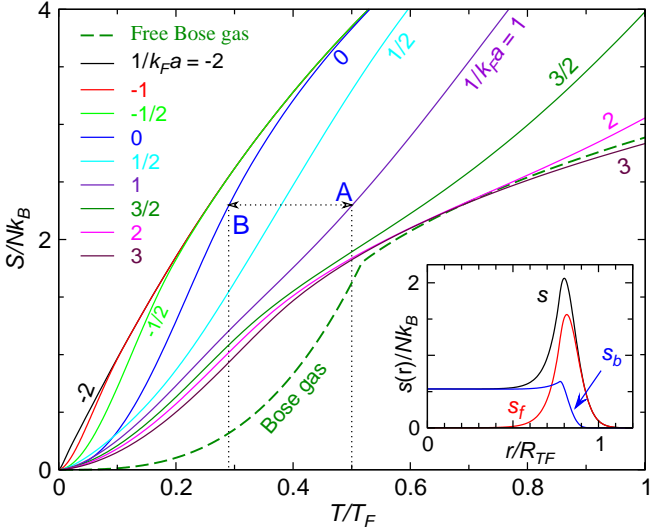}}
\caption{ Entropy per atom as a function of $T$ for
  different pairing strengths from weak coupling BCS through strong
  coupling BEC in a harmonic trap.  The dotted lines show an
  isentropic magnetic field sweep between $1/k_Fa=1$ and
  unitarity. For comparison, the dashed line represents the ideal Bose
  gas.  The inset plots the spatial profile of total entropy $s(r)$
  (black curve) and its fermionic ($s_f$, red) and bosonic ($s_b$,
  blue) component contributions at unitarity for $T=T_c/4$.  Here $T_c
  =0.27T_F$. Reproduced from Ref.~\cite{ChenThermo}. }
\label{fig:entropy}
\end{figure}

The effects of a pseudogap or finite momentum pair contributions on
thermodynamics is summarized by Fig.~\ref{fig:entropy}, where the
entropy per particle is shown for a series of pairing strengths from
BCS through BEC for a Fermi gas in a trap \cite{noteon2channel}. The
black curve for $1/k_Fa = -2$ is close to a noninteracting Fermi gas,
exhibiting a linear $T$ dependence at low $T$. In the opposite strong
coupling BEC regime, the $1/k_Fa = 3$ curve is close to the ideal Bose
gas curve, above the BEC asymptote for $T_c/T_F$, $0.518$. At high $T$
(but $\ll T^*$), it is easy to guesstimate from the figure that the
entropy in the deep BEC regime is given roughly by half that for a
free Fermi gas. The existence of finite momentum pairs allows a
continuous evolution from the Fermi gas limit through the Bose gas
limit, as the pairing strength increases. The presence of the trap
inhomogeneity inevitably makes the situation more complicated that its
homogeneous counterpart, leading to a power law $T$ dependence at low $T$
for all pairing strengths.

At unitarity, the distributions of the fermionic and bosonic
components of the entropy are shown in the inset. At low $T$ in the
broken symmetry, superfluid phase, the bosonic contribution $s_b(r)$
(blue curve) is nearly flat inside the superfluid core, and decays
outside the core. On the other hand, the fermionic part, $s_f(r)$,
comes mainly from the edge of the Fermi gas cloud, where the pairing
gap becomes very small. The sum $s(r)$ has a peak at the trap edge as
well. Considering the phase space factor $r^2$ in the trap integral,
the behavior of the entropy $S$ at unitarity is dominated by the
fermionic component at the trap edge.  As the system evolves deep into
the BEC regime, the fermionic part becomes negligible so that the
bosonic part eventually dominates.

The above thermodynamics behavior has an immediate consequence. It can
be used as a thermometry. It is well known that the temperature
measurement in a Fermi gas is notoriously difficult. It is essentially
impossible to measure the temperature at an arbitrary interaction
strength. Measurement of the temperature in a Fermi gas without a
population imbalance has been done successfully only in the BCS limit,
deep BEC limit and at unitarity. In practice, it is convenient to
connect the actual temperature at a given $1/k_Fa$ with the
temperature in the non-interacting limit, using an adiabatic,
isentropic magnetic field sweep. In other words, one can use the
entropy in place of the temperature.  As an example, the dotted lines
in Fig.~\ref{fig:entropy} shows how to connect the temperatures at
$1/k_Fa$ and at unitarity.

\begin{figure}
%\centerline{\includegraphics[width=3.4in,clip]{Fig3n.eps}
\centerline{\includegraphics[width=3.4in,clip]{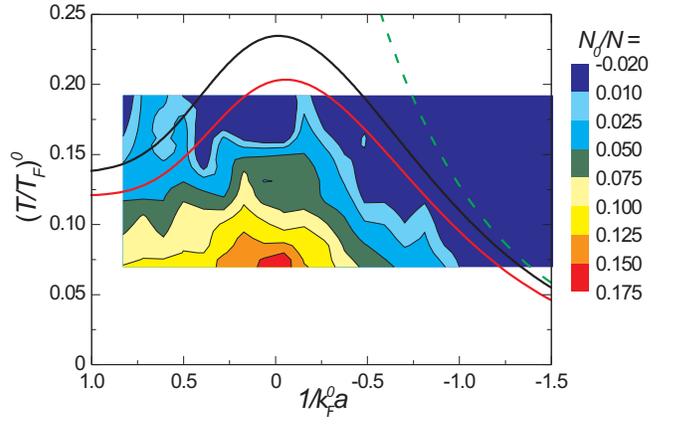}
}
\caption{ Phase diagram of $^{40}$K. A contour plot of
  the measured condensate fraction $N_0/N$ as a function of $1/k_F^0a$
  and effective temperature $(T/T_F)^0$ in the noninteracting limit is
  compared with theoretically calculated contour lines at $N_s/N=0$
  (at $T_c$, black curve) and 0.01 (red curve).  Despite the large
  uncertainty in experimental data, the overall trend of the
  experimental contour of $N_0/N=0.01$ and the theoretical line for
  $N_s/N=0.01$ are in good agreement.  The dashed line represents the
  naive BCS result $T_c/T_F^0 \approx 0.615 e^{\pi/2k_F^0a}$.  Here
  $k_F^0 \equiv k_F$ and $T_F^0\equiv T_F$ are the global Fermi
  momentum and Fermi temperature, respectively. Reproduced from
  Ref.~\cite{Jin_us}. }
\label{fig:phasediagram}
\end{figure}

As an application of the above pseudogap related thermometry, in
Fig.~\ref{fig:phasediagram} we plot the theoretically calculated phase
diagram of $^{40}$K in a trap with an effective temperature
$(T/T_F)^0$ measured adiabatically in the non-interacting limit, and
compare with the experimental phase diagram from Jin's group
\cite{Jin4,Jin_us}. The black and red curves are $T_c$ and the
$N_s/N=0.01$ contour line, respectively. The experimental data show
the contour plot of the condensate fraction. Given the large error bar
in the data, we note that the overall trend of the experimental
contour of $N_0/N=0.01$ and the theoretical line for $N_s/N=0.01$ are
in good agreement \cite{noteoncontourplot}. One may also compare the
$T_c$ curve with that shown in Fig.~\ref{fig:3DTrap} to see directly
the difference between $T_c/T_F$ and $(T_c/T_F)^0$ as an effect of the
adiabatic isentropic sweep.

\section{Experimental evidence of the pseudogap in atomic Fermi gases}

The concept of pseudogap was first introduced into atomic Fermi gases
in Ref.~\cite{JS2}. It was not accepted and well understood by the AMO
community, until more and more indirect and direct experimental probes
provided evidence for its existence. In this section, we shall present
evidence of the pseudogap in atomic Fermi gases from various
experiments, especially in the unitary regime.

Due to the extreme low $T$ and extreme small size as well as charge
neutrality, the choice of experimental probes to ascertain the
existence of the pseudogap is very limited for trapped Fermi
gases. Typical condensed matter probes such as resistivity
measurement, optical conductivity, penetration depth measurement, and
angle-resolved photoemission spectroscopy (in the conventional sense)
are not available. Therefore, one often has to resort to indirect
measurements.

\begin{figure}
%\centerline{\includegraphics[width=3.2in,clip]{AllT4.eps}
\centerline{\includegraphics[width=3.2in,clip]{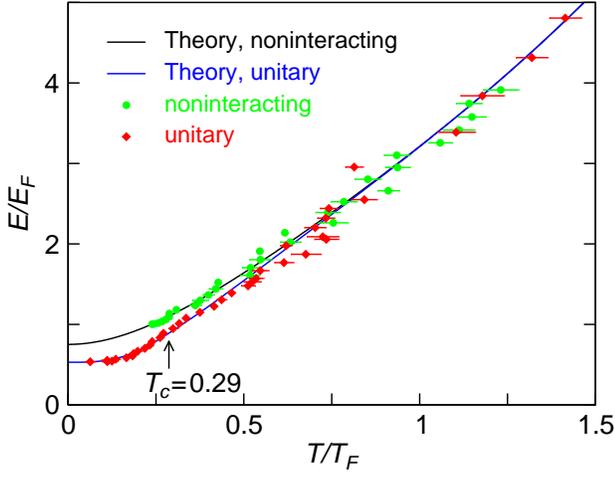}
}
\caption{ Comparison between theoretical calculations
  (lines) and experimental measurements of the energy per atom as a
  function of $T/T_F$ for noninteracting and unitary $^6$Li
  gases. Here the temperature for the unitary case involves a
  temperature calibration \cite{ThermoScience-full}. The unitary and
  the noninteracting energy data do not merge until about $T^* \approx
  0.6 T_F$. Reproduced from Ref.~\cite{ThermoScience-full}. }
\label{fig:Energy}
\end{figure}

\begin{figure}[b]
%\centerline{\includegraphics[width=3.2in,clip]{Profiles20140825.eps}
\centerline{\includegraphics[width=3.2in,clip]{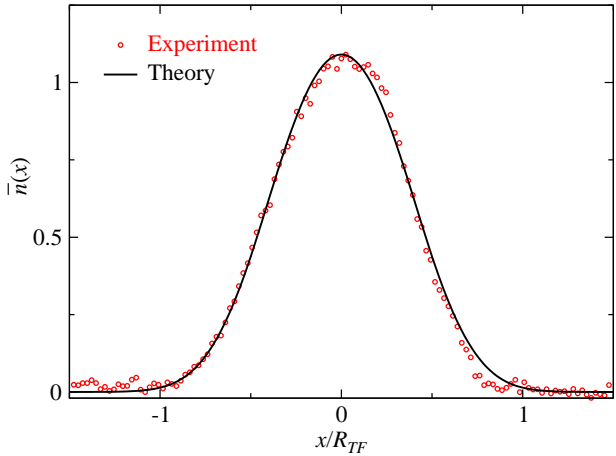}
}
\caption{ Comparison between theoretical calculations
  (black solid line) and experimental measurements (red circles) of
  the one-dimensional density profile of a unitary $^6$Li gas at $T =
  0.19T_F\approx 0.7T_c$, in the superfluid phase.  Reproduced from
  Ref.~\cite{JS5}. }
\label{fig:Profile}
\end{figure}

\subsection{Thermodynamics and density profiles}

Shown in Fig.~\ref{fig:Energy} is the energy per atom for a unitary
$^6$Li Fermi gas. The lines are calculations of the present pairing
fluctuation theory, while the symbols are experimental data from the
Thomas group \cite{ThermoScience-full} at Duke University. The result
for noninteracting Fermi gases serves as a calibration of the
experimental measurement, where the Thomas-Fermi (TF) approximation
works well. It is evident that the theory and experiment agree very
well. It is worth mentioning that, in both the noninteracting and
unitary cases, a finite trap depth as given by the experiment was used
in order to arrive at the good agreement at high $T$. One of the most
important messages one can read off the figure is that the unitary
energy curve does not rise to that of the noninteracting curve until
$T^*\approx 0.6 T_F \gg T_c \approx 0.29T_F$ from the theory. This is
a manifestation of the existence of a pseudogap above $T_c$ at
unitarity, which helps to lower the energy.

If the energy curve $E(T)$ provides a signature of the pseudogap above
$T_c$ at unitarity, the spatial density profile below $T_c$ may serve
as \emph{indirect} evidence of the pseudogap below $T_c$. In
Fig.~\ref{fig:Profile} we present a comparison of the one-dimensional
density profile $\bar{n}(x) = \int \d z\d y\, n(r)$ of a unitary Fermi
gas between theory and experiment, at a temperature substantially
below $T_c$. The agreement is good. There is no sign of the kink
behavior at the edge of the superfluid core in the data
\cite{noteon1Ddensity}. We stress that such a good agreement is not
expected for a mean-field theory \cite{JasonHo,Chiofalo} or a theory
that exhibits non-monotonic dependence in radius or temperature
\cite{Strinati4}. As shown in Fig.~\ref{fig:DensityDecomp}, the
pseudogap or finite momentum pair contributions are essential in
arriving at such a smooth density profile.

\begin{figure}
%\centerline{\includegraphics[width=3.2in,clip]{Nk2dU-2.44g62K30Tf0.67_comp-T-kFa3c.eps}}
\centerline{\includegraphics[width=3.in,clip]{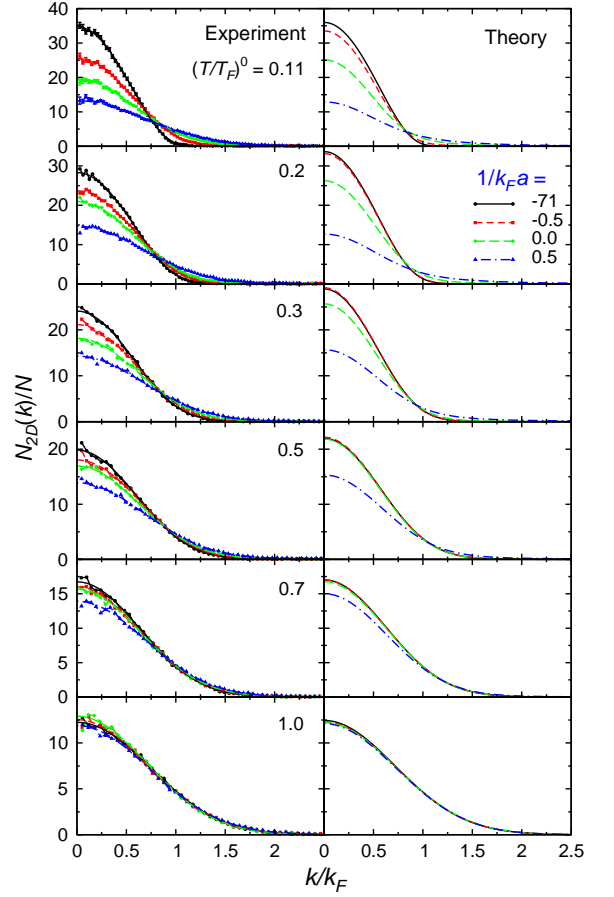}}
\caption{ Temperature evolution of the 2D momentum
  distribution $N_{2D}(k)$ of a $^{40}$K gas in a trap with different
  pairing strengths from noninteracting through near-BEC cases. The
  effective temperature in the noninteracting limit, $(T/T_F)^0$, is
  labeled. Reproduced from Ref.~\cite{Jin2_us}. }
\label{fig:Nk}
\end{figure}

\subsection{Momentum distribution}

The presence of a pseudogap necessarily has an important consequence
on the momentum distribution $N_k$ of the Fermi gases. Without a
pseudogap, $N_k$ would behave very much like a noninteracting Fermi
gas at a given temperature. In contrast, it will be spread to a larger
range in the momentum space, according to 
\begin{eqnarray}
N_{2D}(k) &=& \int\frac{\d k_z}{2\pi}\, N_k = \int \frac{\d k_z}{2\pi}\,\d^3 r\, n_k(r)\,,\nonumber\\  n_k(r)
&=& 1-\frac{\xik}{\Ek} + 2\frac{\xik}{\Ek} f(\Ek) \,.
\label{eq:N2dk}
\end{eqnarray}
In Fig.~\ref{fig:Nk}, we present the 2D momentum distribution
$N_{2D}(k)$ at a series of temperatures from below to far above $T_c$
for different pairing strengths from BCS through (near-)BEC
regimes. The left and right columns are from experiment and theory,
respectively. The comparison reveals a good agreement between
them. The most important message here is that
the $N_{2D}(k)$ curves differs significantly for different pairing
strengths at $(T/T_F)^0 = 0.3$ and 0.5, above $T_c$, and they do not
merge until  $(T/T_F)^0 > 0.7$ where the (pseudo-) excitation gap
disappears. Like the $(T/T_F)^0 = 0.11$ case, the difference between
the curves for different pairing strengths is caused by the presence of
the (pseudo-)gap.

\begin{figure}
%\centerline{\includegraphics[width=3.2in,clip]{RFTransitions2.eps}
\centerline{\includegraphics[width=3.2in,clip]{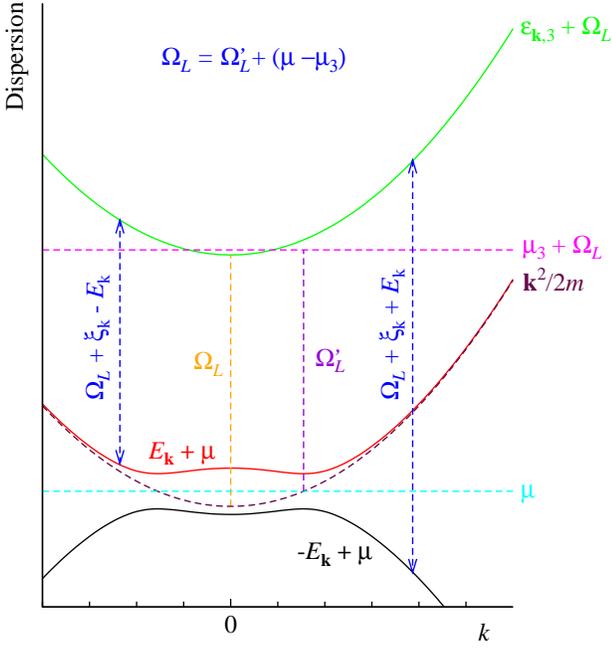}
}
\caption{ Energy levels in an RF transition. $\Omega_L$
  is the RF frequency for exciting a free atom from hyperfine level 2
  (maroon dashed curve) to level 3 (green solid
  curve). $\Omega_L^\prime$ is the same energy but measured relative
  to the respective chemical potentials. The black and red solid
  curves are the dispersion of the condensed and thermally excited
  quasiparticle branches of a paired atom in level 2, with energy
  level given by $\mp \Ek + \mu$, respectively. Reproduced from
  Ref.~\cite{Chen_RPP}. }
\label{fig:RFTransitions}
\end{figure}

\subsection{(Momentum integrated) radio frequency spectroscopy}

Among various experimental techniques, radio-frequency (RF)
spectroscopy \cite{Grimm4,Ketterlepairsize,KetterleRF} is arguably the most
direct probe for the existence of an excitation gap. The basic physics
of a RF process is shown in Fig.~\ref{fig:RFTransitions}. Atoms in
hyperfine levels 1 and 2 are subject to the pairing interaction,
whereas atoms in level 3 are free of such pairing. Therefore, by
exciting an atom from level 2 to an unoccupied level 3, one can tell
how much extra energy is needed in addition to the hyperfine level
splitting. This extra energy, referred to as detuning, provides a
measure of the ``binding energy'' of the level 2 atoms due to
interactions. To the lowest order, the RF current is given by
\begin{equation}
I(\nu)=\left. -\frac{1}{2\pi}\sumk A(\mathbf{k},
\omega)f(\omega)\right|_{\omega = \xik-\nu} \,,
\label{eq:RFcurrent}
\end{equation}
where $\nu$ is the detuning, $A(\mathbf{k},\omega)$ is the spectral
function for level 2 atoms, and we have set the RF matrix element to
unity. As is well known, when level 2 atoms are paired with level 1
atoms, the spectral function $A(\mathbf{k},\omega)$ consists of two
branches, the condensed and thermally excited quasiparticle branches,
represented in Fig.~\ref{fig:RFTransitions} by the black and red solid
curves, respectively. The thermal branch corresponds to negative
detuning, and will not be observable for either very low or very high
$T$. The former case will be suppressed by the Fermi function
$f(\omega)$, while the high temperature will destroy the pairing in
the later case.  When interactions exist between level 3 atoms and
level 1 atoms, then we have a final state effect, which will also
affects the RF spectrum \cite{Baym2,Baym3,Punk,Strinati7,Basu,ourRF3}.

\begin{figure}
%\centerline{\includegraphics[width=2.4in,clip]{Chin2.eps}
\centerline{\includegraphics[width=2.4in,clip]{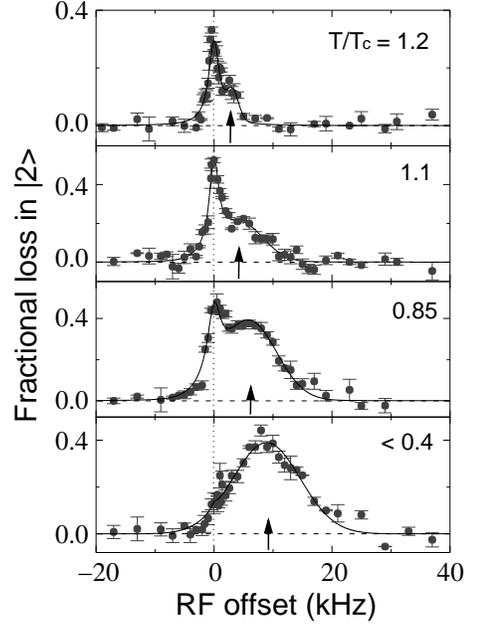}
}
\caption{ RF spectra of a unitary $^6$Li gas for
  different temperatures. The solid lines are fits to guide the
  eye. The vertical dotted line marks the atomic transition, and the
  arrows indicate the peak location of the pairing signal. The
  original effective temperature $T'/T_F$ has been converted
  isentropically to the real (reduced) temperature $T/T_c$ using the
  entropy data shown in Fig.~\ref{fig:entropy}. Reproduced from
  Ref.~\cite{Grimm4}. }
\label{fig:RFChin}
\end{figure}

Shown in Fig.~\ref{fig:RFChin} is the earliest report on the RF
spectroscopy measurement of a unitary $^6$Li gas at different
temperatures. The fractional loss is proportional to
$A(\mathbf{k},\omega)$ and the RF offset is the detuning $\nu$. The
effective temperature $T'/T_F$, measured in the BEC limit after an
isentropic sweep, has been converted to the real temperature $T/T_c$,
using the calculated entropy data shown in Fig.~\ref{fig:entropy}. The
important feature of this figure is the double peak structure, with
one narrow sharp peak at zero detuning, and a broad peak with a
positive detuning. The narrow peak can be easily attributed to the
transition from the free level 2 atoms, found at the edge of the
trap. On the other hand, the broad peak has been associated with
paired level 2 atoms. The trap inhomogeneity necessarily leads to a
distribution in the pairing gap, and thus the broadness of the RF
signal. The phase space factor $r^2$ in the trap integral determines
that the pairing signal will be peaked at an intermediate radius. At
unitarity, for a given gap $\Delta$, the detuning would be a momentum
average of
\begin{equation}
\nu = \Ek + \xik  \geq \sqrt{\mu^2 + \Delta^2} - \mu < \Delta\,,
\end{equation}
with the spectral weight given by the integrand of
Eq.~(\ref{eq:RFcurrent}), namely, the momentum dependent RF
current. While quantitatively, the location of the broad pairing peak
does not give directly the pairing gap, qualitatively, its presence is
a signature of pairing. As revealed by Fig.~\ref{fig:RFChin}, the
broad peak can already be seen above $T_c$ at $T/T_c =1.2$.  The total
spectral weight under the broad peak as well as the detuning for this
peak increases as $T$ decreases further. Deep in the superfluid phase
at  $T/T_c < 0.4$, the free atom peak is essentially gone; all level 2
atoms are paired, and the pairing peak detuning reaches its maximum.

\begin{figure}
%\centerline{\includegraphics[clip,width=2.6in,height=2.8in]{822G.eps}}
\centerline{\includegraphics[clip,width=2.6in,height=2.8in]{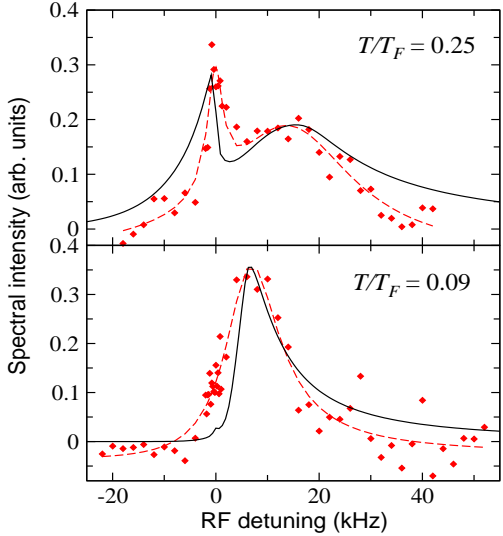}}
\caption{Comparison of calculated RF spectra (solid curves, $T_c
  \approx 0.29T_F$) with experiment (symbols) in a harmonic trap
  calculated at 822 G for the two lower temperatures. The temperatures
  were chosen based on Ref.~\cite{Grimm4}.  The dashed lines are a
  guide to the eye. Reproduced from Ref.~\cite{heyan}.}
\label{fig:822G}
\end{figure}

Recent experiment\cite{Zwierlein2011} and QMC results
\cite{TroyerPRL2006,TroyerPRL2008,Wingate} suggest that for $^6$Li at
unitarity, the transition temperature $T_c/T_F \approx 0.17$,
substantially lower than $0.29$ predicted in the present theory. This
further substantiates the existence of a pseudogap above $T_c$.

The experimental result of Ref.~\cite{Grimm4} was interpreted
successfully \cite{Torma2,heyan} using the present pairing fluctuation
theory soon after its publication. In Fig.~\ref{fig:822G}, we present
a comparison of calculated RF spectra (solid curve, $T_c \approx
0.29T_F$) with experiment (symbols) in a harmonic trap calculated for
$^6$Li at 822~G (on the BEC side of the Feshbach resonance at 834~G)
for the two lower experimental temperatures. The overall agreement is
satisfactory, which can be further improved by including the final
state effect \cite{Chen_RPP}. While our focus remains on qualitative
evidence of the pseudogap, we shall not go into details about the
final state effect. Interested readers may find further information in
Refs.~\cite{ourRF3,Chen_RPP,Basu,Baym2,Punk,Strinati7}.

\subsection{Momentum resolved radio frequency spectroscopy}

Despite the very intuitive picture about the double peak structure in
the RF spectra, the momentum integration has caused some disputes
regarding the origin of the double peak structure
\cite{Ketterlepairsize,Stoof3}, and thus the physical interpretation about
the RF spectroscopy measurements. This has a lot to do with the final
state effect, first noticed by Muller and coworkers \cite{Basu}. 
The lack of simple relation between the pairing gap size and the pairing
peak location in the RF spectra has made it difficult to extract the
gap from the data quantitatively.

\begin{figure}
%\centerline{\includegraphics[clip,width=3.3in]{mu0.62PG0.7gm0.1T0.5_contour.eps}}
\centerline{\includegraphics[clip,width=3.3in]{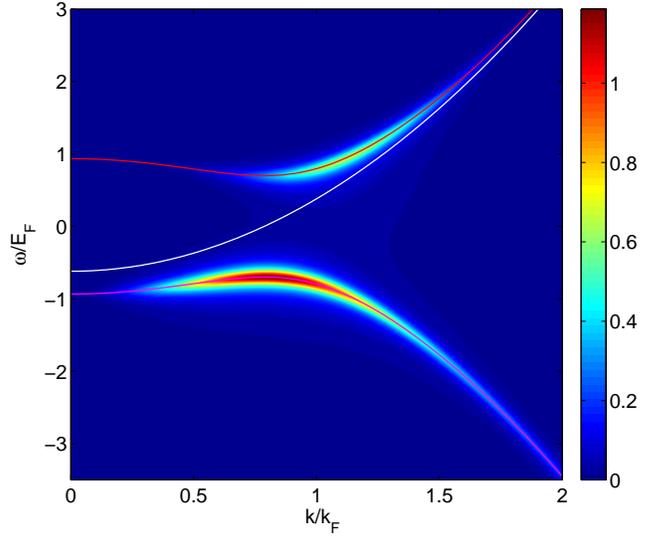}}
\caption{. Contour plots of the occupied spectral
  intensity at unitarity in a homogeneous Fermi gas for $T/T_c \approx
  1.9$. The population of the two branches are determined
  self-consistently. The white curve represents the dispersion of
  unpaired atom.  Reproduced from Ref.~\cite{Chen_MRRF}.}
\label{fig:RFhomo}
\end{figure}

A great step forward was made by Jin and coworkers \cite{Jin6}, who
performed momentum resolved RF (MRRF) spectroscopy measurement for the
first time, in a $^{40}$K gas. The RF current is given by the
integrand of Eq.~(\ref{eq:RFcurrent}). It turns out that the MRRF
spectroscopy is equivalent to the ARPES
\cite{arpesanl1,arpesstanford}, which is a very important and useful
tool in condensed matter physics. Further simplification comes from
the fact that there is no final state effect in a $^{40}$K gas. This
makes the interpretation of the MRRF spectra relatively simple and
unambiguous. In Fig.~\ref{fig:RFhomo}, we present the contour plot of
the occupied spectral intensity in the $\omega$~--~$k$ plane, for a
homogeneous 3D Fermi gas at unitarity at $T/T_F=0.5$. The two branches
corresponding to the condensed and thermally excited quasiparticles
shown in Fig.~\ref{fig:RFTransitions} are clearly visible. For
comparison, the white curve shows the dispersion of a free atom, with
the same chemical potential. The particle-hole mixing as a pairing
effect is evident, as manifested by the avoided crossing and
back-bending of both the lower and upper branches. This back-bending
takes place at $k=\sqrt{2m\mu} < k_F$.  To see the upper branch
clearly, one needs to have relatively high temperature which is
comparable with $\Delta$.

Ideally, one would like to have a homogeneous system. Unfortunately, a
trap potential is necessary in order to hold the Fermi gas
together. This complicates the otherwise very simple interpretation of
the RF spectra.

\begin{figure}
%\centerline{\includegraphics[clip,width=1.6in]{IntensityContour_sml2.eps}\
%  \ 
%\includegraphics[clip,width=1.65in,height=1.6in]{UnitarySg0.35-0.02gm0.35-0.02R0.3T1.1contour_3.eps}}
\centerline{\includegraphics[clip]{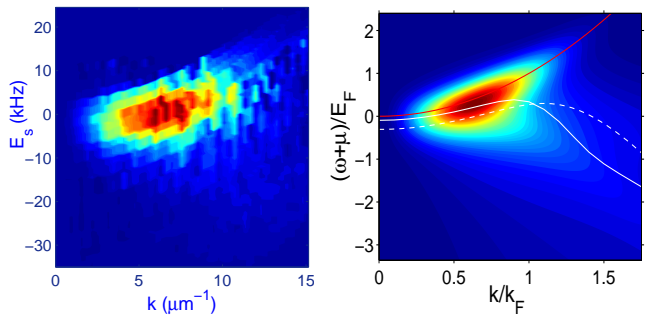}}
\caption{ Comparison between experiment (left) and
  theory (right) of the contour plots of the occupied spectral
  intensity in a unitary Fermi gas in a trap at $T/T_c \approx
  1.1$. The red curve represents the free atom dispersion.  A large
  amount of spectral weight has been shifted from a free atom peak to
  a paired atom peak (downward dispersing curve). Here the intensity
  increases from dark blue for 0 to dark red for the maximum, and the
  solid and dashed white lines indicate the loci of the peaks in the
  energy distribution curves from theory and experiment,
  respectively. The single particle energy $E_s$ is equivalent to
  $\omega + \mu$, the energy measured relative to the bottom of the
  band.  Reproduced from Refs.~\cite{Jin6,Chen_MRRF}.}
\label{fig:MRRF}
\end{figure}

In Fig.~\ref{fig:MRRF}, we present a comparison of the key result on
the spectral intensity map between experiment (left) and theory
(right) for a unitary Fermi gas above $T_c$ at $T/T_c = 1.1$. The
similarity between the two panels is obvious. As can be seen, a large
fraction of the spectral weight has been shifted from the free atom
branch to the paired atom branch. Indeed, at high $T$ where pairing
effect is negligible, the spectral weight concentrates on the free
particle dispersion (not shown). As the temperature decreases, a
second (downward dispersing) branch emerges.  This lower branch is
associated with the breaking of a pair and necessarily contains trap
averaging effects. With decreasing temperature, the intensity map
first bifurcates and eventually at very low $T$ becomes dominated by
the lower branch, when essentially all atoms are paired.

From Fig.~\ref{fig:MRRF}, such bifurcation and downward dispersion
already take place above $T_c$, indicating unambiguously that a
pseudogap exists in the unitary Fermi gas. To extract this downward
dispersion, the energy distribution curves (EDCs) have been fitted to
a single Gaussian function experimentally. This leads to the white
dashed curve in the right panel. The white solid dispersion curve is
obtained theoretically following the same procedure.  A BCS-like fit
to this dispersion can be used to determine the pairing gap, as has
been done in Ref.~\cite{Jin6}, as $E_s = \mu -\sqrt{\xik^2
  +\Delta^2}$. The best fit to the experimental dispersion yields
$\Delta = 9.5$~kHz, comparable to $E_F$. The agreement between
experiment and theory is reasonably good, despite the trap
inhomogeneity. The white dashed experimental curve back bends at
$k>k_F$, in contradiction to what is expected on physical ground at
unitarity. This is mainly caused by the low experimental resolution
and the incorrect single-Gaussian function fitting procedure. Our
theory predicts double peaks in the EDC curves for $k\geq k_F$, and
this has been confirmed by careful inspection of the experimental data
\cite{Chen_MRRF}.  Further improved experimental data
\cite{JinPrivate} have led to a dispersion much closer to the
theoretical result. The observation of a pseudogap has also been
confirmed by Refs.~\cite{JinStrinati_nphys,StrinatiJin}.

It should be mentioned that via the simple approximation
Eq.~(\ref{eq:pgsimplification}), the present pairing fluctuation
theory has demonstrated in a simple, analytic way that a pseudogap
necessarily exists when the pairing interaction is strong. For
NSR-based theories, due to the inconsistency between the gap equation,
which contains no pairing fluctuation contributions, and the number
equation, one would have to extract the pseudogap from the
renormalized spectral function in a cumbersome way. In this sense, the
numerical route of Strinati \textit{et al} \cite{Strinaticuprates,Strinati2}
can be viewed simply as a confirmation of our analytically result.

\subsection{Population imbalanced Fermi gases}

In this subsection, we provide evidence for the existence of a
pseudogap in population imbalanced Fermi gases. For extension of the
present pairing fluctuation theory to the case of population
imbalance, we refer the readers to
Refs.~\cite{Chen2007PRB,Chien06,ChienPRL}

\begin{figure}
%\centerline{\includegraphics[clip,width=3.2in]{TP_Trap_Ia0_2_MIT.eps}}
\centerline{\includegraphics[clip,width=3.2in]{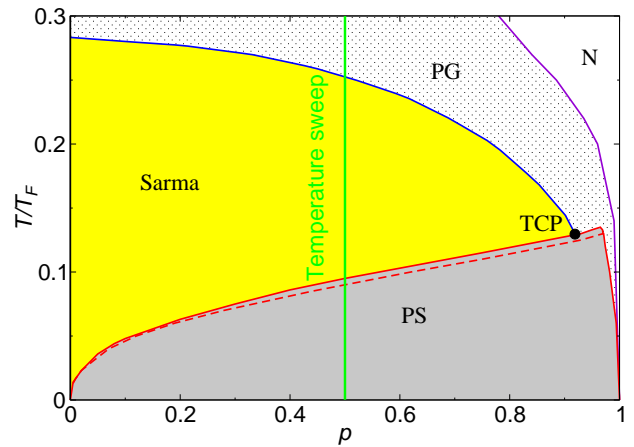}}
\caption{ Calculated phase diagram of a population
  imbalanced Fermi gas in a trap at unitarity.  Here ``PS'' stands for
  phase separation, ``Sarma'' for polarized Sarma superfluid, ``PG''
  for pseudogapped normal state, ``N'' for unpaired normal state, and
  ``TCP'' for tricritical point.  Reproduced from
  Ref.~\cite{ChienPRL}.}
\label{fig:PhasePopImb}
\end{figure}

Shown in Fig.~\ref{fig:PhasePopImb} is the calculated phase diagram of
a population imbalanced Fermi gas in a trap at unitarity. The overall
polarization, $p = (N_\uparrow - N_\downarrow)/N$, is different from
its local counterpart. At low $T$, phase separation (PS) takes place,
where a BCS superfluid core of an equal spin mixture at the trap
center is surrounded by polarized Fermi gases. Above the PS phase,
there exist intermediate temperature superfluids, which is referred to
as Sarma superfluid, for which the local spin polarization penetrates
all the way into the trap center. Above the Sarma phase, there is a
pseudogap phase (PG) where pseudogap exists without superfluidity,
before the system becomes unpaired normal state (N) at high $T$.

\begin{figure}
%\centerline{\includegraphics[clip,width=2.8in]{TP_Trap_Ia0_2_MIT.eps}}
%\centerline{\includegraphics[clip,width=3.in]{3p_Ia0_n1n2_new2.eps}\;\;}
%\centerline{\includegraphics[clip,width=3.4in]{PRL97_030401_Fig7c.eps}}
\centerline{\includegraphics[clip,width=3.3in]{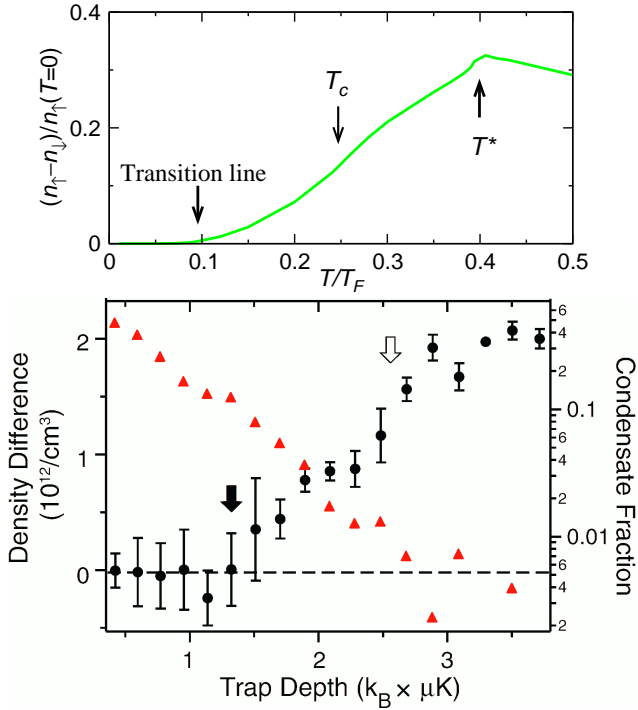}}
\caption{ Upper panel: Relative density difference at
  the trap center at low $T$ as a function of $T/T_F$ at $p=0.5$,
  i.e., along the vertical green line in
  Fig.~\ref{fig:PhasePopImb}. The three arrows indicate the PS/Sarma
  boundary, the Sarma superfluid/PG transition $T_c$, and the PG/N
  crossover temperature $T^*$, respectively. Lower panel: Experimental
  data from MIT, showing the density difference (black solid circles,
  left axis) at the trap center of a unitary Fermi gas at $p=0.5$,
  along with the condensate fraction (red triangles, right axis), as a
  function of trap depth, which is proportional to $T$. The solid and
  empty arrows indicate the PS/Sarma transition temperature and
  superfluid $T_c$, respectively.  Reproduced from
  Refs.~\cite{ChienPRL,MITPRL06}.}
\label{fig:PopImb}
\end{figure}

The behavior of the polarization at the trap center, $(n_\uparrow -
n_\downarrow)/n_\uparrow(T=0)$, in a temperature sweep at $p=0.5$, is
shown in the upper panel of Fig.~\ref{fig:PopImb}. An important
feature here is that its evolution across $T_c$ is smooth, without
a clear signature of the superfluid transition. A downturn is
predicted at the crossover temperature $T^*$, where the pseudogap
becomes negligible. We emphasize that this smooth evolution across
$T_c$ is a consequence of the fact that the total excitation gap is
continuous across $T_c$. This feature has been verified by
experimental data from the Ketterle group \cite{MITPRL06}, as shown in
the lower panel. Note that the experimental trap depth is proportional
to the temperature. The agreement between experiment and theory is
remarkable. Therefore, we conclude that the experimental data have
provided strong support for the existence of a pseudogap above $T_c$.

\subsection{Dispute against the existence of a pseudogap}

Despite the definitive evidence of the pseudogap from various
experiments, especially the MRRF spectroscopy measurements, there have
still been some disputes against the existence of the pseudogap from
thermodynamics measurement
\cite{Salomon2010Nature,SalomonPRL106}. Especially, in
Ref.~\cite{Salomon2010Nature}, Salomon and coworkers reported ``a
$T^2$ dependence of the pressure with temperature'', and thus claimed
that ``This behavior is reminiscent of a Fermi liquid, and indicates
that pseudogap effects expected for strongly interacting Fermi
superfluids do not show up at the thermodynamic level within our
experimental precision.'' \emph{However, this cannot be used as
  evidence against the existence of a pseudogap}, because the
macroscopic quantity pressure used in their equation of state (EOS)
involves a trap and momentum integration over all microscopic
states. There are many possible microscopic states which can produce
the same macroscopic thermodynamic quantities after integration. This
is a many-to-one mapping. For example, within the BCS mean-field
theory, the relation between pressure and energy of a Fermi gas with a
contact potential at unitarity is given by $p/E = 2/3$, exactly the
same as that for a noninteracting Fermi gas (which exhibits an ideal
Fermi liquid behavior). In fact, their key experimental data were
taken at $(T/\mu)^2 > 0.1$, or equivalently, $T/\mu > 0.3$. This is
far from being a low $T$ regime, where one can talk about power law
dependence. At such a high temperature, it is not particularly useful
to extract its power law dependence on $T$. The pressure calculated
with a pseudogap would follow a similar $T$ dependence in this
temperature regime, just like that of the energy (per particle),
$E(T)$.

\section{Where to look further for the pseudogap}

\subsection{Effects of particle-hole fluctuations}

As in most other theories of BCS-BEC crossover, e.g., the NSR-based
theories, the particle-hole channel has been dropped in the treatment
of the present pairing fluctuation theory. This is justified in the
context of superconductivity, where the particle-hole channel mainly
contributes to a change in the chemical potential, which can be taken
from experiment. In addition, superfluidity and pairing concerns
primarily the particle-particle channel.  In many cases, the pairing
interaction strength is not precisely known, and thus may be treated
as a fitting parameter. Nevertheless, when the pairing interaction
strength is indeed known precisely, one may need to consider the
effect of particle-hole fluctuations. 

In the weak coupling limit, the contribution of particle-hole
fluctuations was first considered by Gor'kov and Melik-Barkhudarov
(GMB) \cite{GMB} to the leading order. They found that both $T_c$ and
zero temperature gap are suppressed by a \emph{big} factor of
$(4e)^{1/3}\approx 2.22$. A few others have recently considered
particle-hole fluctuations within the context of Fermi gases and
BCS-BEC crossover
\cite{Heiselberg2000,KimTorma2009,TormaPethick2009,Yin2009}.  In
Ref.~\cite{ParticleHoleChannel}, the present pairing fluctuation
theory is extended to include the particle-hole channel in such a
fashion that the $T$-matrices of both the particle-particle channel
and the particle-hole channel intertwine with each other and are
treated self-consistently. The main result is that in the BCS through
unitary regime where the chemical potential $\mu >0$, particle-hole
fluctuations cause an effective reduction of the pairing strength. In
particular, the unitary limit is shifted towards the BEC regime, to
$1/k_Fa \approx 0.35$. The original maximum $T_c$ at unitarity (see
Fig.~\ref{fig:3D}) has now occurred at the new location. This seems to
be in good agreement with the QMC result from
Ref.~\cite{TroyerPRL2008}, which reported a maximum $T_c/E_F\approx
0.25$ around $1/k_Fa = 0.47$. Depending how the interaction parameter
is determined, this seems to suggest that one may need to consider
looking for the pseudogap around the new unitary limit in future
experiments.  Further details of the effect of particle-hole
fluctuations may be found in Ref.~\cite{ParticleHoleChannel}.

\begin{figure}
%\centerline{\includegraphics[clip,width=3.2in]{BtTrapU5g800K50nu0_Nb-T3.eps}}
\centerline{\includegraphics[clip,width=3.2in]{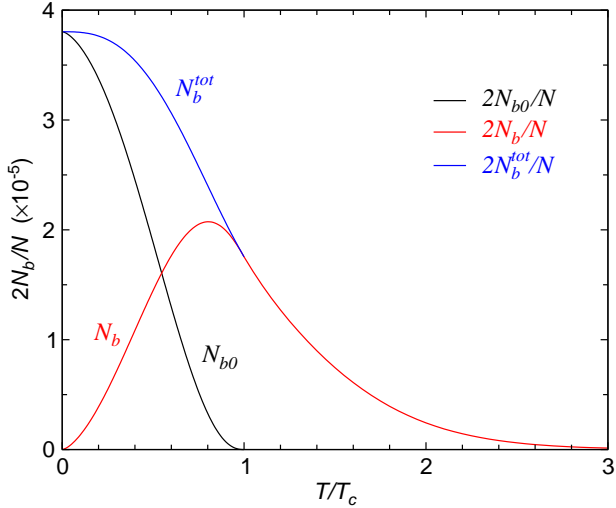}}
\caption{ Closed-channel fraction as a function of
  $T/T_c$ at unitarity for $T_F = 0.4\mu$K for $^6$Li in a harmonic
  trap. The black, red, and blue curves are the condensed
  ($2N_{b0}/N$), noncondensed ($2N_b/N$) and total ($2N_b^{tot}/N$)
  fractions, respectively. Here $T_c=0.273T_F$.  Reproduced from
  Ref.~\cite{ChenClosed}.}
\label{fig:Closed}
\end{figure}

\subsection{Widespread pseudogap phenomena}

There are widespread pseudogap related phenomena in ultracold Fermi
gases. In this section, we shall only name a few examples, instead of
giving a full search. 

For a wide Feshbach resonance such as the widely studied resonances in
$^6$Li and $^{40}$K, the closed channel fraction has turned out to be
closely related to the pseudogap. In Fig.~\ref{fig:Closed}, we show
the closed-channel fraction as a function of $T$ for a unitary Fermi
gas, calculated using a two-channel version of the present pairing
fluctuation theory \cite{ChenClosed}.  Here the black ($N_{b0}$) and
red ($N_b$) curves stand for the condensed and thermal part of the
closed-channel molecules, while the blue curve ($N_b^{tot}$) stands
for the sum.  They are proportional to $\Delta_{sc}^2$,
$\Delta_{pg}^2$, and $\Delta^2$, respectively. At low $T$, the
calculated fraction $2N_b^{tot}/N$ as a function of pairing strength
is in good quantitative agreement with experiment
\cite{ChenClosed,Hulet4}. It is known that at unitarity, pairing can
exist only due to many-body effect. Above $T_c$, should there be no
pairing (or equivalently, pseudogap), the closed-channel fraction
would drop to zero due to the inter-channel coupling. Therefore,
detection of the closed-channel fraction above $T_c$ should provide a
direct measurement of the pseudogap.

\begin{figure}
%\centerline{\includegraphics[clip,width=3.2in] {Unitary_BEC_M6.40.eps}}
\centerline{\includegraphics[clip,width=3.2in] {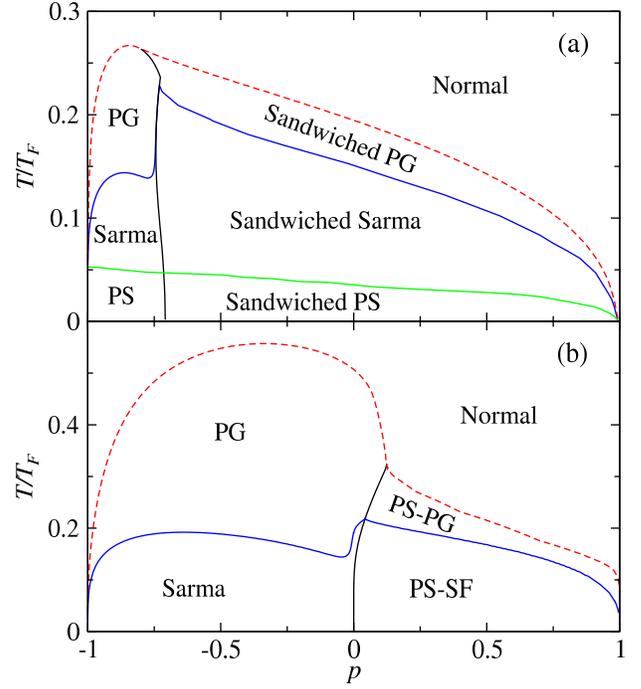}}
\caption{$T$--$p$ Phase diagram of a $^{6}$Li-$^{40}$K
  mixture in a harmonic trap at (a) unitarity and (b) $1/k_Fa = 0.5$,
  with $\omega_{\uparrow}=\omega_{\downarrow}$.  The solid lines
  separate different phases, and the (red) dashed line is approximated
  by mean field calculations. We choose the population imbalance $p>0$
  when $^{40}$K is the majority. Here ``PG" and ``PS" indicate
  pseudogapped normal state and phase separation, respectively, and
  ``SF'' stands for superfluid. The PS-PG and PS-SF phase has an
  ``inverted'' two-shell structure, with a normal gas of the majority
  heavy atoms at the trap center, surrounded by a superfluid or
  pseudogapped paired Fermi-Fermi mixture. Reproduced from
  Ref.~\cite{Wang2013}.}
\label{fig:MassImb}
\end{figure}

The pseudogap phenomena can be found not only in equal mass two
component Fermi gases, but also in a Fermi-Fermi mixture, with a
strong mass imbalance. In Fig.~\ref{fig:MassImb}, we present a phase
diagram for a $^{6}$Li-$^{40}$K mixture in a harmonic trap (a) at
unitarity and (b) in the near-BEC regime $1/k_Fa = 0.5$, with
$\omega_{\uparrow}=\omega_{\downarrow}$. Here $\omega_\sigma$ is the
angular frequency of the spin dependent trapping potential. The
convention is such that $p>0$ when the heavy species is the
majority. The ``PS'', ``Sarma'' and ``PG'' phase in
Fig.~\ref{fig:MassImb}(a) at unitarity is similar to that in
Fig.~\ref{fig:PhasePopImb}. As usual, at the highest temperature, the
system behaves like a mix of uncorrelated normal Fermi
gases. Otherwise, the dominant part of the phase diagram at unitarity
is a sandwiched three-layer shell structure, for which the inner and
outer shells are normal Fermi gases, while the mid-shell is either a
BCS, Sarma superfluid or a PG normal state, for a temperature from low
through high. In the near BEC case in Fig.~\ref{fig:MassImb}(b), the
PS phase for $p<0$ is no longer stable and the PG and Sarma regions
expands substantially. For $p>0$, the outer shell of a majority Fermi
gas has disappeared so that the system becomes an ``inverted''
two-shell structure, with a majority Fermi gas at the trap center,
surrounded by a superfluid or pseudogapped paired Fermi-Fermi mixture
in the outer shell.  Such two-shell or three-shell structures and the
local density profiles may be probed using the \emph{in situ}
phase-contrast imaging and 3D image reconstruction technique as in
Ref.~\cite{MITPRL06}. One may also use vortex lattices
\cite{KetterleV} to detect the (sandwiched) PS and Sarma state, so as
to distinguish the superfluid and pseudogapped phases.  The paired
state may also be detected using a Bragg spectroscopy technique
\cite{GuoPRL105,ValePRL112}, which may also be able to distinguish condensed
versus noncondensed pairs.

It should be emphasized that the sandwiched PG and the PS-PG phases
are both very unusual and very interesting. The associated phase
separation involves pseudogapped normal state rather than a superfluid
state. Such phase separations have never been predicted or reported
before in the literature.

There are many other experiments and physical quantities which exhibit
pseudogap phenomena. For example, atomic Fermi gases on optical
lattices will be another realm to search for the pseudogap in the
future, despite that experiment on optical lattices falls far behind
theory. Another realm is 2D Fermi gases, since low dimensionality
intrinsically enhances fluctuations, including the pseudogap related
pairing fluctuations.  It is expected that more support for the
existence of a pseudogap will come up as new experiments become
available.

\section{Summary}

In summary, we have given a review of the study of the pseudogap
physics in atomic Fermi gases, and presented a series of experimental
evidence of the existence of a pseudogap in Fermi gases, especially in
the unitary regime. In this context, we have introduced a pairing
fluctuation theory, and have shown that it thus far has addressed
successfully multiple atomic Fermi gas experiments. In particular, the
momentum resolved radio frequency spectroscopy measurement has
provided the most direct probe and the most convincing evidence of the
pseudogap.  Since the existence of a pseudogap is a natural
consequence of the present theory, and most competing theories do not
have a pseudogap in their fermion self energy in a self-consistent
fashion, the experimental evidence of a pseudogap can be viewed as a
strong support for this theory. Given the analogy between
superfluidity in Fermi gases and superconductivity in high $T_c$
superconductivity, we argue that the present pairing fluctuation
theory for the pseudogap is also a strong candidate for high $T_c$
superconductivity.

\acknowledgments

This work is supported by the National Basic Research Program of China
(under grants No. 2011CB921303 and No. 2012CB927404), the Natural Science
Foundation of China (No. 10974173 and No. 11274267), and of Zhejiang
Province (No. LZ13A040001).
%, and of Zhejiang University of Technology (No. 2011XY031).

%\vspace*{-1ex}

%\bibliographystyle{apsrmp} 
%\bibliographystyle{apsrev4-2} 
%\bibliography{Review3}

\begin{thebibliography}{194}%
\makeatletter
\providecommand \@ifxundefined [1]{%
 \@ifx{#1\undefined}
}%
\providecommand \@ifnum [1]{%
 \ifnum #1\expandafter \@firstoftwo
 \else \expandafter \@secondoftwo
 \fi
}%
\providecommand \@ifx [1]{%
 \ifx #1\expandafter \@firstoftwo
 \else \expandafter \@secondoftwo
 \fi
}%
\providecommand \natexlab [1]{#1}%
\providecommand \enquote  [1]{ #1 }%
\providecommand \bibnamefont  [1]{#1}%
\providecommand \bibfnamefont [1]{#1}%
\providecommand \citenamefont [1]{#1}%
\providecommand \href@noop [0]{\@secondoftwo}%
\providecommand \href [0]{\begingroup \@sanitize@url \@href}%
\providecommand \@href[1]{\@@startlink{#1}\@@href}%
\providecommand \@@href[1]{\endgroup#1\@@endlink}%
\providecommand \@sanitize@url [0]{\catcode `\\12\catcode `\$12\catcode
  `\&12\catcode `\#12\catcode `\^12\catcode `\_12\catcode `\%12\relax}%
\providecommand \@@startlink[1]{}%
\providecommand \@@endlink[0]{}%
\providecommand \url  [0]{\begingroup\@sanitize@url \@url }%
\providecommand \@url [1]{\endgroup\@href {#1}{\urlprefix }}%
\providecommand \urlprefix  [0]{URL }%
\providecommand \Eprint [0]{\href }%
\providecommand \doibase [0]{http://dx.doi.org/}%
\providecommand \selectlanguage [0]{\@gobble}%
\providecommand \bibinfo  [0]{\@secondoftwo}%
\providecommand \bibfield  [0]{\@secondoftwo}%
\providecommand \translation [1]{[#1]}%
\providecommand \BibitemOpen [0]{}%
\providecommand \bibitemStop [0]{}%
\providecommand \bibitemNoStop [0]{.\EOS\space}%
\providecommand \EOS [0]{\spacefactor3000\relax}%
\providecommand \BibitemShut  [1]{\csname bibitem#1\endcsname}%
\let\auto@bib@innerbib\@empty
%</preamble>
\bibitem [{\citenamefont {Chen}\ \emph
  {et~al.}(2005{\natexlab{a}})\citenamefont {Chen}, \citenamefont {Stajic},
  \citenamefont {Tan},\ and\ \citenamefont {Levin}}]{ourreview}%
  \BibitemOpen
  \bibfield  {author} {\bibinfo {author} {\bibfnamefont {Q.~J.}\ \bibnamefont
  {Chen}}, \bibinfo {author} {\bibfnamefont {J.}~\bibnamefont {Stajic}},
  \bibinfo {author} {\bibfnamefont {S.~N.}\ \bibnamefont {Tan}}, \ and\
  \bibinfo {author} {\bibfnamefont {K.}~\bibnamefont {Levin}},\ }\bibfield
  {title} {\enquote {\bibinfo {title} {{BCS-BEC} crossover: From high
  temperature superconductors to ultracold superfluids},}\ }\href@noop {}
  {\bibfield  {journal} {\textit {\bibinfo  {journal} {Phys. Rep.}}}, \bibinfo
  {year} {2005}{\natexlab{a}},\ \bibinfo {volume} {412}:\ \bibinfo {pages}
  {1}}\BibitemShut {NoStop}%
\bibitem [{\citenamefont {Bloch}\ \emph {et~al.}(2008)\citenamefont {Bloch},
  \citenamefont {Dalibard},\ and\ \citenamefont {Zwerger}}]{BlochRMP}%
  \BibitemOpen
  \bibfield  {author} {\bibinfo {author} {\bibfnamefont {I.}~\bibnamefont
  {Bloch}}, \bibinfo {author} {\bibfnamefont {J.}~\bibnamefont {Dalibard}}, \
  and\ \bibinfo {author} {\bibfnamefont {W.}~\bibnamefont {Zwerger}},\
  }\bibfield  {title} {\enquote {\bibinfo {title} {Many-body physics with
  ultracold gases},}\ }\href@noop {} {\bibfield  {journal} {\textit {\bibinfo
  {journal} {\rmp}}}, \bibinfo {year} {2008},\ \bibinfo {volume} {80}:\
  \bibinfo {pages} {885}}\BibitemShut {NoStop}%
\bibitem [{\citenamefont {Maldacena}(1998)}]{Maldacena98}%
  \BibitemOpen
  \bibfield  {author} {\bibinfo {author} {\bibfnamefont {J.~M.}\ \bibnamefont
  {Maldacena}},\ }\bibfield  {title} {\enquote {\bibinfo {title} {The large
  {$N$} limit of superconformal field theories and supergravity},}\ }\href@noop
  {} {\bibfield  {journal} {\textit {\bibinfo  {journal} {Adv. Theor. Math.
  Phys.}}}, \bibinfo {year} {1998},\ \bibinfo {volume} {2}:\ \bibinfo {pages}
  {231}},\ \bibinfo {note} {see also e-print
  arXiv:hep-th/9711200v3}\BibitemShut {NoStop}%
\bibitem [{\citenamefont {Witten}(1998)}]{Witten98}%
  \BibitemOpen
  \bibfield  {author} {\bibinfo {author} {\bibfnamefont {E.}~\bibnamefont
  {Witten}},\ }\bibfield  {title} {\enquote {\bibinfo {title} {Anti de sitter
  space and holography},}\ }\href@noop {} {\bibfield  {journal} {\textit
  {\bibinfo  {journal} {Adv. Theor. Math. Phys.}}}, \bibinfo {year} {1998},\
  \bibinfo {volume} {2}:\ \bibinfo {pages} {253}}\BibitemShut {NoStop}%
\bibitem [{\citenamefont {Aharony}\ \emph {et~al.}(2000)\citenamefont
  {Aharony}, \citenamefont {Gubser}, \citenamefont {Maldacena}, \citenamefont
  {Ooguri},\ and\ \citenamefont {Oz}}]{MaldacenaPhysRep}%
  \BibitemOpen
  \bibfield  {author} {\bibinfo {author} {\bibfnamefont {O.}~\bibnamefont
  {Aharony}}, \bibinfo {author} {\bibfnamefont {S.~S.}\ \bibnamefont {Gubser}},
  \bibinfo {author} {\bibfnamefont {J.}~\bibnamefont {Maldacena}}, \bibinfo
  {author} {\bibfnamefont {H.}~\bibnamefont {Ooguri}}, \ and\ \bibinfo {author}
  {\bibfnamefont {Y.}~\bibnamefont {Oz}},\ }\bibfield  {title} {\enquote
  {\bibinfo {title} {Large {$N$} field theories, string theory and gravity},}\
  }\href@noop {} {\bibfield  {journal} {\textit {\bibinfo  {journal} {Phys.
  Rep.}}}, \bibinfo {year} {2000},\ \bibinfo {volume} {323}:\ \bibinfo {pages}
  {183}}\BibitemShut {NoStop}%
\bibitem [{\citenamefont {Cubrovi\'c}\ \emph {et~al.}(2009)\citenamefont
  {Cubrovi\'c}, \citenamefont {Zaanen},\ and\ \citenamefont {Schalm}}]{Zaanen}%
  \BibitemOpen
  \bibfield  {author} {\bibinfo {author} {\bibfnamefont {M.}~\bibnamefont
  {Cubrovi\'c}}, \bibinfo {author} {\bibfnamefont {J.}~\bibnamefont {Zaanen}},
  \ and\ \bibinfo {author} {\bibfnamefont {K.}~\bibnamefont {Schalm}},\
  }\bibfield  {title} {\enquote {\bibinfo {title} {String theory, quantum phase
  transitions, and the emergent fermi liquid},}\ }\href@noop {} {\bibfield
  {journal} {\textit {\bibinfo  {journal} {Science}}}, \bibinfo {year} {2009},\
  \bibinfo {volume} {325}:\ \bibinfo {pages} {439}}\BibitemShut {NoStop}%
\bibitem [{\citenamefont {Timusk}\ and\ \citenamefont {Statt}(1999)}]{Timusk}%
  \BibitemOpen
  \bibfield  {author} {\bibinfo {author} {\bibfnamefont {T.}~\bibnamefont
  {Timusk}}\ and\ \bibinfo {author} {\bibfnamefont {B.}~\bibnamefont {Statt}},\
  }\bibfield  {title} {\enquote {\bibinfo {title} {The pseudogap in
  high-temperature superconductors: An experimental survey.}}\ }\href@noop {}
  {\bibfield  {journal} {\textit {\bibinfo  {journal} {Rep. Prog. Phys.}}},
  \bibinfo {year} {1999},\ \bibinfo {volume} {62}:\ \bibinfo {pages}
  {61}}\BibitemShut {NoStop}%
\bibitem [{\citenamefont {Schrieffer}(1983)}]{Schrieffer}%
  \BibitemOpen
  \bibfield  {author} {\bibinfo {author} {\bibfnamefont {J.~R.}\ \bibnamefont
  {Schrieffer}},\ }\href@noop {} {\emph {\bibinfo {title} {Theory of
  Superconductivity}}},\ \bibinfo {edition} {3rd}\ ed.\ (\bibinfo  {publisher}
  {Perseus Books},\ \bibinfo {address} {Reading, MA},\ \bibinfo {year}
  {1983})\BibitemShut {NoStop}%
\bibitem [{\citenamefont {Bose}(1924)}]{Bose}%
  \BibitemOpen
  \bibfield  {author} {\bibinfo {author} {\bibfnamefont {S.~N.}\ \bibnamefont
  {Bose}},\ }\bibfield  {title} {\enquote {\bibinfo {title} {Plancks gesetz und
  lichtquantenhypothese},}\ }\href@noop {} {\bibfield  {journal} {\textit
  {\bibinfo  {journal} {Z. Phys.}}}, \bibinfo {year} {1924},\ \bibinfo {volume}
  {26}:\ \bibinfo {pages} {178}}\BibitemShut {NoStop}%
\bibitem [{\citenamefont {Einstein}(1925)}]{Einstein}%
  \BibitemOpen
  \bibfield  {author} {\bibinfo {author} {\bibfnamefont {A.}~\bibnamefont
  {Einstein}},\ }\bibfield  {title} {\enquote {\bibinfo {title} {Quantentheorie
  des einatomigen idealen gases},}\ }\href@noop {} {\bibfield  {journal}
  {\textit {\bibinfo  {journal} {Sitzungsberichte der Preussischen Akademie der
  Wissenschaften}}}, \bibinfo {year} {1925},\ \bibinfo {volume} {1}:\ \bibinfo
  {pages} {3}}\BibitemShut {NoStop}%
\bibitem [{\citenamefont {Pitaevskii}\ and\ \citenamefont
  {Stringari}(2003)}]{PSB03}%
  \BibitemOpen
  \bibfield  {author} {\bibinfo {author} {\bibfnamefont {L.}~\bibnamefont
  {Pitaevskii}}\ and\ \bibinfo {author} {\bibfnamefont {S.}~\bibnamefont
  {Stringari}},\ }\href@noop {} {\emph {\bibinfo {title} {Bose-Einstein
  Condensation}}}\ (\bibinfo  {publisher} {Oxford},\ \bibinfo {address} {New
  York},\ \bibinfo {year} {2003})\BibitemShut {NoStop}%
\bibitem [{\citenamefont {Pethick}\ and\ \citenamefont {Smith}(2002)}]{PS02}%
  \BibitemOpen
  \bibfield  {author} {\bibinfo {author} {\bibfnamefont {C.~J.}\ \bibnamefont
  {Pethick}}\ and\ \bibinfo {author} {\bibfnamefont {H.}~\bibnamefont
  {Smith}},\ }\href@noop {} {\emph {\bibinfo {title} {Bose-Einstein
  Condensation in Dilute Gases}}}\ (\bibinfo  {publisher} {Cambridge University
  Press},\ \bibinfo {address} {Cambridge},\ \bibinfo {year} {2002})\BibitemShut
  {NoStop}%
\bibitem [{\citenamefont {Eagles}(1969)}]{Eagles}%
  \BibitemOpen
  \bibfield  {author} {\bibinfo {author} {\bibfnamefont {D.~M.}\ \bibnamefont
  {Eagles}},\ }\bibfield  {title} {\enquote {\bibinfo {title} {Possible pairing
  without superconductivity at low carrier concentrations in bulk and thin-film
  superconducting semiconductors},}\ }\href@noop {} {\bibfield  {journal}
  {\textit {\bibinfo  {journal} {Phys. Rev.}}}, \bibinfo {year} {1969},\
  \bibinfo {volume} {186}:\ \bibinfo {pages} {456}}\BibitemShut {NoStop}%
\bibitem [{\citenamefont {Leggett}(1980)}]{Leggett}%
  \BibitemOpen
  \bibfield  {author} {\bibinfo {author} {\bibfnamefont {A.~J.}\ \bibnamefont
  {Leggett}},\ }\bibfield  {title} {\enquote {\bibinfo {title} {Diatomic
  molecules and {Cooper} pairs},}\ }in\ \href@noop {} {\emph {\bibinfo
  {booktitle} {Modern Trends in the Theory of Condensed Matter}}}\ (\bibinfo
  {publisher} {Springer-Verlag},\ \bibinfo {address} {Berlin},\ \bibinfo {year}
  {1980})\ pp.\ \bibinfo {pages} {13--27}\BibitemShut {NoStop}%
\bibitem [{\citenamefont {Nozi\`{e}res}\ and\ \citenamefont
  {Schmitt-Rink}(1985)}]{NSR}%
  \BibitemOpen
  \bibfield  {author} {\bibinfo {author} {\bibfnamefont {P.}~\bibnamefont
  {Nozi\`{e}res}}\ and\ \bibinfo {author} {\bibfnamefont {S.}~\bibnamefont
  {Schmitt-Rink}},\ }\bibfield  {title} {\enquote {\bibinfo {title} {Bose
  condensation in an attractive fermion gas: {F}rom weak to strong coupling
  superconductivity},}\ }\href@noop {} {\bibfield  {journal} {\textit {\bibinfo
   {journal} {J. Low Temp. Phys.}}}, \bibinfo {year} {1985},\ \bibinfo {volume}
  {59}:\ \bibinfo {pages} {195}}\BibitemShut {NoStop}%
\bibitem [{\citenamefont {Friedberg}\ and\ \citenamefont
  {Lee}(1989{\natexlab{a}})}]{TDLee1}%
  \BibitemOpen
  \bibfield  {author} {\bibinfo {author} {\bibfnamefont {R.}~\bibnamefont
  {Friedberg}}\ and\ \bibinfo {author} {\bibfnamefont {T.~D.}\ \bibnamefont
  {Lee}},\ }\bibfield  {title} {\enquote {\bibinfo {title} {Boson-fermion model
  of superconductivity},}\ }\href@noop {} {\bibfield  {journal} {\textit
  {\bibinfo  {journal} {Phys. Lett. A}}}, \bibinfo {year}
  {1989}{\natexlab{a}},\ \bibinfo {volume} {138}:\ \bibinfo {pages}
  {423}}\BibitemShut {NoStop}%
\bibitem [{\citenamefont {Friedberg}\ and\ \citenamefont
  {Lee}(1989{\natexlab{b}})}]{TDLee2}%
  \BibitemOpen
  \bibfield  {author} {\bibinfo {author} {\bibfnamefont {T.}~\bibnamefont
  {Friedberg}}\ and\ \bibinfo {author} {\bibfnamefont {T.~D.}\ \bibnamefont
  {Lee}},\ }\bibfield  {title} {\enquote {\bibinfo {title} {Gap energy and
  long-range order in the boson-fermion model of superconductivity.}}\
  }\href@noop {} {\bibfield  {journal} {\textit {\bibinfo  {journal} {Phys.
  Rev. B}}}, \bibinfo {year} {1989}{\natexlab{b}},\ \bibinfo {volume} {40}:\
  \bibinfo {pages} {6745}}\BibitemShut {NoStop}%
\bibitem [{\citenamefont {S\'a~de Melo}\ \emph {et~al.}(1993)\citenamefont
  {S\'a~de Melo}, \citenamefont {Randeria},\ and\ \citenamefont
  {Engelbrecht}}]{SadeMelo}%
  \BibitemOpen
  \bibfield  {author} {\bibinfo {author} {\bibfnamefont {C.~A.~R.}\
  \bibnamefont {S\'a~de Melo}}, \bibinfo {author} {\bibfnamefont
  {M.}~\bibnamefont {Randeria}}, \ and\ \bibinfo {author} {\bibfnamefont
  {J.~R.}\ \bibnamefont {Engelbrecht}},\ }\bibfield  {title} {\enquote
  {\bibinfo {title} {Crossover from bcs to bose superconductivity- transition -
  temperature and time dependent ginzburg-landau theory},}\ }\href@noop {}
  {\bibfield  {journal} {\textit {\bibinfo  {journal} {Phys. Rev. Lett.}}},
  \bibinfo {year} {1993},\ \bibinfo {volume} {71}:\ \bibinfo {pages}
  {3202}}\BibitemShut {NoStop}%
\bibitem [{\citenamefont {Randeria}(1995)}]{Randeriareview}%
  \BibitemOpen
  \bibfield  {author} {\bibinfo {author} {\bibfnamefont {M.}~\bibnamefont
  {Randeria}},\ }\bibfield  {title} {\enquote {\bibinfo {title} {Crossover from
  {BCS} theory to {Bose-Einstein} condensation},}\ }in\ \href@noop {} {\emph
  {\bibinfo {booktitle} {Bose Einstein Condensation}}},\ \bibinfo {editor}
  {edited by\ \bibinfo {editor} {\bibfnamefont {A.}~\bibnamefont {Griffin}},
  \bibinfo {editor} {\bibfnamefont {D.}~\bibnamefont {Snoke}}, \ and\ \bibinfo
  {editor} {\bibfnamefont {S.}~\bibnamefont {Stringari}}}\ (\bibinfo
  {publisher} {Cambridge Univ. Press},\ \bibinfo {address} {Cambridge},\
  \bibinfo {year} {1995})\ pp.\ \bibinfo {pages} {355--92}\BibitemShut
  {NoStop}%
\bibitem [{\citenamefont {Jank\'o}\ \emph {et~al.}(1997)\citenamefont
  {Jank\'o}, \citenamefont {Maly},\ and\ \citenamefont {Levin}}]{Janko}%
  \BibitemOpen
  \bibfield  {author} {\bibinfo {author} {\bibfnamefont {B.}~\bibnamefont
  {Jank\'o}}, \bibinfo {author} {\bibfnamefont {J.}~\bibnamefont {Maly}}, \
  and\ \bibinfo {author} {\bibfnamefont {K.}~\bibnamefont {Levin}},\ }\bibfield
   {title} {\enquote {\bibinfo {title} {Pseudogap effects induced by resonant
  pair scattering.}}\ }\href@noop {} {\bibfield  {journal} {\textit {\bibinfo
  {journal} {Phys. Rev. B}}}, \bibinfo {year} {1997},\ \bibinfo {volume} {56}:\
  \bibinfo {pages} {R11407}}\BibitemShut {NoStop}%
\bibitem [{\citenamefont {Maly}\ \emph
  {et~al.}(1999{\natexlab{a}})\citenamefont {Maly}, \citenamefont {Jank\'o},\
  and\ \citenamefont {Levin}}]{Maly1}%
  \BibitemOpen
  \bibfield  {author} {\bibinfo {author} {\bibfnamefont {J.}~\bibnamefont
  {Maly}}, \bibinfo {author} {\bibfnamefont {B.}~\bibnamefont {Jank\'o}}, \
  and\ \bibinfo {author} {\bibfnamefont {K.}~\bibnamefont {Levin}},\ }\bibfield
   {title} {\enquote {\bibinfo {title} {Numerical studies of the $s$-wave
  pseudogap state and related {$T_c$}: the ``pairing approximation'' theory},}\
  }\href@noop {} {\bibfield  {journal} {\textit {\bibinfo  {journal} {Physica
  C}}}, \bibinfo {year} {1999}{\natexlab{a}},\ \bibinfo {volume} {321}:\
  \bibinfo {pages} {113}}\BibitemShut {NoStop}%
\bibitem [{\citenamefont {Maly}\ \emph
  {et~al.}(1999{\natexlab{b}})\citenamefont {Maly}, \citenamefont {Jank\'o},\
  and\ \citenamefont {Levin}}]{Maly2}%
  \BibitemOpen
  \bibfield  {author} {\bibinfo {author} {\bibfnamefont {J.}~\bibnamefont
  {Maly}}, \bibinfo {author} {\bibfnamefont {B.}~\bibnamefont {Jank\'o}}, \
  and\ \bibinfo {author} {\bibfnamefont {K.}~\bibnamefont {Levin}},\ }\bibfield
   {title} {\enquote {\bibinfo {title} {Superconductivity from a pseudogapped
  normal state: {A} mode-coupling approach to precursor superconductivity.}}\
  }\href@noop {} {\bibfield  {journal} {\textit {\bibinfo  {journal} {Phys.
  Rev. B}}}, \bibinfo {year} {1999}{\natexlab{b}},\ \bibinfo {volume} {59}:\
  \bibinfo {pages} {1354}}\BibitemShut {NoStop}%
\bibitem [{\citenamefont {Chen}\ \emph {et~al.}(1998)\citenamefont {Chen},
  \citenamefont {Kosztin}, \citenamefont {Jank\'o},\ and\ \citenamefont
  {Levin}}]{Chen2}%
  \BibitemOpen
  \bibfield  {author} {\bibinfo {author} {\bibfnamefont {Q.~J.}\ \bibnamefont
  {Chen}}, \bibinfo {author} {\bibfnamefont {I.}~\bibnamefont {Kosztin}},
  \bibinfo {author} {\bibfnamefont {B.}~\bibnamefont {Jank\'o}}, \ and\
  \bibinfo {author} {\bibfnamefont {K.}~\bibnamefont {Levin}},\ }\bibfield
  {title} {\enquote {\bibinfo {title} {Pairing fluctuation theory of
  superconducting properties in underdoped to overdoped cuprates.}}\
  }\href@noop {} {\bibfield  {journal} {\textit {\bibinfo  {journal} {Phys.
  Rev. Lett.}}}, \bibinfo {year} {1998},\ \bibinfo {volume} {81}:\ \bibinfo
  {pages} {4708}}\BibitemShut {NoStop}%
\bibitem [{\citenamefont {Chen}\ \emph {et~al.}(1999)\citenamefont {Chen},
  \citenamefont {Kosztin}, \citenamefont {Jank\'o},\ and\ \citenamefont
  {Levin}}]{Chen1}%
  \BibitemOpen
  \bibfield  {author} {\bibinfo {author} {\bibfnamefont {Q.~J.}\ \bibnamefont
  {Chen}}, \bibinfo {author} {\bibfnamefont {I.}~\bibnamefont {Kosztin}},
  \bibinfo {author} {\bibfnamefont {B.}~\bibnamefont {Jank\'o}}, \ and\
  \bibinfo {author} {\bibfnamefont {K.}~\bibnamefont {Levin}},\ }\bibfield
  {title} {\enquote {\bibinfo {title} {Superconducting transitions from the
  pseudogap state: $d$-wave symmetry, lattice, and low-dimensional effects.}}\
  }\href@noop {} {\bibfield  {journal} {\textit {\bibinfo  {journal} {Phys.
  Rev. B}}}, \bibinfo {year} {1999},\ \bibinfo {volume} {59}:\ \bibinfo {pages}
  {7083}}\BibitemShut {NoStop}%
\bibitem [{\citenamefont {Kosztin}\ \emph {et~al.}(1998)\citenamefont
  {Kosztin}, \citenamefont {Chen}, \citenamefont {Jank\'o},\ and\ \citenamefont
  {Levin}}]{Kosztin1}%
  \BibitemOpen
  \bibfield  {author} {\bibinfo {author} {\bibfnamefont {I.}~\bibnamefont
  {Kosztin}}, \bibinfo {author} {\bibfnamefont {Q.~J.}\ \bibnamefont {Chen}},
  \bibinfo {author} {\bibfnamefont {B.}~\bibnamefont {Jank\'o}}, \ and\
  \bibinfo {author} {\bibfnamefont {K.}~\bibnamefont {Levin}},\ }\bibfield
  {title} {\enquote {\bibinfo {title} {Relationship between the pseudo- and
  superconducting gaps: {Effects} of residual pairing correlations below
  {$T_c$}.}}\ }\href@noop {} {\bibfield  {journal} {\textit {\bibinfo
  {journal} {Phys. Rev. B}}}, \bibinfo {year} {1998},\ \bibinfo {volume} {58}:\
  \bibinfo {pages} {R5936}}\BibitemShut {NoStop}%
\bibitem [{\citenamefont {Micnas}\ \emph {et~al.}(1990)\citenamefont {Micnas},
  \citenamefont {Ranninger},\ and\ \citenamefont {Robaszkiewicz}}]{MicnasRMP}%
  \BibitemOpen
  \bibfield  {author} {\bibinfo {author} {\bibfnamefont {R.}~\bibnamefont
  {Micnas}}, \bibinfo {author} {\bibfnamefont {J.}~\bibnamefont {Ranninger}}, \
  and\ \bibinfo {author} {\bibfnamefont {S.}~\bibnamefont {Robaszkiewicz}},\
  }\bibfield  {title} {\enquote {\bibinfo {title} {Superconductivity in
  narrow-band systems with local nonretarded attractive interactions.}}\
  }\href@noop {} {\bibfield  {journal} {\textit {\bibinfo  {journal} {Rev. Mod.
  Phys.}}}, \bibinfo {year} {1990},\ \bibinfo {volume} {62}:\ \bibinfo {pages}
  {113}}\BibitemShut {NoStop}%
\bibitem [{\citenamefont {Micnas}\ and\ \citenamefont
  {Robaszkiewicz}(1998)}]{Micnas1}%
  \BibitemOpen
  \bibfield  {author} {\bibinfo {author} {\bibfnamefont {R.}~\bibnamefont
  {Micnas}}\ and\ \bibinfo {author} {\bibfnamefont {S.}~\bibnamefont
  {Robaszkiewicz}},\ }\bibfield  {title} {\enquote {\bibinfo {title}
  {Superconductivity in systems with local attractive interactions.}}\
  }\href@noop {} {\bibfield  {journal} {\textit {\bibinfo  {journal} {Cond.
  Matt. Phys.}}}, \bibinfo {year} {1998},\ \bibinfo {volume} {1}:\ \bibinfo
  {pages} {89}}\BibitemShut {NoStop}%
\bibitem [{\citenamefont {Micnas}\ \emph {et~al.}(1995)\citenamefont {Micnas},
  \citenamefont {Pedersen}, \citenamefont {Schafroth}, \citenamefont
  {Schneider}, \citenamefont {Rodriguez-Nunez},\ and\ \citenamefont
  {Beck}}]{Micnas95}%
  \BibitemOpen
  \bibfield  {author} {\bibinfo {author} {\bibfnamefont {R.}~\bibnamefont
  {Micnas}}, \bibinfo {author} {\bibfnamefont {M.~H.}\ \bibnamefont
  {Pedersen}}, \bibinfo {author} {\bibfnamefont {S.}~\bibnamefont {Schafroth}},
  \bibinfo {author} {\bibfnamefont {T.}~\bibnamefont {Schneider}}, \bibinfo
  {author} {\bibfnamefont {J.}~\bibnamefont {Rodriguez-Nunez}}, \ and\ \bibinfo
  {author} {\bibfnamefont {H.}~\bibnamefont {Beck}},\ }\bibfield  {title}
  {\enquote {\bibinfo {title} {Excitation spectrum of the attractive hubbard
  model.}}\ }\href@noop {} {\bibfield  {journal} {\textit {\bibinfo  {journal}
  {Phys. Rev. B}}}, \bibinfo {year} {1995},\ \bibinfo {volume} {52}:\ \bibinfo
  {pages} {16223}}\BibitemShut {NoStop}%
\bibitem [{\citenamefont {Ranninger}\ and\ \citenamefont
  {Robin}(1996)}]{Ranninger}%
  \BibitemOpen
  \bibfield  {author} {\bibinfo {author} {\bibfnamefont {J.}~\bibnamefont
  {Ranninger}}\ and\ \bibinfo {author} {\bibfnamefont {J.~M.}\ \bibnamefont
  {Robin}},\ }\bibfield  {title} {\enquote {\bibinfo {title} {Manifestations of
  the pseudogap in the boson-fermion model for
  {Bose-Einstein}-condensation-driven superconductivity.}}\ }\href@noop {}
  {\bibfield  {journal} {\textit {\bibinfo  {journal} {Phys. Rev. B}}},
  \bibinfo {year} {1996},\ \bibinfo {volume} {53}:\ \bibinfo {pages}
  {R11961}}\BibitemShut {NoStop}%
\bibitem [{\citenamefont {Drechsler}\ and\ \citenamefont
  {Zwerger}(1992)}]{Drechsler}%
  \BibitemOpen
  \bibfield  {author} {\bibinfo {author} {\bibfnamefont {M.}~\bibnamefont
  {Drechsler}}\ and\ \bibinfo {author} {\bibfnamefont {W.}~\bibnamefont
  {Zwerger}},\ }\bibfield  {title} {\enquote {\bibinfo {title} {Crossover from
  {BCS} superconductivity to {Bose} condensation},}\ }\href@noop {} {\bibfield
  {journal} {\textit {\bibinfo  {journal} {Ann. Physik}}}, \bibinfo {year}
  {1992},\ \bibinfo {volume} {1}:\ \bibinfo {pages} {15}}\BibitemShut {NoStop}%
\bibitem [{\citenamefont {Haussmann}(1993)}]{Haussmann}%
  \BibitemOpen
  \bibfield  {author} {\bibinfo {author} {\bibfnamefont {R.}~\bibnamefont
  {Haussmann}},\ }\bibfield  {title} {\enquote {\bibinfo {title} {Crossover
  from {BCS} superconductivity to {Bose-Einstein} condensation: a
  self-consistent theory.}}\ }\href@noop {} {\bibfield  {journal} {\textit
  {\bibinfo  {journal} {Z. Phys. B}}}, \bibinfo {year} {1993},\ \bibinfo
  {volume} {91}:\ \bibinfo {pages} {291}}\BibitemShut {NoStop}%
\bibitem [{\citenamefont {Haussmann}(1994)}]{Haussmann2}%
  \BibitemOpen
  \bibfield  {author} {\bibinfo {author} {\bibfnamefont {R.}~\bibnamefont
  {Haussmann}},\ }\bibfield  {title} {\enquote {\bibinfo {title} {Properties of
  a {Fermi} liquid at the superfluid transition in the crossover region between
  {BCS} superconductivity and {Bose-Einstein} condensation},}\ }\href@noop {}
  {\bibfield  {journal} {\textit {\bibinfo  {journal} {Phys. Rev. B}}},
  \bibinfo {year} {1994},\ \bibinfo {volume} {49}:\ \bibinfo {pages}
  {12975}}\BibitemShut {NoStop}%
\bibitem [{\citenamefont {Tchernyshyov}(1997)}]{Tchern}%
  \BibitemOpen
  \bibfield  {author} {\bibinfo {author} {\bibfnamefont {O.}~\bibnamefont
  {Tchernyshyov}},\ }\bibfield  {title} {\enquote {\bibinfo {title}
  {Noninteracting {Cooper} pairs inside a pseudogap.}}\ }\href@noop {}
  {\bibfield  {journal} {\textit {\bibinfo  {journal} {Phys. Rev. B}}},
  \bibinfo {year} {1997},\ \bibinfo {volume} {56}:\ \bibinfo {pages}
  {3372}}\BibitemShut {NoStop}%
\bibitem [{\citenamefont {Gorbar}\ \emph {et~al.}(1996)\citenamefont {Gorbar},
  \citenamefont {Loktev},\ and\ \citenamefont {Sharapov}}]{Gorbar}%
  \BibitemOpen
  \bibfield  {author} {\bibinfo {author} {\bibfnamefont {E.~V.}\ \bibnamefont
  {Gorbar}}, \bibinfo {author} {\bibfnamefont {V.~M.}\ \bibnamefont {Loktev}},
  \ and\ \bibinfo {author} {\bibfnamefont {S.~G.}\ \bibnamefont {Sharapov}},\
  }\bibfield  {title} {\enquote {\bibinfo {title} {Crossover from {BCS} to
  composite-boson (local-pair) superconductivity in quasi-{2D} systems},}\
  }\href@noop {} {\bibfield  {journal} {\textit {\bibinfo  {journal} {Physica
  C}}}, \bibinfo {year} {1996},\ \bibinfo {volume} {257}:\ \bibinfo {pages}
  {355}}\BibitemShut {NoStop}%
\bibitem [{\citenamefont {Gusynin}\ \emph {et~al.}(1997)\citenamefont
  {Gusynin}, \citenamefont {Loktev},\ and\ \citenamefont {Sharapov}}]{Gusynin}%
  \BibitemOpen
  \bibfield  {author} {\bibinfo {author} {\bibfnamefont {V.~P.}\ \bibnamefont
  {Gusynin}}, \bibinfo {author} {\bibfnamefont {V.~M.}\ \bibnamefont {Loktev}},
  \ and\ \bibinfo {author} {\bibfnamefont {S.~G.}\ \bibnamefont {Sharapov}},\
  }\bibfield  {title} {\enquote {\bibinfo {title} {Phase diagram of a {2D}
  metal system with a variable number of carriers},}\ }\href@noop {} {\bibfield
   {journal} {\textit {\bibinfo  {journal} {JETP Lett.}}}, \bibinfo {year}
  {1997},\ \bibinfo {volume} {65}:\ \bibinfo {pages} {182}}\BibitemShut
  {NoStop}%
\bibitem [{\citenamefont {Marini}\ \emph {et~al.}(1998)\citenamefont {Marini},
  \citenamefont {Pistolesi},\ and\ \citenamefont {Strinati}}]{Marini}%
  \BibitemOpen
  \bibfield  {author} {\bibinfo {author} {\bibfnamefont {M.}~\bibnamefont
  {Marini}}, \bibinfo {author} {\bibfnamefont {F.}~\bibnamefont {Pistolesi}}, \
  and\ \bibinfo {author} {\bibfnamefont {G.~C.}\ \bibnamefont {Strinati}},\
  }\bibfield  {title} {\enquote {\bibinfo {title} {Evolution from {BCS}
  superconductivity to {Bose} condensation: {Analytic} results for the
  crossover in three dimensions},}\ }\href@noop {} {\bibfield  {journal}
  {\textit {\bibinfo  {journal} {Eur. Phys. J. B}}}, \bibinfo {year} {1998},\
  \bibinfo {volume} {1}:\ \bibinfo {pages} {151}}\BibitemShut {NoStop}%
\bibitem [{\citenamefont {DeMarco}\ and\ \citenamefont {Jin}(1999)}]{Jin}%
  \BibitemOpen
  \bibfield  {author} {\bibinfo {author} {\bibfnamefont {B.}~\bibnamefont
  {DeMarco}}\ and\ \bibinfo {author} {\bibfnamefont {D.~S.}\ \bibnamefont
  {Jin}},\ }\bibfield  {title} {\enquote {\bibinfo {title} {Onset of fermi
  degeneracy in a trapped atomic gas},}\ }\href@noop {} {\bibfield  {journal}
  {\textit {\bibinfo  {journal} {Science}}}, \bibinfo {year} {1999},\ \bibinfo
  {volume} {285}:\ \bibinfo {pages} {1703}}\BibitemShut {NoStop}%
\bibitem [{\citenamefont {Anderson}\ \emph {et~al.}(1995)\citenamefont
  {Anderson}, \citenamefont {Ensher}, \citenamefont {Matthews}, \citenamefont
  {Wieman},\ and\ \citenamefont {Cornell}}]{CornellBEC}%
  \BibitemOpen
  \bibfield  {author} {\bibinfo {author} {\bibfnamefont {M.~H.}\ \bibnamefont
  {Anderson}}, \bibinfo {author} {\bibfnamefont {J.~R.}\ \bibnamefont
  {Ensher}}, \bibinfo {author} {\bibfnamefont {M.~R.}\ \bibnamefont
  {Matthews}}, \bibinfo {author} {\bibfnamefont {C.~E.}\ \bibnamefont
  {Wieman}}, \ and\ \bibinfo {author} {\bibfnamefont {E.~A.}\ \bibnamefont
  {Cornell}},\ }\bibfield  {title} {\enquote {\bibinfo {title} {Observation of
  {Bose-Einstein} condensation in a dilute atomic vapor},}\ }\href@noop {}
  {\bibfield  {journal} {\textit {\bibinfo  {journal} {Science}}}, \bibinfo
  {year} {1995},\ \bibinfo {volume} {269}:\ \bibinfo {pages} {198}}\BibitemShut
  {NoStop}%
\bibitem [{\citenamefont {Bradley}\ \emph {et~al.}(1995)\citenamefont
  {Bradley}, \citenamefont {Sackett}, \citenamefont {Tollett},\ and\
  \citenamefont {Hulet}}]{HuletBEC1}%
  \BibitemOpen
  \bibfield  {author} {\bibinfo {author} {\bibfnamefont {C.~C.}\ \bibnamefont
  {Bradley}}, \bibinfo {author} {\bibfnamefont {C.~A.}\ \bibnamefont
  {Sackett}}, \bibinfo {author} {\bibfnamefont {J.~J.}\ \bibnamefont
  {Tollett}}, \ and\ \bibinfo {author} {\bibfnamefont {R.~G.}\ \bibnamefont
  {Hulet}},\ }\bibfield  {title} {\enquote {\bibinfo {title} {Evidence of
  bose-einstein condensation in an atomic gas with attractive interactions},}\
  }\href@noop {} {\bibfield  {journal} {\textit {\bibinfo  {journal} {\prl}}},
  \bibinfo {year} {1995},\ \bibinfo {volume} {75}:\ \bibinfo {pages}
  {1687}}\BibitemShut {NoStop}%
\bibitem [{\citenamefont {Bradley}\ \emph {et~al.}(1997)\citenamefont
  {Bradley}, \citenamefont {Sackett}, \citenamefont {Tollett},\ and\
  \citenamefont {Hulet}}]{HuletBEC2}%
  \BibitemOpen
  \bibfield  {author} {\bibinfo {author} {\bibfnamefont {C.~C.}\ \bibnamefont
  {Bradley}}, \bibinfo {author} {\bibfnamefont {C.~A.}\ \bibnamefont
  {Sackett}}, \bibinfo {author} {\bibfnamefont {J.~J.}\ \bibnamefont
  {Tollett}}, \ and\ \bibinfo {author} {\bibfnamefont {R.~G.}\ \bibnamefont
  {Hulet}},\ }\bibfield  {title} {\enquote {\bibinfo {title} {Evidence of
  {Bose-Einstein} condensation in an atomic gas with attractive interactions
  [\textit{Phys. Rev. Lett.} 75, 1687 (1995)]},}\ }\href@noop {} {\bibfield
  {journal} {\textit {\bibinfo  {journal} {\prl}}}, \bibinfo {year} {1997},\
  \bibinfo {volume} {79}:\ \bibinfo {pages} {1170}}\BibitemShut {NoStop}%
\bibitem [{\citenamefont {Davis}\ \emph {et~al.}(1995)\citenamefont {Davis},
  \citenamefont {Mewes}, \citenamefont {Andrews}, \citenamefont {van Druten},
  \citenamefont {Durfee}, \citenamefont {Kurn},\ and\ \citenamefont
  {Ketterle}}]{KetterleBEC}%
  \BibitemOpen
  \bibfield  {author} {\bibinfo {author} {\bibfnamefont {K.~B.}\ \bibnamefont
  {Davis}}, \bibinfo {author} {\bibfnamefont {M.~O.}\ \bibnamefont {Mewes}},
  \bibinfo {author} {\bibfnamefont {M.~R.}\ \bibnamefont {Andrews}}, \bibinfo
  {author} {\bibfnamefont {N.~J.}\ \bibnamefont {van Druten}}, \bibinfo
  {author} {\bibfnamefont {D.~S.}\ \bibnamefont {Durfee}}, \bibinfo {author}
  {\bibfnamefont {D.~M.}\ \bibnamefont {Kurn}}, \ and\ \bibinfo {author}
  {\bibfnamefont {W.}~\bibnamefont {Ketterle}},\ }\bibfield  {title} {\enquote
  {\bibinfo {title} {{Bose-Einstein} condensation in a gas of sodium atoms},}\
  }\href@noop {} {\bibfield  {journal} {\textit {\bibinfo  {journal} {\prl}}},
  \bibinfo {year} {1995},\ \bibinfo {volume} {75}:\ \bibinfo {pages}
  {3969}}\BibitemShut {NoStop}%
\bibitem [{\citenamefont {Greiner}\ \emph {et~al.}(2003)\citenamefont
  {Greiner}, \citenamefont {Regal},\ and\ \citenamefont {Jin}}]{Jin3}%
  \BibitemOpen
  \bibfield  {author} {\bibinfo {author} {\bibfnamefont {M.}~\bibnamefont
  {Greiner}}, \bibinfo {author} {\bibfnamefont {C.~A.}\ \bibnamefont {Regal}},
  \ and\ \bibinfo {author} {\bibfnamefont {D.~S.}\ \bibnamefont {Jin}},\
  }\bibfield  {title} {\enquote {\bibinfo {title} {Emergence of a molecular
  {Bose-Einstein} condensate from a {Fermi} gas},}\ }\href@noop {} {\bibfield
  {journal} {\textit {\bibinfo  {journal} {Nature}}}, \bibinfo {year} {2003},\
  \bibinfo {volume} {426}:\ \bibinfo {pages} {537}}\BibitemShut {NoStop}%
\bibitem [{\citenamefont {Jochim}\ \emph {et~al.}(2003)\citenamefont {Jochim},
  \citenamefont {Bartenstein}, \citenamefont {Altmeyer}, \citenamefont {Hendl},
  \citenamefont {Riedl}, \citenamefont {Chin}, \citenamefont {Denschlag},\ and\
  \citenamefont {Grimm}}]{Grimm}%
  \BibitemOpen
  \bibfield  {author} {\bibinfo {author} {\bibfnamefont {S.}~\bibnamefont
  {Jochim}}, \bibinfo {author} {\bibfnamefont {M.}~\bibnamefont {Bartenstein}},
  \bibinfo {author} {\bibfnamefont {A.}~\bibnamefont {Altmeyer}}, \bibinfo
  {author} {\bibfnamefont {G.}~\bibnamefont {Hendl}}, \bibinfo {author}
  {\bibfnamefont {S.}~\bibnamefont {Riedl}}, \bibinfo {author} {\bibfnamefont
  {C.}~\bibnamefont {Chin}}, \bibinfo {author} {\bibfnamefont {J.~H.}\
  \bibnamefont {Denschlag}}, \ and\ \bibinfo {author} {\bibfnamefont
  {R.}~\bibnamefont {Grimm}},\ }\bibfield  {title} {\enquote {\bibinfo {title}
  {{Bose-Einstein} condensation of molecules},}\ }\href@noop {} {\bibfield
  {journal} {\textit {\bibinfo  {journal} {Science}}}, \bibinfo {year} {2003},\
  \bibinfo {volume} {302}:\ \bibinfo {pages} {2101}}\BibitemShut {NoStop}%
\bibitem [{\citenamefont {Zwierlein}\ \emph {et~al.}(2003)\citenamefont
  {Zwierlein} \emph {et~al.}}]{Ketterle2}%
  \BibitemOpen
  \bibfield  {author} {\bibinfo {author} {\bibfnamefont {M.~W.}\ \bibnamefont
  {Zwierlein}} \emph {et~al.},\ }\bibfield  {title} {\enquote {\bibinfo {title}
  {Observation of {Bose-Einstein} condensation of molecules},}\ }\href@noop {}
  {\bibfield  {journal} {\textit {\bibinfo  {journal} {Phys. Rev. Lett.}}},
  \bibinfo {year} {2003},\ \bibinfo {volume} {91}:\ \bibinfo {pages}
  {250401}}\BibitemShut {NoStop}%
\bibitem [{\citenamefont {Regal}\ \emph {et~al.}(2004)\citenamefont {Regal},
  \citenamefont {Greiner},\ and\ \citenamefont {Jin}}]{Jin4}%
  \BibitemOpen
  \bibfield  {author} {\bibinfo {author} {\bibfnamefont {C.~A.}\ \bibnamefont
  {Regal}}, \bibinfo {author} {\bibfnamefont {M.}~\bibnamefont {Greiner}}, \
  and\ \bibinfo {author} {\bibfnamefont {D.~S.}\ \bibnamefont {Jin}},\
  }\bibfield  {title} {\enquote {\bibinfo {title} {Observation of resonance
  condensation of fermionic atom pairs},}\ }\href@noop {} {\bibfield  {journal}
  {\textit {\bibinfo  {journal} {Phys. Rev. Lett.}}}, \bibinfo {year} {2004},\
  \bibinfo {volume} {92}:\ \bibinfo {pages} {040403}}\BibitemShut {NoStop}%
\bibitem [{\citenamefont {Bartenstein}\ \emph {et~al.}(2004)\citenamefont
  {Bartenstein}, \citenamefont {Altmeyer}, \citenamefont {Riedl}, \citenamefont
  {Jochim}, \citenamefont {Chin}, \citenamefont {Denschlag},\ and\
  \citenamefont {Grimm}}]{Grimm2}%
  \BibitemOpen
  \bibfield  {author} {\bibinfo {author} {\bibfnamefont {M.}~\bibnamefont
  {Bartenstein}}, \bibinfo {author} {\bibfnamefont {A.}~\bibnamefont
  {Altmeyer}}, \bibinfo {author} {\bibfnamefont {S.}~\bibnamefont {Riedl}},
  \bibinfo {author} {\bibfnamefont {S.}~\bibnamefont {Jochim}}, \bibinfo
  {author} {\bibfnamefont {C.}~\bibnamefont {Chin}}, \bibinfo {author}
  {\bibfnamefont {J.~H.}\ \bibnamefont {Denschlag}}, \ and\ \bibinfo {author}
  {\bibfnamefont {R.}~\bibnamefont {Grimm}},\ }\bibfield  {title} {\enquote
  {\bibinfo {title} {Crossover from a molecular {B}ose-{E}instein condensate to
  a degenerate {F}ermi gas.}}\ }\href@noop {} {\bibfield  {journal} {\textit
  {\bibinfo  {journal} {Phys. Rev. Lett.}}}, \bibinfo {year} {2004},\ \bibinfo
  {volume} {92}:\ \bibinfo {pages} {120401}}\BibitemShut {NoStop}%
\bibitem [{\citenamefont {Chin}\ \emph {et~al.}(2004)\citenamefont {Chin},
  \citenamefont {Bartenstein}, \citenamefont {Altmeyer}, \citenamefont {Riedl},
  \citenamefont {Jochim}, \citenamefont {Hecker-Denschlag},\ and\ \citenamefont
  {Grimm}}]{Grimm4}%
  \BibitemOpen
  \bibfield  {author} {\bibinfo {author} {\bibfnamefont {C.}~\bibnamefont
  {Chin}}, \bibinfo {author} {\bibfnamefont {M.}~\bibnamefont {Bartenstein}},
  \bibinfo {author} {\bibfnamefont {A.}~\bibnamefont {Altmeyer}}, \bibinfo
  {author} {\bibfnamefont {S.}~\bibnamefont {Riedl}}, \bibinfo {author}
  {\bibfnamefont {S.}~\bibnamefont {Jochim}}, \bibinfo {author} {\bibfnamefont
  {J.}~\bibnamefont {Hecker-Denschlag}}, \ and\ \bibinfo {author}
  {\bibfnamefont {R.}~\bibnamefont {Grimm}},\ }\bibfield  {title} {\enquote
  {\bibinfo {title} {Observation of the pairing gap in a strongly interacting
  {F}ermi gas superfluids},}\ }\href@noop {} {\bibfield  {journal} {\textit
  {\bibinfo  {journal} {Science}}}, \bibinfo {year} {2004},\ \bibinfo {volume}
  {305}:\ \bibinfo {pages} {1128}}\BibitemShut {NoStop}%
\bibitem [{\citenamefont {Zwierlein}\ \emph {et~al.}(2004)\citenamefont
  {Zwierlein}, \citenamefont {Stan}, \citenamefont {Schunck}, \citenamefont
  {Raupach}, \citenamefont {Kerman},\ and\ \citenamefont
  {Ketterle}}]{Ketterle3}%
  \BibitemOpen
  \bibfield  {author} {\bibinfo {author} {\bibfnamefont {M.~W.}\ \bibnamefont
  {Zwierlein}}, \bibinfo {author} {\bibfnamefont {C.~A.}\ \bibnamefont {Stan}},
  \bibinfo {author} {\bibfnamefont {C.~H.}\ \bibnamefont {Schunck}}, \bibinfo
  {author} {\bibfnamefont {S.~M.~F.}\ \bibnamefont {Raupach}}, \bibinfo
  {author} {\bibfnamefont {A.~J.}\ \bibnamefont {Kerman}}, \ and\ \bibinfo
  {author} {\bibfnamefont {W.}~\bibnamefont {Ketterle}},\ }\bibfield  {title}
  {\enquote {\bibinfo {title} {Condensation of pairs of fermionic atoms near a
  {F}eshbach resonance.}}\ }\href@noop {} {\bibfield  {journal} {\textit
  {\bibinfo  {journal} {Phys. Rev. Lett.}}}, \bibinfo {year} {2004},\ \bibinfo
  {volume} {92}:\ \bibinfo {pages} {120403}}\BibitemShut {NoStop}%
\bibitem [{\citenamefont {Kinast}\ \emph {et~al.}(2005)\citenamefont {Kinast},
  \citenamefont {Turlapov}, \citenamefont {Thomas}, \citenamefont {Chen},
  \citenamefont {Stajic},\ and\ \citenamefont {Levin}}]{ThermoScience-full}%
  \BibitemOpen
  \bibfield  {author} {\bibinfo {author} {\bibfnamefont {J.}~\bibnamefont
  {Kinast}}, \bibinfo {author} {\bibfnamefont {A.}~\bibnamefont {Turlapov}},
  \bibinfo {author} {\bibfnamefont {J.~E.}\ \bibnamefont {Thomas}}, \bibinfo
  {author} {\bibfnamefont {Q.~J.}\ \bibnamefont {Chen}}, \bibinfo {author}
  {\bibfnamefont {J.}~\bibnamefont {Stajic}}, \ and\ \bibinfo {author}
  {\bibfnamefont {K.}~\bibnamefont {Levin}},\ }\bibfield  {title} {\enquote
  {\bibinfo {title} {Heat capacity of a strongly-interacting {Fermi} gas},}\
  }\href@noop {} {\bibfield  {journal} {\textit {\bibinfo  {journal}
  {Science}}}, \bibinfo {year} {2005},\ \bibinfo {volume} {307}:\ \bibinfo
  {pages} {1296}},\ \bibinfo {note} {published online 27 January 2005;
  doi:10.1126/science.1109220}\BibitemShut {NoStop}%
\bibitem [{\citenamefont {Zwierlein}\ \emph {et~al.}(2005)\citenamefont
  {Zwierlein}, \citenamefont {Abo-Shaeer}, \citenamefont {Schirotzek},\ and\
  \citenamefont {Ketterle}}]{KetterleV}%
  \BibitemOpen
  \bibfield  {author} {\bibinfo {author} {\bibfnamefont {M.~W.}\ \bibnamefont
  {Zwierlein}}, \bibinfo {author} {\bibfnamefont {J.~R.}\ \bibnamefont
  {Abo-Shaeer}}, \bibinfo {author} {\bibfnamefont {A.}~\bibnamefont
  {Schirotzek}}, \ and\ \bibinfo {author} {\bibfnamefont {W.}~\bibnamefont
  {Ketterle}},\ }\bibfield  {title} {\enquote {\bibinfo {title} {Vortices and
  superfluidity in a strongly interacting {Fermi} gas},}\ }\href@noop {}
  {\bibfield  {journal} {\textit {\bibinfo  {journal} {Nature}}}, \bibinfo
  {year} {2005},\ \bibinfo {volume} {435}:\ \bibinfo {pages}
  {170404}}\BibitemShut {NoStop}%
\bibitem [{\citenamefont {Zwierlein}\ \emph {et~al.}(2006)\citenamefont
  {Zwierlein}, \citenamefont {Schirotzek}, \citenamefont {Schunck},\ and\
  \citenamefont {Ketterle}}]{ZSSK06}%
  \BibitemOpen
  \bibfield  {author} {\bibinfo {author} {\bibfnamefont {M.~W.}\ \bibnamefont
  {Zwierlein}}, \bibinfo {author} {\bibfnamefont {A.}~\bibnamefont
  {Schirotzek}}, \bibinfo {author} {\bibfnamefont {C.~H.}\ \bibnamefont
  {Schunck}}, \ and\ \bibinfo {author} {\bibfnamefont {W.}~\bibnamefont
  {Ketterle}},\ }\bibfield  {title} {\enquote {\bibinfo {title} {Fermionic
  superfluidity with imbalanced spin populations},}\ }\href@noop {} {\bibfield
  {journal} {\textit {\bibinfo  {journal} {Science}}}, \bibinfo {year} {2006},\
  \bibinfo {volume} {311}:\ \bibinfo {pages} {492}}\BibitemShut {NoStop}%
\bibitem [{\citenamefont {Partridge}\ \emph {et~al.}(2006)\citenamefont
  {Partridge}, \citenamefont {Li}, \citenamefont {Kamar}, \citenamefont
  {Liao},\ and\ \citenamefont {Hulet}}]{Rice1}%
  \BibitemOpen
  \bibfield  {author} {\bibinfo {author} {\bibfnamefont {G.~B.}\ \bibnamefont
  {Partridge}}, \bibinfo {author} {\bibfnamefont {W.}~\bibnamefont {Li}},
  \bibinfo {author} {\bibfnamefont {R.~I.}\ \bibnamefont {Kamar}}, \bibinfo
  {author} {\bibfnamefont {Y.~A.}\ \bibnamefont {Liao}}, \ and\ \bibinfo
  {author} {\bibfnamefont {R.~G.}\ \bibnamefont {Hulet}},\ }\bibfield  {title}
  {\enquote {\bibinfo {title} {Pairing and phase separation in a polarized
  {F}ermi gas},}\ }\href@noop {} {\bibfield  {journal} {\textit {\bibinfo
  {journal} {Science}}}, \bibinfo {year} {2006},\ \bibinfo {volume} {311}:\
  \bibinfo {pages} {503}}\BibitemShut {NoStop}%
\bibitem [{\citenamefont {Fulde}\ and\ \citenamefont {Ferrell}(1964)}]{FF}%
  \BibitemOpen
  \bibfield  {author} {\bibinfo {author} {\bibfnamefont {P.}~\bibnamefont
  {Fulde}}\ and\ \bibinfo {author} {\bibfnamefont {R.~A.}\ \bibnamefont
  {Ferrell}},\ }\bibfield  {title} {\enquote {\bibinfo {title}
  {Superconductivity in a strong spin-exchange field},}\ }\href@noop {}
  {\bibfield  {journal} {\textit {\bibinfo  {journal} {Phys. Rev.}}}, \bibinfo
  {year} {1964},\ \bibinfo {volume} {135}:\ \bibinfo {pages}
  {A550}}\BibitemShut {NoStop}%
\bibitem [{\citenamefont {Larkin}\ and\ \citenamefont
  {Ovchinnikov}(1964)}]{LO_ru}%
  \BibitemOpen
  \bibfield  {author} {\bibinfo {author} {\bibfnamefont {A.~I.}\ \bibnamefont
  {Larkin}}\ and\ \bibinfo {author} {\bibfnamefont {Y.~N.}\ \bibnamefont
  {Ovchinnikov}},\ }\bibfield  {title} {\enquote {\bibinfo {title}
  {Neodnorodnoe sostoyanie sverkhprovodnikov},}\ }\href@noop {} {\bibfield
  {journal} {\textit {\bibinfo  {journal} {Zh. Eksp. Teor. Fiz.}}}, \bibinfo
  {year} {1964},\ \bibinfo {volume} {47}:\ \bibinfo {pages} {1136}}\BibitemShut
  {NoStop}%
\bibitem [{\citenamefont {Larkin}\ and\ \citenamefont
  {Ovchinnikov}(1965)}]{LO}%
  \BibitemOpen
  \bibfield  {author} {\bibinfo {author} {\bibfnamefont {A.~I.}\ \bibnamefont
  {Larkin}}\ and\ \bibinfo {author} {\bibfnamefont {Y.~N.}\ \bibnamefont
  {Ovchinnikov}},\ }\bibfield  {title} {\enquote {\bibinfo {title}
  {Inhomogeneous state of superconductors},}\ }\href@noop {} {\bibfield
  {journal} {\textit {\bibinfo  {journal} {Sov. Phys. JETP}}}, \bibinfo {year}
  {1965},\ \bibinfo {volume} {20}:\ \bibinfo {pages} {762}}\BibitemShut
  {NoStop}%
\bibitem [{\citenamefont {Chen}(2000)}]{ChenPhD}%
  \BibitemOpen
  \bibfield  {author} {\bibinfo {author} {\bibfnamefont {Q.~J.}\ \bibnamefont
  {Chen}},\ }\emph {\bibinfo {title} {Generalization of {BCS} theory to short
  coherence length superconductors: {A BCS-Bose-Einstein} crossover
  scenario}},\ \href@noop {} {Ph.D. thesis},\ \bibinfo  {school} {University of
  Chicago}, \bibinfo {year} {2000},,\ \bibinfo {note} {(available in the
  ProQuest Dissertations \& Theses Database online).}\BibitemShut {Stop}%
\bibitem [{\citenamefont {Chen}\ \emph
  {et~al.}(2006{\natexlab{a}})\citenamefont {Chen}, \citenamefont {Stajic},\
  and\ \citenamefont {Levin}}]{ReviewLTP-Full}%
  \BibitemOpen
  \bibfield  {author} {\bibinfo {author} {\bibfnamefont {Q.~J.}\ \bibnamefont
  {Chen}}, \bibinfo {author} {\bibfnamefont {J.}~\bibnamefont {Stajic}}, \ and\
  \bibinfo {author} {\bibfnamefont {K.}~\bibnamefont {Levin}},\ }\bibfield
  {title} {\enquote {\bibinfo {title} {Applying {BCS-BEC} crossover theory to
  high temperature superconductors and ultracold atomic fermi gases},}\
  }\href@noop {} {\bibfield  {journal} {\textit {\bibinfo  {journal} {Low Temp.
  Phys.}}}, \bibinfo {year} {2006}{\natexlab{a}},\ \bibinfo {volume} {32}:\
  \bibinfo {pages} {406}},\ \bibinfo {note} {\textit{Fiz. Nizk. Temp.}, 2006,
  {32}: 538.}\BibitemShut {Stop}%
\bibitem [{\citenamefont {Giorgini}\ \emph {et~al.}(2008)\citenamefont
  {Giorgini}, \citenamefont {Pitaevskii},\ and\ \citenamefont
  {Stringari}}]{GiorginiRMP}%
  \BibitemOpen
  \bibfield  {author} {\bibinfo {author} {\bibfnamefont {S.}~\bibnamefont
  {Giorgini}}, \bibinfo {author} {\bibfnamefont {L.~P.}\ \bibnamefont
  {Pitaevskii}}, \ and\ \bibinfo {author} {\bibfnamefont {S.}~\bibnamefont
  {Stringari}},\ }\bibfield  {title} {\enquote {\bibinfo {title} {Theory of
  ultracold atomic {Fermi} gases},}\ }\href@noop {} {\bibfield  {journal}
  {\textit {\bibinfo  {journal} {\rmp}}}, \bibinfo {year} {2008},\ \bibinfo
  {volume} {80}:\ \bibinfo {pages} {1215}}\BibitemShut {NoStop}%
\bibitem [{\citenamefont {Chin}\ \emph {et~al.}(2010)\citenamefont {Chin},
  \citenamefont {Grimm}, \citenamefont {Julienne},\ and\ \citenamefont
  {Tiesinga}}]{ChinRMP}%
  \BibitemOpen
  \bibfield  {author} {\bibinfo {author} {\bibfnamefont {C.}~\bibnamefont
  {Chin}}, \bibinfo {author} {\bibfnamefont {R.}~\bibnamefont {Grimm}},
  \bibinfo {author} {\bibfnamefont {P.}~\bibnamefont {Julienne}}, \ and\
  \bibinfo {author} {\bibfnamefont {E.}~\bibnamefont {Tiesinga}},\ }\bibfield
  {title} {\enquote {\bibinfo {title} {Feshbach resonances in ultracold
  gases},}\ }\href@noop {} {\bibfield  {journal} {\textit {\bibinfo  {journal}
  {\rmp}}}, \bibinfo {year} {2010},\ \bibinfo {volume} {82}:\ \bibinfo {pages}
  {1225}}\BibitemShut {NoStop}%
\bibitem [{\citenamefont {Inguscio}\ \emph {et~al.}(2008)\citenamefont
  {Inguscio}, \citenamefont {Ketterle},\ and\ \citenamefont
  {Salomon}}]{VarennaProc}%
  \BibitemOpen
  \bibinfo {editor} {\bibfnamefont {M.}~\bibnamefont {Inguscio}}, \bibinfo
  {editor} {\bibfnamefont {W.}~\bibnamefont {Ketterle}}, \ and\ \bibinfo
  {editor} {\bibfnamefont {C.}~\bibnamefont {Salomon}},\ eds.,\ \href@noop {}
  {\emph {\bibinfo {title} {Ultracold Fermi gases, Proceedings of the
  International School of Physics ``Enrico Fermi,"}}},\ Vol.\ \bibinfo {volume}
  {CLXIV, Varenna, 2006},\ \bibinfo {organization} {Societ\`a Italiana di
  Fisica Bologna, Italy}\ (\bibinfo  {publisher} {ISO Press},\ \bibinfo
  {address} {Amsterdam},\ \bibinfo {year} {2008})\BibitemShut {NoStop}%
\bibitem [{\citenamefont {Ding}\ \emph {et~al.}(1996)\citenamefont {Ding},
  \citenamefont {Yokoya}, \citenamefont {Campuzano}, \citenamefont {Takahashi},
  \citenamefont {Randeria}, \citenamefont {Norman}, \citenamefont {Mochiku},
  \citenamefont {Hadowaki},\ and\ \citenamefont {Giapintzakis}}]{arpesanl1}%
  \BibitemOpen
  \bibfield  {author} {\bibinfo {author} {\bibfnamefont {H.}~\bibnamefont
  {Ding}}, \bibinfo {author} {\bibfnamefont {T.}~\bibnamefont {Yokoya}},
  \bibinfo {author} {\bibfnamefont {J.~C.}\ \bibnamefont {Campuzano}}, \bibinfo
  {author} {\bibfnamefont {T.}~\bibnamefont {Takahashi}}, \bibinfo {author}
  {\bibfnamefont {M.}~\bibnamefont {Randeria}}, \bibinfo {author}
  {\bibfnamefont {M.~R.}\ \bibnamefont {Norman}}, \bibinfo {author}
  {\bibfnamefont {T.}~\bibnamefont {Mochiku}}, \bibinfo {author} {\bibfnamefont
  {K.}~\bibnamefont {Hadowaki}}, \ and\ \bibinfo {author} {\bibfnamefont
  {J.}~\bibnamefont {Giapintzakis}},\ }\bibfield  {title} {\enquote {\bibinfo
  {title} {Spectroscopic evidence for a pseudogap in the normal state of
  underdoped high-{$T_c$} superconductors.}}\ }\href@noop {} {\bibfield
  {journal} {\textit {\bibinfo  {journal} {Nature}}}, \bibinfo {year} {1996},\
  \bibinfo {volume} {382}:\ \bibinfo {pages} {51}}\BibitemShut {NoStop}%
\bibitem [{\citenamefont {Renner}\ \emph
  {et~al.}(1998{\natexlab{a}})\citenamefont {Renner}, \citenamefont {Revaz},
  \citenamefont {Kadowaki}, \citenamefont {Maggio-Aprile},\ and\ \citenamefont
  {Fischer}}]{Renner}%
  \BibitemOpen
  \bibfield  {author} {\bibinfo {author} {\bibfnamefont {C.}~\bibnamefont
  {Renner}}, \bibinfo {author} {\bibfnamefont {B.}~\bibnamefont {Revaz}},
  \bibinfo {author} {\bibfnamefont {K.}~\bibnamefont {Kadowaki}}, \bibinfo
  {author} {\bibfnamefont {I.}~\bibnamefont {Maggio-Aprile}}, \ and\ \bibinfo
  {author} {\bibfnamefont {O.}~\bibnamefont {Fischer}},\ }\bibfield  {title}
  {\enquote {\bibinfo {title} {Observation of the low temperature pseudogap in
  the vortex cores of {Bi$_2$Sr$_2$CaCu$_2$O$_{8+ \delta}$}.}}\ }\href@noop {}
  {\bibfield  {journal} {\textit {\bibinfo  {journal} {Phys. Rev. Lett.}}},
  \bibinfo {year} {1998}{\natexlab{a}},\ \bibinfo {volume} {80}:\ \bibinfo
  {pages} {3606}}\BibitemShut {NoStop}%
\bibitem [{\citenamefont {Renner}\ \emph
  {et~al.}(1998{\natexlab{b}})\citenamefont {Renner}, \citenamefont {Revaz},
  \citenamefont {Genoud}, \citenamefont {Kadowaki},\ and\ \citenamefont
  {Fischer}}]{Renner1998}%
  \BibitemOpen
  \bibfield  {author} {\bibinfo {author} {\bibfnamefont {C.}~\bibnamefont
  {Renner}}, \bibinfo {author} {\bibfnamefont {B.}~\bibnamefont {Revaz}},
  \bibinfo {author} {\bibfnamefont {J.-Y.}\ \bibnamefont {Genoud}}, \bibinfo
  {author} {\bibfnamefont {K.}~\bibnamefont {Kadowaki}}, \ and\ \bibinfo
  {author} {\bibfnamefont {O.}~\bibnamefont {Fischer}},\ }\bibfield  {title}
  {\enquote {\bibinfo {title} {Pseudogap precursor of the superconducting gap
  in under- and overdoped {Bi$_2$Sr$_2$CaCu$_2$O$_{8+\delta}$}.}}\ }\href@noop
  {} {\bibfield  {journal} {\textit {\bibinfo  {journal} {Phys. Rev. Lett.}}},
  \bibinfo {year} {1998}{\natexlab{b}},\ \bibinfo {volume} {80}:\ \bibinfo
  {pages} {149}}\BibitemShut {NoStop}%
\bibitem [{\citenamefont {Krasnov}\ \emph {et~al.}(2000)\citenamefont
  {Krasnov}, \citenamefont {Yurgens}, \citenamefont {Winkler}, \citenamefont
  {Delsing},\ and\ \citenamefont {Claeson}}]{Krasnov2000}%
  \BibitemOpen
  \bibfield  {author} {\bibinfo {author} {\bibfnamefont {V.~M.}\ \bibnamefont
  {Krasnov}}, \bibinfo {author} {\bibfnamefont {A.}~\bibnamefont {Yurgens}},
  \bibinfo {author} {\bibfnamefont {D.}~\bibnamefont {Winkler}}, \bibinfo
  {author} {\bibfnamefont {P.}~\bibnamefont {Delsing}}, \ and\ \bibinfo
  {author} {\bibfnamefont {T.}~\bibnamefont {Claeson}},\ }\bibfield  {title}
  {\enquote {\bibinfo {title} {Evidence for coexistence of the superconducting
  gap and the pseudogap in {Bi}-2212 from intrinsic tunneling spectroscopy},}\
  }\href@noop {} {\bibfield  {journal} {\textit {\bibinfo  {journal} {Phys.
  Rev. Lett.}}}, \bibinfo {year} {2000},\ \bibinfo {volume} {84}:\ \bibinfo
  {pages} {5860}}\BibitemShut {NoStop}%
\bibitem [{\citenamefont {Kugler}\ \emph {et~al.}(2001)\citenamefont {Kugler},
  \citenamefont {Fischer}, \citenamefont {Renner}, \citenamefont {Ono},\ and\
  \citenamefont {Ando}}]{Fischer2}%
  \BibitemOpen
  \bibfield  {author} {\bibinfo {author} {\bibfnamefont {M.}~\bibnamefont
  {Kugler}}, \bibinfo {author} {\bibfnamefont {O.}~\bibnamefont {Fischer}},
  \bibinfo {author} {\bibfnamefont {C.}~\bibnamefont {Renner}}, \bibinfo
  {author} {\bibfnamefont {S.}~\bibnamefont {Ono}}, \ and\ \bibinfo {author}
  {\bibfnamefont {Y.}~\bibnamefont {Ando}},\ }\bibfield  {title} {\enquote
  {\bibinfo {title} {Scanning tunneling spectroscopy of
  {Bi$_2$Sr$_2$CuO$_{6+\delta}$}: {New} evidence for the common origion of the
  pseudogap and superconductivity.}}\ }\href@noop {} {\bibfield  {journal}
  {\textit {\bibinfo  {journal} {Phys. Rev. Lett.}}}, \bibinfo {year} {2001},\
  \bibinfo {volume} {86}:\ \bibinfo {pages} {4911}}\BibitemShut {NoStop}%
\bibitem [{\citenamefont {Loram}\ \emph {et~al.}(1994)\citenamefont {Loram},
  \citenamefont {Mirza}, \citenamefont {Cooper}, \citenamefont {Liang},\ and\
  \citenamefont {Wade}}]{Loram}%
  \BibitemOpen
  \bibfield  {author} {\bibinfo {author} {\bibfnamefont {J.~W.}\ \bibnamefont
  {Loram}}, \bibinfo {author} {\bibfnamefont {K.}~\bibnamefont {Mirza}},
  \bibinfo {author} {\bibfnamefont {J.}~\bibnamefont {Cooper}}, \bibinfo
  {author} {\bibfnamefont {W.}~\bibnamefont {Liang}}, \ and\ \bibinfo {author}
  {\bibfnamefont {J.}~\bibnamefont {Wade}},\ }\bibfield  {title} {\enquote
  {\bibinfo {title} {Electronic specific heat of {YBa$_2$Cu$_3$O$_{6+x}$} from
  1.8 to 300~{K}.}}\ }\href@noop {} {\bibfield  {journal} {\textit {\bibinfo
  {journal} {J. Supercond.}}}, \bibinfo {year} {1994},\ \bibinfo {volume} {7}:\
  \bibinfo {pages} {243}}\BibitemShut {NoStop}%
\bibitem [{\citenamefont {Williams}\ \emph {et~al.}(1998)\citenamefont
  {Williams}, \citenamefont {Haines},\ and\ \citenamefont
  {Tallon}}]{Williams1998}%
  \BibitemOpen
  \bibfield  {author} {\bibinfo {author} {\bibfnamefont {G.~V.~M.}\
  \bibnamefont {Williams}}, \bibinfo {author} {\bibfnamefont {E.~M.}\
  \bibnamefont {Haines}}, \ and\ \bibinfo {author} {\bibfnamefont {J.~L.}\
  \bibnamefont {Tallon}},\ }\bibfield  {title} {\enquote {\bibinfo {title}
  {Pair breaking in the presence of a normal-state pseudogap in high-{$T_c$}
  cuprates},}\ }\href@noop {} {\bibfield  {journal} {\textit {\bibinfo
  {journal} {Phys. Rev. B}}}, \bibinfo {year} {1998},\ \bibinfo {volume} {57}:\
  \bibinfo {pages} {146}}\BibitemShut {NoStop}%
\bibitem [{\citenamefont {Walker}\ \emph {et~al.}(1995)\citenamefont {Walker},
  \citenamefont {Mackenzie},\ and\ \citenamefont {Cooper}}]{Walker1995}%
  \BibitemOpen
  \bibfield  {author} {\bibinfo {author} {\bibfnamefont {D.~J.~C.}\
  \bibnamefont {Walker}}, \bibinfo {author} {\bibfnamefont {A.~P.}\
  \bibnamefont {Mackenzie}}, \ and\ \bibinfo {author} {\bibfnamefont {J.~R.}\
  \bibnamefont {Cooper}},\ }\bibfield  {title} {\enquote {\bibinfo {title}
  {Transport properties of zinc-doped {YB}$_2${Cu}$_3${O}$_{7-\delta}$ thin
  films},}\ }\href@noop {} {\bibfield  {journal} {\textit {\bibinfo  {journal}
  {Phys. Rev. B}}}, \bibinfo {year} {1995},\ \bibinfo {volume} {51}:\ \bibinfo
  {pages} {15653}}\BibitemShut {NoStop}%
\bibitem [{\citenamefont {Graf}\ \emph {et~al.}(1995)\citenamefont {Graf},
  \citenamefont {Lawrence}, \citenamefont {Hundley}, \citenamefont {Thompson},
  \citenamefont {Lacerda}, \citenamefont {Haanappel}, \citenamefont
  {Torikachcili}, \citenamefont {Fisk},\ and\ \citenamefont
  {Canfield}}]{Graf1995}%
  \BibitemOpen
  \bibfield  {author} {\bibinfo {author} {\bibfnamefont {T.}~\bibnamefont
  {Graf}}, \bibinfo {author} {\bibfnamefont {J.~M.}\ \bibnamefont {Lawrence}},
  \bibinfo {author} {\bibfnamefont {M.~F.}\ \bibnamefont {Hundley}}, \bibinfo
  {author} {\bibfnamefont {J.~D.}\ \bibnamefont {Thompson}}, \bibinfo {author}
  {\bibfnamefont {A.}~\bibnamefont {Lacerda}}, \bibinfo {author} {\bibfnamefont
  {E.}~\bibnamefont {Haanappel}}, \bibinfo {author} {\bibfnamefont {M.~S.}\
  \bibnamefont {Torikachcili}}, \bibinfo {author} {\bibfnamefont
  {Z.}~\bibnamefont {Fisk}}, \ and\ \bibinfo {author} {\bibfnamefont {P.~C.}\
  \bibnamefont {Canfield}},\ }\bibfield  {title} {\enquote {\bibinfo {title}
  {Resistivity, magnetization, and specific heat of {YbAgCu}$_4$ in high
  magnetic fields.}}\ }\href@noop {} {\bibfield  {journal} {\textit {\bibinfo
  {journal} {Phys. Rev. B}}}, \bibinfo {year} {1995},\ \bibinfo {volume} {51}:\
  \bibinfo {pages} {15053}}\BibitemShut {NoStop}%
\bibitem [{\citenamefont {Yan}\ \emph {et~al.}(1995)\citenamefont {Yan},
  \citenamefont {Matl}, \citenamefont {Harris},\ and\ \citenamefont
  {Ong}}]{Yan1995}%
  \BibitemOpen
  \bibfield  {author} {\bibinfo {author} {\bibfnamefont {Y.~F.}\ \bibnamefont
  {Yan}}, \bibinfo {author} {\bibfnamefont {P.}~\bibnamefont {Matl}}, \bibinfo
  {author} {\bibfnamefont {J.~M.}\ \bibnamefont {Harris}}, \ and\ \bibinfo
  {author} {\bibfnamefont {N.~P.}\ \bibnamefont {Ong}},\ }\bibfield  {title}
  {\enquote {\bibinfo {title} {Negative magnetoresistance in the $c$-axis
  resistivity of {Bi$_2$Sr$_2$CaCu$_2$O$_{8+ \delta}$} and
  {YBa}$_2${Cu}$_3${O}$_{6+x}$},}\ }\href@noop {} {\bibfield  {journal}
  {\textit {\bibinfo  {journal} {Phys. Rev. B}}}, \bibinfo {year} {1995},\
  \bibinfo {volume} {52}:\ \bibinfo {pages} {R751}}\BibitemShut {NoStop}%
\bibitem [{\citenamefont {Williams}\ \emph {et~al.}(1996)\citenamefont
  {Williams}, \citenamefont {Tallon}, \citenamefont {Dupree},\ and\
  \citenamefont {Michalak}}]{Williams1996}%
  \BibitemOpen
  \bibfield  {author} {\bibinfo {author} {\bibfnamefont {G.}~\bibnamefont
  {Williams}}, \bibinfo {author} {\bibfnamefont {J.~L.}\ \bibnamefont
  {Tallon}}, \bibinfo {author} {\bibfnamefont {R.}~\bibnamefont {Dupree}}, \
  and\ \bibinfo {author} {\bibfnamefont {R.}~\bibnamefont {Michalak}},\
  }\bibfield  {title} {\enquote {\bibinfo {title} {Transport and {NMR} studies
  of the effect of {Ni} substitution on superconductivity and the normal-state
  pseudogap in {YBa}$_2${Cu}$_4${O}$_8$},}\ }\href@noop {} {\bibfield
  {journal} {\textit {\bibinfo  {journal} {Phys. Rev. B}}}, \bibinfo {year}
  {1996},\ \bibinfo {volume} {54}:\ \bibinfo {pages} {9532}}\BibitemShut
  {NoStop}%
\bibitem [{\citenamefont {Williams}\ \emph {et~al.}(1997)\citenamefont
  {Williams}, \citenamefont {Tallon}, \citenamefont {Haines}, \citenamefont
  {Michalak},\ and\ \citenamefont {Dupree}}]{Williams1997}%
  \BibitemOpen
  \bibfield  {author} {\bibinfo {author} {\bibfnamefont {G.}~\bibnamefont
  {Williams}}, \bibinfo {author} {\bibfnamefont {J.~L.}\ \bibnamefont
  {Tallon}}, \bibinfo {author} {\bibfnamefont {E.~M.}\ \bibnamefont {Haines}},
  \bibinfo {author} {\bibfnamefont {R.}~\bibnamefont {Michalak}}, \ and\
  \bibinfo {author} {\bibfnamefont {R.}~\bibnamefont {Dupree}},\ }\bibfield
  {title} {\enquote {\bibinfo {title} {NMR evidence for a $d$-wave normal-state
  pseudogap.}}\ }\href@noop {} {\bibfield  {journal} {\textit {\bibinfo
  {journal} {Phys. Rev. Lett.}}}, \bibinfo {year} {1997},\ \bibinfo {volume}
  {78}:\ \bibinfo {pages} {721}}\BibitemShut {NoStop}%
\bibitem [{\citenamefont {Magishi}\ \emph {et~al.}(1996)\citenamefont
  {Magishi}, \citenamefont {Kituoka}, \citenamefont {Zheng}, \citenamefont
  {Asayama}, \citenamefont {Kondo}, \citenamefont {Shimakawa}, \citenamefont
  {Manako},\ and\ \citenamefont {Kubo}}]{Magishi1996}%
  \BibitemOpen
  \bibfield  {author} {\bibinfo {author} {\bibfnamefont {K.}~\bibnamefont
  {Magishi}}, \bibinfo {author} {\bibfnamefont {Y.}~\bibnamefont {Kituoka}},
  \bibinfo {author} {\bibfnamefont {G.-Q.}\ \bibnamefont {Zheng}}, \bibinfo
  {author} {\bibfnamefont {K.}~\bibnamefont {Asayama}}, \bibinfo {author}
  {\bibfnamefont {T.}~\bibnamefont {Kondo}}, \bibinfo {author} {\bibfnamefont
  {Y.}~\bibnamefont {Shimakawa}}, \bibinfo {author} {\bibfnamefont
  {T.}~\bibnamefont {Manako}}, \ and\ \bibinfo {author} {\bibfnamefont
  {Y.}~\bibnamefont {Kubo}},\ }\bibfield  {title} {\enquote {\bibinfo {title}
  {Spin-gap behavior in underdoped {TlSr$_2$(Lu$_{0.7}$Ca$_{0.3}$)Cu$_2$O}$_y$:
  $^{63}${Cu} and $^{205}${Tl} {NMR} studies.}}\ }\href@noop {} {\bibfield
  {journal} {\textit {\bibinfo  {journal} {Phys. Rev. B}}}, \bibinfo {year}
  {1996},\ \bibinfo {volume} {54}:\ \bibinfo {pages} {3070}}\BibitemShut
  {NoStop}%
\bibitem [{\citenamefont {Goto}\ \emph {et~al.}(1997)\citenamefont {Goto},
  \citenamefont {Yasuoka}, \citenamefont {Otzschi},\ and\ \citenamefont
  {Ueda}}]{Goto1997}%
  \BibitemOpen
  \bibfield  {author} {\bibinfo {author} {\bibfnamefont {A.}~\bibnamefont
  {Goto}}, \bibinfo {author} {\bibfnamefont {H.}~\bibnamefont {Yasuoka}},
  \bibinfo {author} {\bibfnamefont {K.}~\bibnamefont {Otzschi}}, \ and\
  \bibinfo {author} {\bibfnamefont {Y.}~\bibnamefont {Ueda}},\ }\bibfield
  {title} {\enquote {\bibinfo {title} {Phase diagram for the spin pseudogap in
  {LaBa$_2$Cu$_3$O$_y$} studied by {NMR}},}\ }\href@noop {} {\bibfield
  {journal} {\textit {\bibinfo  {journal} {Phys. Rev. B}}}, \bibinfo {year}
  {1997},\ \bibinfo {volume} {55}:\ \bibinfo {pages} {12736}}\BibitemShut
  {NoStop}%
\bibitem [{\citenamefont {Bobroff}\ \emph {et~al.}(1997)\citenamefont
  {Bobroff}, \citenamefont {Alloul}, \citenamefont {Mendels}, \citenamefont
  {Viallet}, \citenamefont {Marucco},\ and\ \citenamefont
  {Colson}}]{Bobroff1997}%
  \BibitemOpen
  \bibfield  {author} {\bibinfo {author} {\bibfnamefont {J.}~\bibnamefont
  {Bobroff}}, \bibinfo {author} {\bibfnamefont {H.}~\bibnamefont {Alloul}},
  \bibinfo {author} {\bibfnamefont {P.}~\bibnamefont {Mendels}}, \bibinfo
  {author} {\bibfnamefont {V.}~\bibnamefont {Viallet}}, \bibinfo {author}
  {\bibfnamefont {J.-F.}\ \bibnamefont {Marucco}}, \ and\ \bibinfo {author}
  {\bibfnamefont {D.}~\bibnamefont {Colson}},\ }\bibfield  {title} {\enquote
  {\bibinfo {title} {$^{17}${O} {NMR} evidence for a pseudogap in the monolayer
  {HgBa$_2$CuO$_{4+ \delta}$}},}\ }\href@noop {} {\bibfield  {journal} {\textit
  {\bibinfo  {journal} {Phys. Rev. Lett.}}}, \bibinfo {year} {1997},\ \bibinfo
  {volume} {78}:\ \bibinfo {pages} {3757}}\BibitemShut {NoStop}%
\bibitem [{\citenamefont {Ishida}\ \emph {et~al.}(1998)\citenamefont {Ishida},
  \citenamefont {Yoshida}, \citenamefont {Mito}, \citenamefont {Tokumaga},
  \citenamefont {Kitaoka}, \citenamefont {Asayama}, \citenamefont {Nakayama},
  \citenamefont {Shimoyama},\ and\ \citenamefont {Kishio}}]{Ishida1998}%
  \BibitemOpen
  \bibfield  {author} {\bibinfo {author} {\bibfnamefont {K.}~\bibnamefont
  {Ishida}}, \bibinfo {author} {\bibfnamefont {K.}~\bibnamefont {Yoshida}},
  \bibinfo {author} {\bibfnamefont {T.}~\bibnamefont {Mito}}, \bibinfo {author}
  {\bibfnamefont {Y.}~\bibnamefont {Tokumaga}}, \bibinfo {author}
  {\bibfnamefont {Y.}~\bibnamefont {Kitaoka}}, \bibinfo {author} {\bibfnamefont
  {K.}~\bibnamefont {Asayama}}, \bibinfo {author} {\bibfnamefont
  {Y.}~\bibnamefont {Nakayama}}, \bibinfo {author} {\bibfnamefont
  {J.}~\bibnamefont {Shimoyama}}, \ and\ \bibinfo {author} {\bibfnamefont
  {K.}~\bibnamefont {Kishio}},\ }\bibfield  {title} {\enquote {\bibinfo {title}
  {Pseudogap behavior in single-crystal {Bi$_2$Sr$_2$CaCu$_2$O$_{8+ \delta}$}
  probed by {Cu} {NMR}},}\ }\href@noop {} {\bibfield  {journal} {\textit
  {\bibinfo  {journal} {Phys. Rev. B}}}, \bibinfo {year} {1998},\ \bibinfo
  {volume} {58}:\ \bibinfo {pages} {R5960}}\BibitemShut {NoStop}%
\bibitem [{\citenamefont {Puchkov}\ \emph {et~al.}(1996)\citenamefont
  {Puchkov}, \citenamefont {N},\ and\ \citenamefont {T}}]{Puchkov}%
  \BibitemOpen
  \bibfield  {author} {\bibinfo {author} {\bibfnamefont {A.~V.}\ \bibnamefont
  {Puchkov}}, \bibinfo {author} {\bibfnamefont {B.~D.}\ \bibnamefont {N}}, \
  and\ \bibinfo {author} {\bibfnamefont {T.}~\bibnamefont {T}},\ }\bibfield
  {title} {\enquote {\bibinfo {title} {The pseudogap state in high {$T_c$}
  superconductors: {An} infrared study.}}\ }\href@noop {} {\bibfield  {journal}
  {\textit {\bibinfo  {journal} {J. of Phys. Cond. Matter}}}, \bibinfo {year}
  {1996},\ \bibinfo {volume} {8}:\ \bibinfo {pages} {10049}}\BibitemShut
  {NoStop}%
\bibitem [{\citenamefont {Basov}\ \emph {et~al.}(1996)\citenamefont {Basov},
  \citenamefont {Liang}, \citenamefont {Dabrowski}, \citenamefont {Bonn},
  \citenamefont {Hardy},\ and\ \citenamefont {Timusk}}]{Basov}%
  \BibitemOpen
  \bibfield  {author} {\bibinfo {author} {\bibfnamefont {D.~N.}\ \bibnamefont
  {Basov}}, \bibinfo {author} {\bibfnamefont {R.~X.}\ \bibnamefont {Liang}},
  \bibinfo {author} {\bibfnamefont {B.}~\bibnamefont {Dabrowski}}, \bibinfo
  {author} {\bibfnamefont {D.~A.}\ \bibnamefont {Bonn}}, \bibinfo {author}
  {\bibfnamefont {W.~N.}\ \bibnamefont {Hardy}}, \ and\ \bibinfo {author}
  {\bibfnamefont {T.}~\bibnamefont {Timusk}},\ }\bibfield  {title} {\enquote
  {\bibinfo {title} {Pseudogap and charge dynamics in {CuO$_2$} planes in
  {YBCO}},}\ }\href@noop {} {\bibfield  {journal} {\textit {\bibinfo  {journal}
  {Phys. Rev. Lett.}}}, \bibinfo {year} {1996},\ \bibinfo {volume} {77}:\
  \bibinfo {pages} {4090}}\BibitemShut {NoStop}%
\bibitem [{\citenamefont {Basov}\ \emph {et~al.}(2001)\citenamefont {Basov},
  \citenamefont {Homes}, \citenamefont {Singley}, \citenamefont {Strongin},
  \citenamefont {Timusk}, \citenamefont {Blumberg},\ and\ \citenamefont
  {van~der Marel}}]{Basov2}%
  \BibitemOpen
  \bibfield  {author} {\bibinfo {author} {\bibfnamefont {D.~N.}\ \bibnamefont
  {Basov}}, \bibinfo {author} {\bibfnamefont {C.~C.}\ \bibnamefont {Homes}},
  \bibinfo {author} {\bibfnamefont {E.}~\bibnamefont {Singley}}, \bibinfo
  {author} {\bibfnamefont {M.}~\bibnamefont {Strongin}}, \bibinfo {author}
  {\bibfnamefont {T.}~\bibnamefont {Timusk}}, \bibinfo {author} {\bibfnamefont
  {G.}~\bibnamefont {Blumberg}}, \ and\ \bibinfo {author} {\bibfnamefont
  {D.}~\bibnamefont {van~der Marel}},\ }\bibfield  {title} {\enquote {\bibinfo
  {title} {Unconventional energetics of the pseudogap state and superconducting
  state in high-{$T_c$} cuprates},}\ }\href@noop {} {\bibfield  {journal}
  {\textit {\bibinfo  {journal} {Phys. Rev. B}}}, \bibinfo {year} {2001},\
  \bibinfo {volume} {63}:\ \bibinfo {pages} {134514}}\BibitemShut {NoStop}%
\bibitem [{\citenamefont {Tranquada}\ \emph {et~al.}(1992)\citenamefont
  {Tranquada}, \citenamefont {Gehring}, \citenamefont {Shirane}, \citenamefont
  {Shamoto},\ and\ \citenamefont {Sato}}]{Tranquada}%
  \BibitemOpen
  \bibfield  {author} {\bibinfo {author} {\bibfnamefont {J.~M.}\ \bibnamefont
  {Tranquada}}, \bibinfo {author} {\bibfnamefont {P.~M.}\ \bibnamefont
  {Gehring}}, \bibinfo {author} {\bibfnamefont {G.}~\bibnamefont {Shirane}},
  \bibinfo {author} {\bibfnamefont {S.}~\bibnamefont {Shamoto}}, \ and\
  \bibinfo {author} {\bibfnamefont {M.}~\bibnamefont {Sato}},\ }\bibfield
  {title} {\enquote {\bibinfo {title} {Neutron-scattering study of the
  dynamical spin susceptibility in {YBa$_2$Cu$_3$O$_{6.6}$}.}}\ }\href@noop {}
  {\bibfield  {journal} {\textit {\bibinfo  {journal} {Phys. Rev. B}}},
  \bibinfo {year} {1992},\ \bibinfo {volume} {46}:\ \bibinfo {pages}
  {5561}}\BibitemShut {NoStop}%
\bibitem [{\citenamefont {Dai}\ \emph {et~al.}(2000)\citenamefont {Dai},
  \citenamefont {Mook}, \citenamefont {Hayden},\ and\ \citenamefont
  {Dogan}}]{MookH}%
  \BibitemOpen
  \bibfield  {author} {\bibinfo {author} {\bibfnamefont {P.~C.}\ \bibnamefont
  {Dai}}, \bibinfo {author} {\bibfnamefont {H.~A.}\ \bibnamefont {Mook}},
  \bibinfo {author} {\bibfnamefont {S.~M.}\ \bibnamefont {Hayden}}, \ and\
  \bibinfo {author} {\bibfnamefont {F.}~\bibnamefont {Dogan}},\ }\bibfield
  {title} {\enquote {\bibinfo {title} {The connection between superconducting
  phase correlations and spin excitations in {YBa$_2$Cu$_3$O$_{6.6}$}: A
  magnetic field study},}\ }\href@noop {} {\bibfield  {journal} {\textit
  {\bibinfo  {journal} {Nature}}}, \bibinfo {year} {2000},\ \bibinfo {volume}
  {406}:\ \bibinfo {pages} {965}}\BibitemShut {NoStop}%
\bibitem [{\citenamefont {Lake}\ \emph {et~al.}(1999)\citenamefont {Lake},
  \citenamefont {Aeppli}, \citenamefont {Mason}, \citenamefont {Schroeder},
  \citenamefont {McMorrow}, \citenamefont {Lefmann}, \citenamefont {Isshiki},
  \citenamefont {Nohara}, \citenamefont {Takagi},\ and\ \citenamefont
  {Hayden}}]{Aeppli3}%
  \BibitemOpen
  \bibfield  {author} {\bibinfo {author} {\bibfnamefont {B.}~\bibnamefont
  {Lake}}, \bibinfo {author} {\bibfnamefont {G.}~\bibnamefont {Aeppli}},
  \bibinfo {author} {\bibfnamefont {T.~E.}\ \bibnamefont {Mason}}, \bibinfo
  {author} {\bibfnamefont {A.}~\bibnamefont {Schroeder}}, \bibinfo {author}
  {\bibfnamefont {D.~F.}\ \bibnamefont {McMorrow}}, \bibinfo {author}
  {\bibfnamefont {K.}~\bibnamefont {Lefmann}}, \bibinfo {author} {\bibfnamefont
  {M.}~\bibnamefont {Isshiki}}, \bibinfo {author} {\bibfnamefont
  {M.}~\bibnamefont {Nohara}}, \bibinfo {author} {\bibfnamefont
  {H.}~\bibnamefont {Takagi}}, \ and\ \bibinfo {author} {\bibfnamefont {S.~M.}\
  \bibnamefont {Hayden}},\ }\bibfield  {title} {\enquote {\bibinfo {title}
  {Spin gap and magnetic coherence in a clean high-temperature
  superconductor},}\ }\href@noop {} {\bibfield  {journal} {\textit {\bibinfo
  {journal} {Nature}}}, \bibinfo {year} {1999},\ \bibinfo {volume} {400}:\
  \bibinfo {pages} {43}}\BibitemShut {NoStop}%
\bibitem [{\citenamefont {Ruani}\ and\ \citenamefont
  {Ricci}(1997)}]{Ruani1997}%
  \BibitemOpen
  \bibfield  {author} {\bibinfo {author} {\bibfnamefont {G.}~\bibnamefont
  {Ruani}}\ and\ \bibinfo {author} {\bibfnamefont {P.}~\bibnamefont {Ricci}},\
  }\bibfield  {title} {\enquote {\bibinfo {title} {Transitions at {$T>T_c$} in
  underdoped crystals of {YBa$_2$Cu$_3$O$_{7-x}$} observed by resonant {R}aman
  scattering},}\ }\href@noop {} {\bibfield  {journal} {\textit {\bibinfo
  {journal} {Phys. Rev. B}}}, \bibinfo {year} {1997},\ \bibinfo {volume} {55}:\
  \bibinfo {pages} {93}}\BibitemShut {NoStop}%
\bibitem [{\citenamefont {Chen}\ \emph {et~al.}(1997)\citenamefont {Chen},
  \citenamefont {Nacini}, \citenamefont {Hewitt}, \citenamefont {Irwin},
  \citenamefont {Liang},\ and\ \citenamefont {Hardy}}]{Chen1997}%
  \BibitemOpen
  \bibfield  {author} {\bibinfo {author} {\bibfnamefont {X.~K.}\ \bibnamefont
  {Chen}}, \bibinfo {author} {\bibfnamefont {J.~G.}\ \bibnamefont {Nacini}},
  \bibinfo {author} {\bibfnamefont {K.~C.}\ \bibnamefont {Hewitt}}, \bibinfo
  {author} {\bibfnamefont {J.~C.}\ \bibnamefont {Irwin}}, \bibinfo {author}
  {\bibfnamefont {R.}~\bibnamefont {Liang}}, \ and\ \bibinfo {author}
  {\bibfnamefont {W.~N.}\ \bibnamefont {Hardy}},\ }\bibfield  {title} {\enquote
  {\bibinfo {title} {Electronic {R}aman scattering in underdoped
  {YBa$_2$Cu$_3$O$_{6.5}$}},}\ }\href@noop {} {\bibfield  {journal} {\textit
  {\bibinfo  {journal} {Phys. Rev. B}}}, \bibinfo {year} {1997},\ \bibinfo
  {volume} {56}:\ \bibinfo {pages} {R513}}\BibitemShut {NoStop}%
\bibitem [{\citenamefont {Nemetschek}\ \emph {et~al.}(1997)\citenamefont
  {Nemetschek}, \citenamefont {Opel}, \citenamefont {Hoffmann}, \citenamefont
  {Muller}, \citenamefont {Hackl}, \citenamefont {Berger}, \citenamefont
  {Forro}, \citenamefont {Er},\ and\ \citenamefont {Walker}}]{Nemetschek1997}%
  \BibitemOpen
  \bibfield  {author} {\bibinfo {author} {\bibfnamefont {R.}~\bibnamefont
  {Nemetschek}}, \bibinfo {author} {\bibfnamefont {M.}~\bibnamefont {Opel}},
  \bibinfo {author} {\bibfnamefont {C.}~\bibnamefont {Hoffmann}}, \bibinfo
  {author} {\bibfnamefont {P.~F.}\ \bibnamefont {Muller}}, \bibinfo {author}
  {\bibfnamefont {R.}~\bibnamefont {Hackl}}, \bibinfo {author} {\bibfnamefont
  {H.}~\bibnamefont {Berger}}, \bibinfo {author} {\bibfnamefont
  {L.}~\bibnamefont {Forro}}, \bibinfo {author} {\bibfnamefont
  {A.}~\bibnamefont {Er}}, \ and\ \bibinfo {author} {\bibfnamefont
  {E.}~\bibnamefont {Walker}},\ }\bibfield  {title} {\enquote {\bibinfo {title}
  {Pseudogap and superconducting gap in the electronic {R}aman spectra of
  underdoped cuprates.}}\ }\href@noop {} {\bibfield  {journal} {\textit
  {\bibinfo  {journal} {Phys. Rev. Lett.}}}, \bibinfo {year} {1997},\ \bibinfo
  {volume} {78}:\ \bibinfo {pages} {4837}}\BibitemShut {NoStop}%
\bibitem [{\citenamefont {Quilty}\ \emph {et~al.}(1998)\citenamefont {Quilty},
  \citenamefont {Trodahl},\ and\ \citenamefont {Pooke}}]{Quilty1998}%
  \BibitemOpen
  \bibfield  {author} {\bibinfo {author} {\bibfnamefont {J.~W.}\ \bibnamefont
  {Quilty}}, \bibinfo {author} {\bibfnamefont {H.~J.}\ \bibnamefont {Trodahl}},
  \ and\ \bibinfo {author} {\bibfnamefont {D.~M.}\ \bibnamefont {Pooke}},\
  }\bibfield  {title} {\enquote {\bibinfo {title} {Electronic raman scattering
  from {Bi$_2$Sr$_2$CaCu$_2$O$_{8+ \delta}$}: {D}oping dependence of the
  pseudogap and anomalous 600 cm$^{-1}$ peak.}}\ }\href@noop {} {\bibfield
  {journal} {\textit {\bibinfo  {journal} {Phys. Rev. B}}}, \bibinfo {year}
  {1998},\ \bibinfo {volume} {57}:\ \bibinfo {pages} {R11097}}\BibitemShut
  {NoStop}%
\bibitem [{\citenamefont {Xu}\ \emph {et~al.}(2000)\citenamefont {Xu},
  \citenamefont {Ong}, \citenamefont {Want}, \citenamefont {Kakeshita},\ and\
  \citenamefont {Uchida}}]{Nernst}%
  \BibitemOpen
  \bibfield  {author} {\bibinfo {author} {\bibfnamefont {Z.~A.}\ \bibnamefont
  {Xu}}, \bibinfo {author} {\bibfnamefont {N.}~\bibnamefont {Ong}}, \bibinfo
  {author} {\bibfnamefont {Y.}~\bibnamefont {Want}}, \bibinfo {author}
  {\bibfnamefont {T.}~\bibnamefont {Kakeshita}}, \ and\ \bibinfo {author}
  {\bibfnamefont {S.}~\bibnamefont {Uchida}},\ }\bibfield  {title} {\enquote
  {\bibinfo {title} {Vortex-like excitations and the onset of superconducting
  phase fluctuation in underdoped {La$_{2-x}$Sr$_{x}$CuO$_{4}$}},}\ }\href@noop
  {} {\bibfield  {journal} {\textit {\bibinfo  {journal} {Nature}}}, \bibinfo
  {year} {2000},\ \bibinfo {volume} {406}:\ \bibinfo {pages} {486}}\BibitemShut
  {NoStop}%
\bibitem [{\citenamefont {Wang}\ \emph {et~al.}(2001)\citenamefont {Wang},
  \citenamefont {Xu}, \citenamefont {Kakeshita}, \citenamefont {Uchida},\ and\
  \citenamefont {Ong}}]{Ong2}%
  \BibitemOpen
  \bibfield  {author} {\bibinfo {author} {\bibfnamefont {Y.}~\bibnamefont
  {Wang}}, \bibinfo {author} {\bibfnamefont {Z.~A.}\ \bibnamefont {Xu}},
  \bibinfo {author} {\bibfnamefont {T.}~\bibnamefont {Kakeshita}}, \bibinfo
  {author} {\bibfnamefont {S.}~\bibnamefont {Uchida}}, \ and\ \bibinfo {author}
  {\bibfnamefont {N.~P.}\ \bibnamefont {Ong}},\ }\bibfield  {title} {\enquote
  {\bibinfo {title} {Onset of the vortexlike {N}ernst signal above {$T_c$} in
  {LaSrCuO} and {BISrLaCuO}},}\ }\href@noop {} {\bibfield  {journal} {\textit
  {\bibinfo  {journal} {Phys. Rev. B}}}, \bibinfo {year} {2001},\ \bibinfo
  {volume} {64}:\ \bibinfo {pages} {224519}}\BibitemShut {NoStop}%
\bibitem [{\citenamefont {Wang}\ \emph {et~al.}(2002)\citenamefont {Wang},
  \citenamefont {Ong}, \citenamefont {Xu}, \citenamefont {Kakeshita},
  \citenamefont {Uchida}, \citenamefont {Bonn}, \citenamefont {Liang},\ and\
  \citenamefont {Hardy}}]{Ong3}%
  \BibitemOpen
  \bibfield  {author} {\bibinfo {author} {\bibfnamefont {Y.~Y.}\ \bibnamefont
  {Wang}}, \bibinfo {author} {\bibfnamefont {N.~P.}\ \bibnamefont {Ong}},
  \bibinfo {author} {\bibfnamefont {Z.~A.}\ \bibnamefont {Xu}}, \bibinfo
  {author} {\bibfnamefont {T.}~\bibnamefont {Kakeshita}}, \bibinfo {author}
  {\bibfnamefont {S.}~\bibnamefont {Uchida}}, \bibinfo {author} {\bibfnamefont
  {D.}~\bibnamefont {Bonn}}, \bibinfo {author} {\bibfnamefont {R.}~\bibnamefont
  {Liang}}, \ and\ \bibinfo {author} {\bibfnamefont {W.}~\bibnamefont
  {Hardy}},\ }\bibfield  {title} {\enquote {\bibinfo {title} {The high-field
  phase diagram of the cuprates derived from the {Nernst} effect.}}\
  }\href@noop {} {\bibfield  {journal} {\textit {\bibinfo  {journal} {\prl}}},
  \bibinfo {year} {2002},\ \bibinfo {volume} {88}:\ \bibinfo {pages}
  {257003}}\BibitemShut {NoStop}%
\bibitem [{\citenamefont {Tan}\ and\ \citenamefont {Levin}(2004)}]{Tan}%
  \BibitemOpen
  \bibfield  {author} {\bibinfo {author} {\bibfnamefont {S.}~\bibnamefont
  {Tan}}\ and\ \bibinfo {author} {\bibfnamefont {K.}~\bibnamefont {Levin}},\
  }\bibfield  {title} {\enquote {\bibinfo {title} {Nernst effect and anomalous
  transport in cuprates: {A} preformed-pair alternative to the vortex
  scenario},}\ }\href@noop {} {\bibfield  {journal} {\textit {\bibinfo
  {journal} {Phys. Rev. B}}}, \bibinfo {year} {2004},\ \bibinfo {volume} {69}:\
  \bibinfo {pages} {064510}}\BibitemShut {NoStop}%
\bibitem [{\citenamefont {Loeser}\ \emph {et~al.}(1996)\citenamefont {Loeser},
  \citenamefont {Shen}, \citenamefont {Dessau}, \citenamefont {Marshall},
  \citenamefont {Park}, \citenamefont {Fournier},\ and\ \citenamefont
  {Kapitulnik}}]{arpesstanford}%
  \BibitemOpen
  \bibfield  {author} {\bibinfo {author} {\bibfnamefont {A.~G.}\ \bibnamefont
  {Loeser}}, \bibinfo {author} {\bibfnamefont {Z.-X.}\ \bibnamefont {Shen}},
  \bibinfo {author} {\bibfnamefont {D.~S.}\ \bibnamefont {Dessau}}, \bibinfo
  {author} {\bibfnamefont {D.~S.}\ \bibnamefont {Marshall}}, \bibinfo {author}
  {\bibfnamefont {C.~H.}\ \bibnamefont {Park}}, \bibinfo {author}
  {\bibfnamefont {P.}~\bibnamefont {Fournier}}, \ and\ \bibinfo {author}
  {\bibfnamefont {A.}~\bibnamefont {Kapitulnik}},\ }\bibfield  {title}
  {\enquote {\bibinfo {title} {Excitation gap in the normal state of underdoped
  {Bi$_2$Sr$_2$CaCu$_2$O$_{8+ \delta}$}.}}\ }\href@noop {} {\bibfield
  {journal} {\textit {\bibinfo  {journal} {Science}}}, \bibinfo {year} {1996},\
  \bibinfo {volume} {273}:\ \bibinfo {pages} {325}}\BibitemShut {NoStop}%
\bibitem [{\citenamefont {Kanigel}\ \emph {et~al.}(2008)\citenamefont
  {Kanigel}, \citenamefont {Chatterjee}, \citenamefont {Randeria},
  \citenamefont {Norman}, \citenamefont {Koren}, \citenamefont {Kadawaki},\
  and\ \citenamefont {Campuzano}}]{ANLPRL}%
  \BibitemOpen
  \bibfield  {author} {\bibinfo {author} {\bibfnamefont {A.}~\bibnamefont
  {Kanigel}}, \bibinfo {author} {\bibfnamefont {U.}~\bibnamefont {Chatterjee}},
  \bibinfo {author} {\bibfnamefont {M.}~\bibnamefont {Randeria}}, \bibinfo
  {author} {\bibfnamefont {M.~R.}\ \bibnamefont {Norman}}, \bibinfo {author}
  {\bibfnamefont {G.}~\bibnamefont {Koren}}, \bibinfo {author} {\bibfnamefont
  {K.}~\bibnamefont {Kadawaki}}, \ and\ \bibinfo {author} {\bibfnamefont
  {J.~C.}\ \bibnamefont {Campuzano}},\ }\bibfield  {title} {\enquote {\bibinfo
  {title} {Evidence for pairing above the transition temperature in the
  electronic dispersion of the pseudogap phase},}\ }\href@noop {} {\bibfield
  {journal} {\textit {\bibinfo  {journal} {Phys. Rev. Lett.}}}, \bibinfo {year}
  {2008},\ \bibinfo {volume} {101}:\ \bibinfo {pages} {137002}}\BibitemShut
  {NoStop}%
\bibitem [{hol()}]{holedopingfootnote}%
  \BibitemOpen
  \bibinfo {note} {For simplicity, here we do not discuss electron doping,
  which is rather similar. Further information can be found in
  Ref.~\cite{Timusk}.}\BibitemShut {Stop}%
\bibitem [{\citenamefont {Chakravarty}\ \emph {et~al.}(2001)\citenamefont
  {Chakravarty}, \citenamefont {Laughlin}, \citenamefont {Morr},\ and\
  \citenamefont {Nayak}}]{Laughlin}%
  \BibitemOpen
  \bibfield  {author} {\bibinfo {author} {\bibfnamefont {S.}~\bibnamefont
  {Chakravarty}}, \bibinfo {author} {\bibfnamefont {R.~B.}\ \bibnamefont
  {Laughlin}}, \bibinfo {author} {\bibfnamefont {D.~K.}\ \bibnamefont {Morr}},
  \ and\ \bibinfo {author} {\bibfnamefont {C.}~\bibnamefont {Nayak}},\
  }\bibfield  {title} {\enquote {\bibinfo {title} {Hidden order in cuprates},}\
  }\href@noop {} {\bibfield  {journal} {\textit {\bibinfo  {journal} {Phys.
  Rev. B}}}, \bibinfo {year} {2001},\ \bibinfo {volume} {63}:\ \bibinfo {pages}
  {094503}}\BibitemShut {NoStop}%
\bibitem [{\citenamefont {Lee}(2000)}]{LeePhysicaC2000}%
  \BibitemOpen
  \bibfield  {author} {\bibinfo {author} {\bibfnamefont {P.~A.}\ \bibnamefont
  {Lee}},\ }\bibfield  {title} {\enquote {\bibinfo {title} {High {$T_c$}
  superconductors as doped mott insulators: Fluctuating current and spin
  chirality},}\ }\href@noop {} {\bibfield  {journal} {\textit {\bibinfo
  {journal} {Physica C}}}, \bibinfo {year} {2000},\ \bibinfo {volume}
  {341-348}:\ \bibinfo {pages} {63}}\BibitemShut {NoStop}%
\bibitem [{\citenamefont {Lee}\ and\ \citenamefont
  {Wen}(2001)}]{LeeWenPRB2001}%
  \BibitemOpen
  \bibfield  {author} {\bibinfo {author} {\bibfnamefont {P.~A.}\ \bibnamefont
  {Lee}}\ and\ \bibinfo {author} {\bibfnamefont {X.-G.}\ \bibnamefont {Wen}},\
  }\bibfield  {title} {\enquote {\bibinfo {title} {Vortex structure in
  underdoped cuprates},}\ }\href@noop {} {\bibfield  {journal} {\textit
  {\bibinfo  {journal} {\prb}}}, \bibinfo {year} {2001},\ \bibinfo {volume}
  {63}:\ \bibinfo {pages} {224517}}\BibitemShut {NoStop}%
\bibitem [{\citenamefont {Honerkamp}\ and\ \citenamefont
  {Lee}(2003)}]{LeePRL2003}%
  \BibitemOpen
  \bibfield  {author} {\bibinfo {author} {\bibfnamefont {C.}~\bibnamefont
  {Honerkamp}}\ and\ \bibinfo {author} {\bibfnamefont {P.~A.}\ \bibnamefont
  {Lee}},\ }\bibfield  {title} {\enquote {\bibinfo {title} {Staggered flux
  fluctuations and the quasiparticle scattering rate in the {SU(2)} gauge
  theory of the {$t-J$} model},}\ }\href@noop {} {\bibfield  {journal} {\textit
  {\bibinfo  {journal} {\prl}}}, \bibinfo {year} {2003},\ \bibinfo {volume}
  {90}:\ \bibinfo {pages} {246402}}\BibitemShut {NoStop}%
\bibitem [{\citenamefont {Varma}(1997)}]{VarmaPRB1997}%
  \BibitemOpen
  \bibfield  {author} {\bibinfo {author} {\bibfnamefont {C.~M.}\ \bibnamefont
  {Varma}},\ }\bibfield  {title} {\enquote {\bibinfo {title} {Non-fermi-liquid
  states and pairing instability of a general model of copper oxide metals},}\
  }\href@noop {} {\bibfield  {journal} {\textit {\bibinfo  {journal} {\prb}}},
  \bibinfo {year} {1997},\ \bibinfo {volume} {55}:\ \bibinfo {pages}
  {14554}}\BibitemShut {NoStop}%
\bibitem [{\citenamefont {Varma}(2006)}]{VarmaPRB2006}%
  \BibitemOpen
  \bibfield  {author} {\bibinfo {author} {\bibfnamefont {C.~M.}\ \bibnamefont
  {Varma}},\ }\bibfield  {title} {\enquote {\bibinfo {title} {Theory of the
  pseudogap state of the cuprates},}\ }\href@noop {} {\bibfield  {journal}
  {\textit {\bibinfo  {journal} {\prb}}}, \bibinfo {year} {2006},\ \bibinfo
  {volume} {73}:\ \bibinfo {pages} {155113}}\BibitemShut {NoStop}%
\bibitem [{\citenamefont {Loram}\ \emph {et~al.}(1998)\citenamefont {Loram},
  \citenamefont {Mirza}, \citenamefont {Cooper},\ and\ \citenamefont
  {Tallon}}]{Loram98}%
  \BibitemOpen
  \bibfield  {author} {\bibinfo {author} {\bibfnamefont {J.~W.}\ \bibnamefont
  {Loram}}, \bibinfo {author} {\bibfnamefont {K.~A.}\ \bibnamefont {Mirza}},
  \bibinfo {author} {\bibfnamefont {J.~R.}\ \bibnamefont {Cooper}}, \ and\
  \bibinfo {author} {\bibfnamefont {J.~L.}\ \bibnamefont {Tallon}},\ }\bibfield
   {title} {\enquote {\bibinfo {title} {Specific heat evidence on the normal
  state pseudogap},}\ }\href@noop {} {\bibfield  {journal} {\textit {\bibinfo
  {journal} {J. Phys. Chem. Solids}}}, \bibinfo {year} {1998},\ \bibinfo
  {volume} {59}:\ \bibinfo {pages} {2091}}\BibitemShut {NoStop}%
\bibitem [{\citenamefont {Tallon}\ and\ \citenamefont
  {Loram}(2001)}]{LoramPhysicaC}%
  \BibitemOpen
  \bibfield  {author} {\bibinfo {author} {\bibfnamefont {J.~L.}\ \bibnamefont
  {Tallon}}\ and\ \bibinfo {author} {\bibfnamefont {J.~W.}\ \bibnamefont
  {Loram}},\ }\bibfield  {title} {\enquote {\bibinfo {title} {The doping
  dependence of {$T^*$} -- What is the real high-{$T_c$} phase diagram?}}\
  }\href@noop {} {\bibfield  {journal} {\textit {\bibinfo  {journal} {Physica
  C}}}, \bibinfo {year} {2001},\ \bibinfo {volume} {349}:\ \bibinfo {pages}
  {53}}\BibitemShut {NoStop}%
\bibitem [{\citenamefont {Chen}\ \emph {et~al.}(2001)\citenamefont {Chen},
  \citenamefont {Levin},\ and\ \citenamefont {Kosztin}}]{Chen4}%
  \BibitemOpen
  \bibfield  {author} {\bibinfo {author} {\bibfnamefont {Q.~J.}\ \bibnamefont
  {Chen}}, \bibinfo {author} {\bibfnamefont {K.}~\bibnamefont {Levin}}, \ and\
  \bibinfo {author} {\bibfnamefont {I.}~\bibnamefont {Kosztin}},\ }\bibfield
  {title} {\enquote {\bibinfo {title} {Superconducting phase coherence in the
  presence of a pseudogap: {R}elation to specific heat, tunneling, and vortex
  core spectroscopies.}}\ }\href@noop {} {\bibfield  {journal} {\textit
  {\bibinfo  {journal} {Phys. Rev. B}}}, \bibinfo {year} {2001},\ \bibinfo
  {volume} {63}:\ \bibinfo {pages} {184519}}\BibitemShut {NoStop}%
\bibitem [{\citenamefont {Anderson}(1987)}]{RVB}%
  \BibitemOpen
  \bibfield  {author} {\bibinfo {author} {\bibfnamefont {P.~W.}\ \bibnamefont
  {Anderson}},\ }\bibfield  {title} {\enquote {\bibinfo {title} {The resonating
  valence bond state in {La$_2$CuO$_4$} and superconductivity},}\ }\href@noop
  {} {\bibfield  {journal} {\textit {\bibinfo  {journal} {Science}}}, \bibinfo
  {year} {1987},\ \bibinfo {volume} {235}:\ \bibinfo {pages}
  {1196}}\BibitemShut {NoStop}%
\bibitem [{\citenamefont {Anderson}\ \emph {et~al.}(2004)\citenamefont
  {Anderson}, \citenamefont {Lee}, \citenamefont {Randeria}, \citenamefont
  {Rice}, \citenamefont {Trivedi},\ and\ \citenamefont {Zhang}}]{Vanilla}%
  \BibitemOpen
  \bibfield  {author} {\bibinfo {author} {\bibfnamefont {P.~W.}\ \bibnamefont
  {Anderson}}, \bibinfo {author} {\bibfnamefont {P.~A.}\ \bibnamefont {Lee}},
  \bibinfo {author} {\bibfnamefont {M.}~\bibnamefont {Randeria}}, \bibinfo
  {author} {\bibfnamefont {T.~M.}\ \bibnamefont {Rice}}, \bibinfo {author}
  {\bibfnamefont {N.}~\bibnamefont {Trivedi}}, \ and\ \bibinfo {author}
  {\bibfnamefont {F.~C.}\ \bibnamefont {Zhang}},\ }\bibfield  {title} {\enquote
  {\bibinfo {title} {The physics behind high-temperature superconducting
  cuprates: {T}he "plain vanilla" version of {RVB}},}\ }\href@noop {}
  {\bibfield  {journal} {\textit {\bibinfo  {journal} {J. Phys. - Condens.
  Matter.}}}, \bibinfo {year} {2004},\ \bibinfo {volume} {16}:\ \bibinfo
  {pages} {R755}}\BibitemShut {NoStop}%
\bibitem [{\citenamefont {Nagaosa}\ and\ \citenamefont
  {Lee}(1992)}]{SpinChargeSeparation}%
  \BibitemOpen
  \bibfield  {author} {\bibinfo {author} {\bibfnamefont {N.}~\bibnamefont
  {Nagaosa}}\ and\ \bibinfo {author} {\bibfnamefont {P.~A.}\ \bibnamefont
  {Lee}},\ }\bibfield  {title} {\enquote {\bibinfo {title} {Ginzburg-landau
  theory of the spin-charge-separated system},}\ }\href@noop {} {\bibfield
  {journal} {\textit {\bibinfo  {journal} {\prb}}}, \bibinfo {year} {1992},\
  \bibinfo {volume} {45}:\ \bibinfo {pages} {966}}\BibitemShut {NoStop}%
\bibitem [{Spi()}]{Spin-ChargeSeparation}%
  \BibitemOpen
  \bibinfo {note} {For a review of spin-charge separation, see P. A. Lee,
  \textit{Physica C}, 1999, 317-318: 194.}\BibitemShut {Stop}%
\bibitem [{\citenamefont {Uemura}\ \emph {et~al.}(1989)\citenamefont {Uemura},
  \citenamefont {Luke}, \citenamefont {Sternlieb}, \citenamefont {Brewer},
  \citenamefont {Carolan}, \citenamefont {Hardy}, \citenamefont {Kadono},
  \citenamefont {Kempton}, \citenamefont {Kiefl}, \citenamefont {Kreitzman},
  \citenamefont {Mulhern}, \citenamefont {Riseman}, \citenamefont {Williams},
  \citenamefont {Yang}, \citenamefont {Uchida}, \citenamefont {Takagi},
  \citenamefont {Gopalakrishnan}, \citenamefont {Sleight}, \citenamefont
  {Subramanian}, \citenamefont {Chien}, \citenamefont {Cieplak}, \citenamefont
  {Xiao}, \citenamefont {Lee}, \citenamefont {Statt}, \citenamefont {Stronach},
  \citenamefont {Kossler},\ and\ \citenamefont {Yu}}]{UemuraPRL89}%
  \BibitemOpen
  \bibfield  {author} {\bibinfo {author} {\bibfnamefont {Y.~J.}\ \bibnamefont
  {Uemura}}, \bibinfo {author} {\bibfnamefont {G.~M.}\ \bibnamefont {Luke}},
  \bibinfo {author} {\bibfnamefont {B.~J.}\ \bibnamefont {Sternlieb}}, \bibinfo
  {author} {\bibfnamefont {J.~H.}\ \bibnamefont {Brewer}}, \bibinfo {author}
  {\bibfnamefont {J.~F.}\ \bibnamefont {Carolan}}, \bibinfo {author}
  {\bibfnamefont {W.~N.}\ \bibnamefont {Hardy}}, \bibinfo {author}
  {\bibfnamefont {R.}~\bibnamefont {Kadono}}, \bibinfo {author} {\bibfnamefont
  {J.~R.}\ \bibnamefont {Kempton}}, \bibinfo {author} {\bibfnamefont {R.~F.}\
  \bibnamefont {Kiefl}}, \bibinfo {author} {\bibfnamefont {S.~R.}\ \bibnamefont
  {Kreitzman}}, \bibinfo {author} {\bibfnamefont {P.}~\bibnamefont {Mulhern}},
  \bibinfo {author} {\bibfnamefont {T.~M.}\ \bibnamefont {Riseman}}, \bibinfo
  {author} {\bibfnamefont {D.~L.}\ \bibnamefont {Williams}}, \bibinfo {author}
  {\bibfnamefont {B.~X.}\ \bibnamefont {Yang}}, \bibinfo {author}
  {\bibfnamefont {S.}~\bibnamefont {Uchida}}, \bibinfo {author} {\bibfnamefont
  {H.}~\bibnamefont {Takagi}}, \bibinfo {author} {\bibfnamefont
  {J.}~\bibnamefont {Gopalakrishnan}}, \bibinfo {author} {\bibfnamefont
  {A.~W.}\ \bibnamefont {Sleight}}, \bibinfo {author} {\bibfnamefont {M.~A.}\
  \bibnamefont {Subramanian}}, \bibinfo {author} {\bibfnamefont {C.~L.}\
  \bibnamefont {Chien}}, \bibinfo {author} {\bibfnamefont {M.~Z.}\ \bibnamefont
  {Cieplak}}, \bibinfo {author} {\bibfnamefont {G.}~\bibnamefont {Xiao}},
  \bibinfo {author} {\bibfnamefont {V.~Y.}\ \bibnamefont {Lee}}, \bibinfo
  {author} {\bibfnamefont {B.~W.}\ \bibnamefont {Statt}}, \bibinfo {author}
  {\bibfnamefont {C.~E.}\ \bibnamefont {Stronach}}, \bibinfo {author}
  {\bibfnamefont {W.~J.}\ \bibnamefont {Kossler}}, \ and\ \bibinfo {author}
  {\bibfnamefont {X.~H.}\ \bibnamefont {Yu}},\ }\bibfield  {title} {\enquote
  {\bibinfo {title} {Universal correlations between {$T_c$} and $n_s/m^*$
  (carrier density over effective mass) in high-{$T_c$} cuprate
  superconductors},}\ }\href@noop {} {\bibfield  {journal} {\textit {\bibinfo
  {journal} {\prl}}}, \bibinfo {year} {1989},\ \bibinfo {volume} {62}:\
  \bibinfo {pages} {2317}}\BibitemShut {NoStop}%
\bibitem [{\citenamefont {Uemura}(1997)}]{Uemura}%
  \BibitemOpen
  \bibfield  {author} {\bibinfo {author} {\bibfnamefont {Y.~J.}\ \bibnamefont
  {Uemura}},\ }\bibfield  {title} {\enquote {\bibinfo {title} {{Bose-Einstein
  to BCS} crossover picture for high-{$T_c$} cuprates.}}\ }\href@noop {}
  {\bibfield  {journal} {\textit {\bibinfo  {journal} {Physica C}}}, \bibinfo
  {year} {1997},\ \bibinfo {volume} {282-287}:\ \bibinfo {pages}
  {194}}\BibitemShut {NoStop}%
\bibitem [{\citenamefont {Mishra}\ \emph {et~al.}(2014)\citenamefont {Mishra},
  \citenamefont {Chatterjee}, \citenamefont {Campuzano},\ and\ \citenamefont
  {Norman}}]{Norman2014}%
  \BibitemOpen
  \bibfield  {author} {\bibinfo {author} {\bibfnamefont {V.}~\bibnamefont
  {Mishra}}, \bibinfo {author} {\bibfnamefont {U.}~\bibnamefont {Chatterjee}},
  \bibinfo {author} {\bibfnamefont {J.~C.}\ \bibnamefont {Campuzano}}, \ and\
  \bibinfo {author} {\bibfnamefont {M.~R.}\ \bibnamefont {Norman}},\ }\bibfield
   {title} {\enquote {\bibinfo {title} {Effect of the pseudogap on the
  transition temperature in the cuprates and implications for its origin},}\
  }\href@noop {} {\bibfield  {journal} {\textit {\bibinfo  {journal} {Nat.
  Phys.}}}, \bibinfo {year} {2014},\ \bibinfo {volume} {10}:\ \bibinfo {pages}
  {357}}\BibitemShut {NoStop}%
\bibitem [{\citenamefont {Emery}\ and\ \citenamefont {Kivelson}(1995)}]{Emery}%
  \BibitemOpen
  \bibfield  {author} {\bibinfo {author} {\bibfnamefont {V.~J.}\ \bibnamefont
  {Emery}}\ and\ \bibinfo {author} {\bibfnamefont {S.~A.}\ \bibnamefont
  {Kivelson}},\ }\bibfield  {title} {\enquote {\bibinfo {title} {Importance of
  phase fluctuations in superconductors with small superfluid density.}}\
  }\href@noop {} {\bibfield  {journal} {\textit {\bibinfo  {journal}
  {Nature}}}, \bibinfo {year} {1995},\ \bibinfo {volume} {374}:\ \bibinfo
  {pages} {434}}\BibitemShut {NoStop}%
\bibitem [{\citenamefont {Franz}\ \emph {et~al.}(2002)\citenamefont {Franz},
  \citenamefont {Tesanovic},\ and\ \citenamefont {Vafek}}]{TesanovicQED3}%
  \BibitemOpen
  \bibfield  {author} {\bibinfo {author} {\bibfnamefont {M.}~\bibnamefont
  {Franz}}, \bibinfo {author} {\bibfnamefont {Z.}~\bibnamefont {Tesanovic}}, \
  and\ \bibinfo {author} {\bibfnamefont {O.}~\bibnamefont {Vafek}},\ }\bibfield
   {title} {\enquote {\bibinfo {title} {QED3 theory of pairing pseudogap in
  cuprates: From $d$-wave superconductor to antiferromagnet via `algebraic' Fermi
  liquid},}\ }\href@noop {} {\bibfield  {journal} {\textit {\bibinfo  {journal}
  {Phys. Rev. B}}}, \bibinfo {year} {2002},\ \bibinfo {volume} {66}:\ \bibinfo
  {pages} {054535}}\BibitemShut {NoStop}%
\bibitem [{\citenamefont {Ussishkin}\ \emph {et~al.}(2002)\citenamefont
  {Ussishkin}, \citenamefont {Sondhi},\ and\ \citenamefont {Huse}}]{Ussishkin}%
  \BibitemOpen
  \bibfield  {author} {\bibinfo {author} {\bibfnamefont {I.}~\bibnamefont
  {Ussishkin}}, \bibinfo {author} {\bibfnamefont {S.~L.}\ \bibnamefont
  {Sondhi}}, \ and\ \bibinfo {author} {\bibfnamefont {D.~A.}\ \bibnamefont
  {Huse}},\ }\bibfield  {title} {\enquote {\bibinfo {title} {Gaussian
  superconducting fluctuations, thermal transport, and the nernst effect},}\
  }\href@noop {} {\bibfield  {journal} {\textit {\bibinfo  {journal} {\prl}}},
  \bibinfo {year} {2002},\ \bibinfo {volume} {89}:\ \bibinfo {pages}
  {287001}}\BibitemShut {NoStop}%
\bibitem [{\citenamefont {Milstein}\ \emph {et~al.}(2002)\citenamefont
  {Milstein}, \citenamefont {Kokkelmans},\ and\ \citenamefont
  {Holland}}]{Milstein}%
  \BibitemOpen
  \bibfield  {author} {\bibinfo {author} {\bibfnamefont {J.~N.}\ \bibnamefont
  {Milstein}}, \bibinfo {author} {\bibfnamefont {S.~J. J. M.~F.}\ \bibnamefont
  {Kokkelmans}}, \ and\ \bibinfo {author} {\bibfnamefont {M.~J.}\ \bibnamefont
  {Holland}},\ }\bibfield  {title} {\enquote {\bibinfo {title} {Resonance
  theory of the crossover from {Bardeen-Cooper-Schrieffer} superfluidity to
  {Bose-Einstein} condensation in a dilute {F}ermi gas},}\ }\href@noop {}
  {\bibfield  {journal} {\textit {\bibinfo  {journal} {Phys. Rev. A}}},
  \bibinfo {year} {2002},\ \bibinfo {volume} {66}:\ \bibinfo {pages}
  {043604}}\BibitemShut {NoStop}%
\bibitem [{\citenamefont {Ohashi}\ and\ \citenamefont
  {Griffin}(2002)}]{Griffin}%
  \BibitemOpen
  \bibfield  {author} {\bibinfo {author} {\bibfnamefont {Y.}~\bibnamefont
  {Ohashi}}\ and\ \bibinfo {author} {\bibfnamefont {A.}~\bibnamefont
  {Griffin}},\ }\bibfield  {title} {\enquote {\bibinfo {title} {{BCS-BEC}
  crossover in a gas of fermi atoms with a feshbach resonance},}\ }\href@noop
  {} {\bibfield  {journal} {\textit {\bibinfo  {journal} {Phys. Rev. Lett.}}},
  \bibinfo {year} {2002},\ \bibinfo {volume} {89}:\ \bibinfo {pages}
  {130402}}\BibitemShut {NoStop}%
\bibitem [{\citenamefont {Andrenacci}\ \emph {et~al.}(2003)\citenamefont
  {Andrenacci}, \citenamefont {Pieri},\ and\ \citenamefont
  {Strinati}}]{Strinati6}%
  \BibitemOpen
  \bibfield  {author} {\bibinfo {author} {\bibfnamefont {N.}~\bibnamefont
  {Andrenacci}}, \bibinfo {author} {\bibfnamefont {P.}~\bibnamefont {Pieri}}, \
  and\ \bibinfo {author} {\bibfnamefont {G.~C.}\ \bibnamefont {Strinati}},\
  }\bibfield  {title} {\enquote {\bibinfo {title} {Evolution from {BCS}
  superconductivity to {B}ose {E}instein condensation: {C}urrent correlation
  function in the broken symmetry phase},}\ }\href@noop {} {\bibfield
  {journal} {\textit {\bibinfo  {journal} {\prb}}}, \bibinfo {year} {2003},\
  \bibinfo {volume} {68}:\ \bibinfo {pages} {144507}}\BibitemShut {NoStop}%
\bibitem [{\citenamefont {Perali}\ \emph {et~al.}(2004)\citenamefont {Perali},
  \citenamefont {Pieri}, \citenamefont {Pisani},\ and\ \citenamefont
  {Strinati}}]{Strinati4}%
  \BibitemOpen
  \bibfield  {author} {\bibinfo {author} {\bibfnamefont {A.}~\bibnamefont
  {Perali}}, \bibinfo {author} {\bibfnamefont {P.}~\bibnamefont {Pieri}},
  \bibinfo {author} {\bibfnamefont {L.}~\bibnamefont {Pisani}}, \ and\ \bibinfo
  {author} {\bibfnamefont {G.~C.}\ \bibnamefont {Strinati}},\ }\bibfield
  {title} {\enquote {\bibinfo {title} {{BCS-BEC} crossover at finite
  temperature for superfluid trapped {F}ermi atoms},}\ }\href@noop {}
  {\bibfield  {journal} {\textit {\bibinfo  {journal} {Phys. Rev. Lett.}}},
  \bibinfo {year} {2004},\ \bibinfo {volume} {92}:\ \bibinfo {pages}
  {220404}}\BibitemShut {NoStop}%
\bibitem [{\citenamefont {Hu}\ \emph {et~al.}(2007)\citenamefont {Hu},
  \citenamefont {Drummond},\ and\ \citenamefont {Liu}}]{Drummond3}%
  \BibitemOpen
  \bibfield  {author} {\bibinfo {author} {\bibfnamefont {H.}~\bibnamefont
  {Hu}}, \bibinfo {author} {\bibfnamefont {P.~D.}\ \bibnamefont {Drummond}}, \
  and\ \bibinfo {author} {\bibfnamefont {X.~J.}\ \bibnamefont {Liu}},\
  }\bibfield  {title} {\enquote {\bibinfo {title} {Universal thermodynamics of
  strongly interacting {F}ermi gases},}\ }\href@noop {} {\bibfield  {journal}
  {\textit {\bibinfo  {journal} {Nat. Phys.}}}, \bibinfo {year} {2007},\
  \bibinfo {volume} {3}:\ \bibinfo {pages} {469}}\BibitemShut {NoStop}%
\bibitem [{\citenamefont {Levin}\ \emph {et~al.}(2010)\citenamefont {Levin},
  \citenamefont {Chen}, \citenamefont {He},\ and\ \citenamefont
  {Chien}}]{Levin_AnnPhys}%
  \BibitemOpen
  \bibfield  {author} {\bibinfo {author} {\bibfnamefont {K.}~\bibnamefont
  {Levin}}, \bibinfo {author} {\bibfnamefont {Q.~J.}\ \bibnamefont {Chen}},
  \bibinfo {author} {\bibfnamefont {Y.}~\bibnamefont {He}}, \ and\ \bibinfo
  {author} {\bibfnamefont {C.-C.}\ \bibnamefont {Chien}},\ }\bibfield  {title}
  {\enquote {\bibinfo {title} {Comparison of different pairing fluctuation
  approaches to {BCS-BEC} crossover},}\ }\href@noop {} {\bibfield  {journal}
  {\textit {\bibinfo  {journal} {Ann. Phys.}}}, \bibinfo {year} {2010},\
  \bibinfo {volume} {325}:\ \bibinfo {pages} {233}}\BibitemShut {NoStop}%
\bibitem [{\citenamefont {Bickers}\ \emph {et~al.}(1989)\citenamefont
  {Bickers}, \citenamefont {Scalapino},\ and\ \citenamefont {White}}]{FLEX1}%
  \BibitemOpen
  \bibfield  {author} {\bibinfo {author} {\bibfnamefont {N.~E.}\ \bibnamefont
  {Bickers}}, \bibinfo {author} {\bibfnamefont {D.~J.}\ \bibnamefont
  {Scalapino}}, \ and\ \bibinfo {author} {\bibfnamefont {S.~R.}\ \bibnamefont
  {White}},\ }\bibfield  {title} {\enquote {\bibinfo {title} {Conserving
  approximations for strongly correlated electron systems: {Bethe-S}alpeter
  equation and dynamics for the two dimensional {H}ubbard model},}\ }\href@noop
  {} {\bibfield  {journal} {\textit {\bibinfo  {journal} {Phys. Rev. Lett.}}},
  \bibinfo {year} {1989},\ \bibinfo {volume} {62}:\ \bibinfo {pages}
  {961}}\BibitemShut {NoStop}%
\bibitem [{\citenamefont {Bickers}\ and\ \citenamefont
  {Scalapino}(1989)}]{FLEX2}%
  \BibitemOpen
  \bibfield  {author} {\bibinfo {author} {\bibfnamefont {N.~E.}\ \bibnamefont
  {Bickers}}\ and\ \bibinfo {author} {\bibfnamefont {D.~J.}\ \bibnamefont
  {Scalapino}},\ }\bibfield  {title} {\enquote {\bibinfo {title} {Conserving
  approximations for strongly fluctuating electron systems {(I)}: {F}ormalism
  and calculational approach},}\ }\href@noop {} {\bibfield  {journal} {\textit
  {\bibinfo  {journal} {Ann. Phys.}}}, \bibinfo {year} {1989},\ \bibinfo
  {volume} {193}:\ \bibinfo {pages} {206}}\BibitemShut {NoStop}%
\bibitem [{\citenamefont {Haussmann}\ \emph {et~al.}(2007)\citenamefont
  {Haussmann}, \citenamefont {Rantner}, \citenamefont {Cerrito},\ and\
  \citenamefont {Zwerger}}]{Zwerger}%
  \BibitemOpen
  \bibfield  {author} {\bibinfo {author} {\bibfnamefont {R.}~\bibnamefont
  {Haussmann}}, \bibinfo {author} {\bibfnamefont {W.}~\bibnamefont {Rantner}},
  \bibinfo {author} {\bibfnamefont {S.}~\bibnamefont {Cerrito}}, \ and\
  \bibinfo {author} {\bibfnamefont {W.}~\bibnamefont {Zwerger}},\ }\bibfield
  {title} {\enquote {\bibinfo {title} {Thermodynamics of the {BCS-BEC}
  crossover},}\ }\href@noop {} {\bibfield  {journal} {\textit {\bibinfo
  {journal} {\pra}}}, \bibinfo {year} {2007},\ \bibinfo {volume} {75}:\
  \bibinfo {pages} {023610}}\BibitemShut {NoStop}%
\bibitem [{\citenamefont {Ryota~Watanabe}(2010)}]{OhashiPRA82}%
  \BibitemOpen
  \bibfield  {author} {\bibinfo {author} {\bibfnamefont {Y.~O.}\ \bibnamefont
  {Ryota~Watanabe}, \bibfnamefont {Shunji~Tsuchiya}},\ }\bibfield  {title}
  {\enquote {\bibinfo {title} {Superfluid density of states and pseudogap
  phenomenon in the {BCS-BEC} crossover regime of a superfluid {F}ermi gas},}\
  }\href@noop {} {\bibfield  {journal} {\textit {\bibinfo  {journal} {Phys.
  Rev. A}}}, \bibinfo {year} {2010},\ \bibinfo {volume} {82}:\ \bibinfo {pages}
  {043630}}\BibitemShut {NoStop}%
\bibitem [{\citenamefont {Magierski}\ \emph {et~al.}(2009)\citenamefont
  {Magierski}, \citenamefont {Wlazlowski}, \citenamefont {Bulgac},\ and\
  \citenamefont {Drut}}]{Bulgac2009_rep}%
  \BibitemOpen
  \bibfield  {author} {\bibinfo {author} {\bibfnamefont {P.}~\bibnamefont
  {Magierski}}, \bibinfo {author} {\bibfnamefont {G.}~\bibnamefont
  {Wlazlowski}}, \bibinfo {author} {\bibfnamefont {A.}~\bibnamefont {Bulgac}},
  \ and\ \bibinfo {author} {\bibfnamefont {J.~E.}\ \bibnamefont {Drut}},\
  }\bibfield  {title} {\enquote {\bibinfo {title} {The finite temperature
  pairing gap of a unitary {F}ermi gas by quantum {Monte Carlo}
  calculations},}\ }\href@noop {} {\bibfield  {journal} {\textit {\bibinfo
  {journal} {Phys. Rev. Lett.}}}, \bibinfo {year} {2009},\ \bibinfo {volume}
  {103}:\ \bibinfo {pages} {210403}}\BibitemShut {NoStop}%
\bibitem [{\citenamefont {Pieri}\ \emph {et~al.}(2011)\citenamefont {Pieri},
  \citenamefont {Perali}, \citenamefont {Strinati}, \citenamefont {Riedl},
  \citenamefont {Wright}, \citenamefont {Altmeyer}, \citenamefont {Kohstall},
  \citenamefont {Guajardo}, \citenamefont {Denschlag},\ and\ \citenamefont
  {Grimm}}]{PieriPRA84}%
  \BibitemOpen
  \bibfield  {author} {\bibinfo {author} {\bibfnamefont {P.}~\bibnamefont
  {Pieri}}, \bibinfo {author} {\bibfnamefont {A.}~\bibnamefont {Perali}},
  \bibinfo {author} {\bibfnamefont {G.~C.}\ \bibnamefont {Strinati}}, \bibinfo
  {author} {\bibfnamefont {S.}~\bibnamefont {Riedl}}, \bibinfo {author}
  {\bibfnamefont {M.~J.}\ \bibnamefont {Wright}}, \bibinfo {author}
  {\bibfnamefont {A.}~\bibnamefont {Altmeyer}}, \bibinfo {author}
  {\bibfnamefont {C.}~\bibnamefont {Kohstall}}, \bibinfo {author}
  {\bibfnamefont {E.~R.~S.}\ \bibnamefont {Guajardo}}, \bibinfo {author}
  {\bibfnamefont {J.~H.}\ \bibnamefont {Denschlag}}, \ and\ \bibinfo {author}
  {\bibfnamefont {R.}~\bibnamefont {Grimm}},\ }\bibfield  {title} {\enquote
  {\bibinfo {title} {Pairing-gap, pseudo-gap, and no-gap phases in the
  radio-frequency spectra of a trapped unitary $^6${Li} gas},}\ }\href@noop {}
  {\bibfield  {journal} {\textit {\bibinfo  {journal} {Phys. Rev. A}}},
  \bibinfo {year} {2011},\ \bibinfo {volume} {84}:\ \bibinfo {pages}
  {011608(R)}}\BibitemShut {NoStop}%
\bibitem [{\citenamefont {Kadanoff}\ and\ \citenamefont
  {Martin}(1961)}]{Kadanoff}%
  \BibitemOpen
  \bibfield  {author} {\bibinfo {author} {\bibfnamefont {L.~P.}\ \bibnamefont
  {Kadanoff}}\ and\ \bibinfo {author} {\bibfnamefont {P.~C.}\ \bibnamefont
  {Martin}},\ }\bibfield  {title} {\enquote {\bibinfo {title} {Theory of
  many-particle systems. {II}. {Superconductivity}.}}\ }\href@noop {}
  {\bibfield  {journal} {\textit {\bibinfo  {journal} {Phys. Rev.}}}, \bibinfo
  {year} {1961},\ \bibinfo {volume} {124}:\ \bibinfo {pages} {670}}\BibitemShut
  {NoStop}%
\bibitem [{\citenamefont {Stajic}\ \emph {et~al.}(2004)\citenamefont {Stajic},
  \citenamefont {Milstein}, \citenamefont {Chen}, \citenamefont {Chiofalo},
  \citenamefont {Holland},\ and\ \citenamefont {Levin}}]{JS2}%
  \BibitemOpen
  \bibfield  {author} {\bibinfo {author} {\bibfnamefont {J.}~\bibnamefont
  {Stajic}}, \bibinfo {author} {\bibfnamefont {J.~N.}\ \bibnamefont
  {Milstein}}, \bibinfo {author} {\bibfnamefont {Q.~J.}\ \bibnamefont {Chen}},
  \bibinfo {author} {\bibfnamefont {M.~L.}\ \bibnamefont {Chiofalo}}, \bibinfo
  {author} {\bibfnamefont {M.~J.}\ \bibnamefont {Holland}}, \ and\ \bibinfo
  {author} {\bibfnamefont {K.}~\bibnamefont {Levin}},\ }\bibfield  {title}
  {\enquote {\bibinfo {title} {The nature of superfluidity in ultracold {F}ermi
  gases near {F}eshbach resonances},}\ }\href@noop {} {\bibfield  {journal}
  {\textit {\bibinfo  {journal} {Phys. Rev. A}}}, \bibinfo {year} {2004},\
  \bibinfo {volume} {69}:\ \bibinfo {pages} {063610}}\BibitemShut {NoStop}%
\bibitem [{sep()}]{separableinteraction}%
  \BibitemOpen
  \bibinfo {note} {While a general interaction $V(\mathbf{k-k'})$ may not be
  separable, it can however be decomposed into different channels as
  $V(\mathbf{k-k'}) = \sum_l \varphi^l_\mathbf{k}\varphi^l_\mathbf{k'}$, where
  $\varphi^l_\mathbf{k}$ represents $s$-, $p$-, $d$-wave channels, etc. In most
  cases, only one channel dominates the superfluid order so that we may neglect
  other channels. In this way, the use of a separable potential is
  justified.}\BibitemShut {Stop}%
\bibitem [{\citenamefont {Kokkelmans}\ \emph {et~al.}(2002)\citenamefont
  {Kokkelmans}, \citenamefont {Milstein}, \citenamefont {Chiofalo},
  \citenamefont {Walser},\ and\ \citenamefont {Holland}}]{Kokkelmans}%
  \BibitemOpen
  \bibfield  {author} {\bibinfo {author} {\bibfnamefont {S.~J. J. M.~F.}\
  \bibnamefont {Kokkelmans}}, \bibinfo {author} {\bibfnamefont {J.~N.}\
  \bibnamefont {Milstein}}, \bibinfo {author} {\bibfnamefont {M.~L.}\
  \bibnamefont {Chiofalo}}, \bibinfo {author} {\bibfnamefont {R.}~\bibnamefont
  {Walser}}, \ and\ \bibinfo {author} {\bibfnamefont {M.~J.}\ \bibnamefont
  {Holland}},\ }\bibfield  {title} {\enquote {\bibinfo {title} {Resonance
  superfluidity: Renormalization of resonance scattering theory},}\ }\href@noop
  {} {\bibfield  {journal} {\textit {\bibinfo  {journal} {Phys. Rev. A}}},
  \bibinfo {year} {2002},\ \bibinfo {volume} {65}:\ \bibinfo {pages}
  {053617}}\BibitemShut {NoStop}%
\bibitem [{not({\natexlab{a}})}]{noteonatomicgases}%
  \BibitemOpen
  \bibinfo {note} {Here we will mainly discuss $s$-wave short range contact
  potential for atomic Fermi gases. At present, $p$-wave superfluids are not
  yet available experimentally in atomic Fermi gases.}\BibitemShut {Stop}%
\bibitem [{\citenamefont {Guo}\ \emph {et~al.}(2009)\citenamefont {Guo},
  \citenamefont {Chien}, \citenamefont {Chen}, \citenamefont {He},\ and\
  \citenamefont {Levin}}]{Guo2009PRA}%
  \BibitemOpen
  \bibfield  {author} {\bibinfo {author} {\bibfnamefont {H.}~\bibnamefont
  {Guo}}, \bibinfo {author} {\bibfnamefont {C.-C.}\ \bibnamefont {Chien}},
  \bibinfo {author} {\bibfnamefont {Q.~J.}\ \bibnamefont {Chen}}, \bibinfo
  {author} {\bibfnamefont {Y.}~\bibnamefont {He}}, \ and\ \bibinfo {author}
  {\bibfnamefont {K.}~\bibnamefont {Levin}},\ }\bibfield  {title} {\enquote
  {\bibinfo {title} {Finite-temperature behavior of an interspecies fermionic
  superfluid with population imbalance},}\ }\href {\doibase
  10.1103/PhysRevA.80.011601} {\bibfield  {journal} {\textit {\bibinfo
  {journal} {Phys. Rev. A}}}, \bibinfo {year} {2009},\ \bibinfo {volume} {80}:\
  \bibinfo {pages} {011601(R)}}\BibitemShut {NoStop}%
\bibitem [{\citenamefont {Wang}\ \emph {et~al.}()\citenamefont {Wang},
  \citenamefont {Che}, \citenamefont {Zhang},\ and\ \citenamefont
  {Chen}}]{WangFFLO}%
  \BibitemOpen
  \bibfield  {author} {\bibinfo {author} {\bibfnamefont {J.~B.}\ \bibnamefont
  {Wang}}, \bibinfo {author} {\bibfnamefont {Y.~M.}\ \bibnamefont {Che}},
  \bibinfo {author} {\bibfnamefont {L.~F.}\ \bibnamefont {Zhang}}, \ and\
  \bibinfo {author} {\bibfnamefont {Q.~J.}\ \bibnamefont {Chen}},\ }\href@noop
  {} {\enquote {\bibinfo {title} {Searching for the elusive exotic
  {Fulde-Ferrell-Larkin-Ovchinnikov} states in {Fermi-Fermi} mixtures of
  ultracold quantum gases},}\ }\bibinfo {note} {E-print
  arXiv:1404.5696}\BibitemShut {NoStop}%
\bibitem [{\citenamefont {Chien}\ \emph {et~al.}(2006)\citenamefont {Chien},
  \citenamefont {Chen}, \citenamefont {He},\ and\ \citenamefont
  {Levin}}]{Chien06}%
  \BibitemOpen
  \bibfield  {author} {\bibinfo {author} {\bibfnamefont {C.~C.}\ \bibnamefont
  {Chien}}, \bibinfo {author} {\bibfnamefont {Q.~J.}\ \bibnamefont {Chen}},
  \bibinfo {author} {\bibfnamefont {Y.}~\bibnamefont {He}}, \ and\ \bibinfo
  {author} {\bibfnamefont {K.}~\bibnamefont {Levin}},\ }\bibfield  {title}
  {\enquote {\bibinfo {title} {Intermediate temperature superfluidity in a
  {F}ermi gas with population imbalance},}\ }\href@noop {} {\bibfield
  {journal} {\textit {\bibinfo  {journal} {Phys. Rev. Lett.}}}, \bibinfo {year}
  {2006},\ \bibinfo {volume} {97}:\ \bibinfo {pages} {090402}}\BibitemShut
  {NoStop}%
\bibitem [{\citenamefont {Chien}\ \emph {et~al.}(2007)\citenamefont {Chien},
  \citenamefont {Chen}, \citenamefont {He},\ and\ \citenamefont
  {Levin}}]{ChienPRL}%
  \BibitemOpen
  \bibfield  {author} {\bibinfo {author} {\bibfnamefont {C.~C.}\ \bibnamefont
  {Chien}}, \bibinfo {author} {\bibfnamefont {Q.~J.}\ \bibnamefont {Chen}},
  \bibinfo {author} {\bibfnamefont {Y.}~\bibnamefont {He}}, \ and\ \bibinfo
  {author} {\bibfnamefont {K.}~\bibnamefont {Levin}},\ }\bibfield  {title}
  {\enquote {\bibinfo {title} {Superfluid phase diagrams of trapped {F}ermi
  gases with population imbalance},}\ }\href@noop {} {\bibfield  {journal}
  {\textit {\bibinfo  {journal} {Phys. Rev. Lett.}}}, \bibinfo {year} {2007},\
  \bibinfo {volume} {98}:\ \bibinfo {pages} {110404}}\BibitemShut {NoStop}%
\bibitem [{\citenamefont {Chen}\ \emph {et~al.}(2007)\citenamefont {Chen},
  \citenamefont {He}, \citenamefont {Chien},\ and\ \citenamefont
  {Levin}}]{Chen2007PRB}%
  \BibitemOpen
  \bibfield  {author} {\bibinfo {author} {\bibfnamefont {Q.~J.}\ \bibnamefont
  {Chen}}, \bibinfo {author} {\bibfnamefont {Y.}~\bibnamefont {He}}, \bibinfo
  {author} {\bibfnamefont {C.-C.}\ \bibnamefont {Chien}}, \ and\ \bibinfo
  {author} {\bibfnamefont {K.}~\bibnamefont {Levin}},\ }\bibfield  {title}
  {\enquote {\bibinfo {title} {Theory of superfluids with population imbalance:
  {F}inite temperature and {BCS-BEC} crossover effects},}\ }\href {\doibase
  10.1103/PhysRevB.75.014521} {\bibfield  {journal} {\textit {\bibinfo
  {journal} {Phys. Rev. B}}}, \bibinfo {year} {2007},\ \bibinfo {volume} {75}:\
  \bibinfo {pages} {014521}}\BibitemShut {NoStop}%
\bibitem [{\citenamefont {Wang}\ \emph {et~al.}(2013)\citenamefont {Wang},
  \citenamefont {Guo},\ and\ \citenamefont {Chen}}]{Wang2013}%
  \BibitemOpen
  \bibfield  {author} {\bibinfo {author} {\bibfnamefont {J.~B.}\ \bibnamefont
  {Wang}}, \bibinfo {author} {\bibfnamefont {H.}~\bibnamefont {Guo}}, \ and\
  \bibinfo {author} {\bibfnamefont {Q.~J.}\ \bibnamefont {Chen}},\ }\bibfield
  {title} {\enquote {\bibinfo {title} {Phase diagrams of {F}ermi gases in a
  trap with mass and population imbalances at finite temperature},}\
  }\href@noop {} {\bibfield  {journal} {\textit {\bibinfo  {journal} {Phys.
  Rev. A}}}, \bibinfo {year} {2013},\ \bibinfo {volume} {87}:\ \bibinfo {pages}
  {041601(R)}}\BibitemShut {NoStop}%
\bibitem [{\citenamefont {O'Hara}\ \emph {et~al.}(2002)\citenamefont {O'Hara},
  \citenamefont {Hemmer}, \citenamefont {Gehm}, \citenamefont {Granade},\ and\
  \citenamefont {Thomas}}]{Thomas}%
  \BibitemOpen
  \bibfield  {author} {\bibinfo {author} {\bibfnamefont {K.~M.}\ \bibnamefont
  {O'Hara}}, \bibinfo {author} {\bibfnamefont {S.~L.}\ \bibnamefont {Hemmer}},
  \bibinfo {author} {\bibfnamefont {M.~E.}\ \bibnamefont {Gehm}}, \bibinfo
  {author} {\bibfnamefont {S.~R.}\ \bibnamefont {Granade}}, \ and\ \bibinfo
  {author} {\bibfnamefont {J.~E.}\ \bibnamefont {Thomas}},\ }\bibfield  {title}
  {\enquote {\bibinfo {title} {Observation of a strongly interacting degenerate
  {F}ermi gas of atoms},}\ }\href@noop {} {\bibfield  {journal} {\textit
  {\bibinfo  {journal} {Science}}}, \bibinfo {year} {2002},\ \bibinfo {volume}
  {298}:\ \bibinfo {pages} {2179}}\BibitemShut {NoStop}%
\bibitem [{\citenamefont {Bourdel}\ \emph {et~al.}(2004)\citenamefont
  {Bourdel}, \citenamefont {Khaykovich}, \citenamefont {Cubizolles},
  \citenamefont {Zhang}, \citenamefont {Chevy}, \citenamefont {Teichmann},
  \citenamefont {Tarruell}, \citenamefont {Kokkelmans},\ and\ \citenamefont
  {Salomon}}]{Salomon3}%
  \BibitemOpen
  \bibfield  {author} {\bibinfo {author} {\bibfnamefont {T.}~\bibnamefont
  {Bourdel}}, \bibinfo {author} {\bibfnamefont {L.}~\bibnamefont {Khaykovich}},
  \bibinfo {author} {\bibfnamefont {J.}~\bibnamefont {Cubizolles}}, \bibinfo
  {author} {\bibfnamefont {J.}~\bibnamefont {Zhang}}, \bibinfo {author}
  {\bibfnamefont {F.}~\bibnamefont {Chevy}}, \bibinfo {author} {\bibfnamefont
  {M.}~\bibnamefont {Teichmann}}, \bibinfo {author} {\bibfnamefont
  {L.}~\bibnamefont {Tarruell}}, \bibinfo {author} {\bibfnamefont {S.~J.}\
  \bibnamefont {Kokkelmans}}, \ and\ \bibinfo {author} {\bibfnamefont
  {C.}~\bibnamefont {Salomon}},\ }\bibfield  {title} {\enquote {\bibinfo
  {title} {Experimental study of the {BCS-BEC} crossover region in
  lithium-6},}\ }\href@noop {} {\bibfield  {journal} {\textit {\bibinfo
  {journal} {Phys. Rev. Lett.}}}, \bibinfo {year} {2004},\ \bibinfo {volume}
  {93}:\ \bibinfo {pages} {050401}}\BibitemShut {NoStop}%
\bibitem [{\citenamefont {Carlson}\ \emph {et~al.}(2003)\citenamefont
  {Carlson}, \citenamefont {Chang}, \citenamefont {Pandharipande},\ and\
  \citenamefont {Schmidt}}]{Carlson3}%
  \BibitemOpen
  \bibfield  {author} {\bibinfo {author} {\bibfnamefont {J.}~\bibnamefont
  {Carlson}}, \bibinfo {author} {\bibfnamefont {S.}~\bibnamefont {Chang}},
  \bibinfo {author} {\bibfnamefont {V.}~\bibnamefont {Pandharipande}}, \ and\
  \bibinfo {author} {\bibfnamefont {K.}~\bibnamefont {Schmidt}},\ }\bibfield
  {title} {\enquote {\bibinfo {title} {Superfluid fermi gases with large
  scattering length},}\ }\href@noop {} {\bibfield  {journal} {\textit {\bibinfo
   {journal} {Phys. Rev. Lett.}}}, \bibinfo {year} {2003},\ \bibinfo {volume}
  {91}:\ \bibinfo {pages} {050401}}\BibitemShut {NoStop}%
\bibitem [{\citenamefont {Kosztin}\ \emph {et~al.}(2000)\citenamefont
  {Kosztin}, \citenamefont {Chen}, \citenamefont {Kao},\ and\ \citenamefont
  {Levin}}]{Kosztin2}%
  \BibitemOpen
  \bibfield  {author} {\bibinfo {author} {\bibfnamefont {I.}~\bibnamefont
  {Kosztin}}, \bibinfo {author} {\bibfnamefont {Q.~J.}\ \bibnamefont {Chen}},
  \bibinfo {author} {\bibfnamefont {Y.-J.}\ \bibnamefont {Kao}}, \ and\
  \bibinfo {author} {\bibfnamefont {K.}~\bibnamefont {Levin}},\ }\bibfield
  {title} {\enquote {\bibinfo {title} {Pair excitations, collective modes and
  gauge invariance in the {BCS -- Bose-Einstein} crossover scenario},}\
  }\href@noop {} {\bibfield  {journal} {\textit {\bibinfo  {journal} {Phys.
  Rev. B}}}, \bibinfo {year} {2000},\ \bibinfo {volume} {61}:\ \bibinfo {pages}
  {11662}}\BibitemShut {NoStop}%
\bibitem [{\citenamefont {Chen}\ \emph
  {et~al.}(2006{\natexlab{b}})\citenamefont {Chen}, \citenamefont {He},
  \citenamefont {Chien},\ and\ \citenamefont {Levin}}]{Stability}%
  \BibitemOpen
  \bibfield  {author} {\bibinfo {author} {\bibfnamefont {Q.~J.}\ \bibnamefont
  {Chen}}, \bibinfo {author} {\bibfnamefont {Y.}~\bibnamefont {He}}, \bibinfo
  {author} {\bibfnamefont {C.-C.}\ \bibnamefont {Chien}}, \ and\ \bibinfo
  {author} {\bibfnamefont {K.}~\bibnamefont {Levin}},\ }\bibfield  {title}
  {\enquote {\bibinfo {title} {Stability conditions and phase diagrams for
  two-component {Fermi} gases with population imbalance},}\ }\href {\doibase
  10.1103/PhysRevA.74.063603} {\bibfield  {journal} {\textit {\bibinfo
  {journal} {Phys. Rev. A}}}, \bibinfo {year} {2006}{\natexlab{b}},\ \bibinfo
  {volume} {74}:\ \bibinfo {pages} {063603}}\BibitemShut {NoStop}%
\bibitem [{foo()}]{footnoteongamma}%
  \BibitemOpen
  \bibinfo {note} {In fact, the parameter $\gamma$ can be taken from
  experiment, as has been done in Ref.~\protect\cite{Chen4}, where one can find
  more details.}\BibitemShut {Stop}%
\bibitem [{\citenamefont {Pieri}\ \emph
  {et~al.}(2004{\natexlab{a}})\citenamefont {Pieri}, \citenamefont {Pisani},\
  and\ \citenamefont {Strinati}}]{Strinati8}%
  \BibitemOpen
  \bibfield  {author} {\bibinfo {author} {\bibfnamefont {P.}~\bibnamefont
  {Pieri}}, \bibinfo {author} {\bibfnamefont {L.}~\bibnamefont {Pisani}}, \
  and\ \bibinfo {author} {\bibfnamefont {G.~C.}\ \bibnamefont {Strinati}},\
  }\bibfield  {title} {\enquote {\bibinfo {title} {{BCS-BEC} crossover at
  finite temperature in the broken-symmetry phase},}\ }\href@noop {} {\bibfield
   {journal} {\textit {\bibinfo  {journal} {\prb}}}, \bibinfo {year}
  {2004}{\natexlab{a}},\ \bibinfo {volume} {70}:\ \bibinfo {pages}
  {094508}}\BibitemShut {NoStop}%
\bibitem [{\citenamefont {Fukushima}\ \emph {et~al.}(2007)\citenamefont
  {Fukushima}, \citenamefont {Ohashi}, \citenamefont {Taylor},\ and\
  \citenamefont {Griffin}}]{Griffingroup2}%
  \BibitemOpen
  \bibfield  {author} {\bibinfo {author} {\bibfnamefont {N.}~\bibnamefont
  {Fukushima}}, \bibinfo {author} {\bibfnamefont {Y.}~\bibnamefont {Ohashi}},
  \bibinfo {author} {\bibfnamefont {E.}~\bibnamefont {Taylor}}, \ and\ \bibinfo
  {author} {\bibfnamefont {A.}~\bibnamefont {Griffin}},\ }\bibfield  {title}
  {\enquote {\bibinfo {title} {Superfluid density and condensate fraction in
  the {BCS-BEC} crossover regime at finite temperatures},}\ }\href@noop {}
  {\bibfield  {journal} {\textit {\bibinfo  {journal} {\pra}}}, \bibinfo {year}
  {2007},\ \bibinfo {volume} {75}:\ \bibinfo {pages} {033609}}\BibitemShut
  {NoStop}%
\bibitem [{\citenamefont {Kosztin}\ and\ \citenamefont
  {Leggett}(1997)}]{Kosztin_nonlocal}%
  \BibitemOpen
  \bibfield  {author} {\bibinfo {author} {\bibfnamefont {I.}~\bibnamefont
  {Kosztin}}\ and\ \bibinfo {author} {\bibfnamefont {A.~J.}\ \bibnamefont
  {Leggett}},\ }\bibfield  {title} {\enquote {\bibinfo {title} {Nonlocal
  effects on the magnetic penetration depth in $d$-wave superconductors},}\
  }\href@noop {} {\bibfield  {journal} {\textit {\bibinfo  {journal} {Phys.
  Rev. Lett.}}}, \bibinfo {year} {1997},\ \bibinfo {volume} {79}:\ \bibinfo
  {pages} {135}}\BibitemShut {NoStop}%
\bibitem [{\citenamefont {Hufner}\ \emph {et~al.}(2008)\citenamefont {Hufner},
  \citenamefont {Hossain}, \citenamefont {Damascelli},\ and\ \citenamefont
  {Sawatzky}}]{Sawatzky}%
  \BibitemOpen
  \bibfield  {author} {\bibinfo {author} {\bibfnamefont {S.}~\bibnamefont
  {Hufner}}, \bibinfo {author} {\bibfnamefont {M.~A.}\ \bibnamefont {Hossain}},
  \bibinfo {author} {\bibfnamefont {A.}~\bibnamefont {Damascelli}}, \ and\
  \bibinfo {author} {\bibfnamefont {G.}~\bibnamefont {Sawatzky}},\ }\bibfield
  {title} {\enquote {\bibinfo {title} {Two gaps make a high-temperature
  superconductor?}}\ }\href@noop {} {\bibfield  {journal} {\textit {\bibinfo
  {journal} {Rep. Prog. Phys.}}}, \bibinfo {year} {2008},\ \bibinfo {volume}
  {71}:\ \bibinfo {pages} {062501}}\BibitemShut {NoStop}%
\bibitem [{And()}]{Anderson87}%
  \BibitemOpen
  \bibinfo {note} {G. Baskaran, Z. Zou, and P. W. Anderson, Solid State Commun.
  \textbf{63}, 973 (1987).}\BibitemShut {Stop}%
\bibitem [{\citenamefont {Miyakawa}\ \emph {et~al.}(1999)\citenamefont
  {Miyakawa}, \citenamefont {Zasadzinski}, \citenamefont {Ozyuzer},
  \citenamefont {Guptasarma}, \citenamefont {Hinks}, \citenamefont
  {Kendziora},\ and\ \citenamefont {Gray}}]{JohnZ}%
  \BibitemOpen
  \bibfield  {author} {\bibinfo {author} {\bibfnamefont {N.}~\bibnamefont
  {Miyakawa}}, \bibinfo {author} {\bibfnamefont {J.}~\bibnamefont
  {Zasadzinski}}, \bibinfo {author} {\bibfnamefont {L.}~\bibnamefont
  {Ozyuzer}}, \bibinfo {author} {\bibfnamefont {P.}~\bibnamefont {Guptasarma}},
  \bibinfo {author} {\bibfnamefont {D.}~\bibnamefont {Hinks}}, \bibinfo
  {author} {\bibfnamefont {C.}~\bibnamefont {Kendziora}}, \ and\ \bibinfo
  {author} {\bibfnamefont {K.}~\bibnamefont {Gray}},\ }\bibfield  {title}
  {\enquote {\bibinfo {title} {Predominantly superconducting origin of large
  energy gaps in underdoped {BISCO} from tunneling spectroscopy},}\ }\href@noop
  {} {\bibfield  {journal} {\textit {\bibinfo  {journal} {Phys. Rev. Lett.}}},
  \bibinfo {year} {1999},\ \bibinfo {volume} {83}:\ \bibinfo {pages}
  {1018}}\BibitemShut {NoStop}%
\bibitem [{\citenamefont {Ho}(2004)}]{JasonHo}%
  \BibitemOpen
  \bibfield  {author} {\bibinfo {author} {\bibfnamefont {T.-L.}\ \bibnamefont
  {Ho}},\ }\bibfield  {title} {\enquote {\bibinfo {title} {Universal
  thermodynamics of degenerate quantum gases in the unitary limit},}\
  }\href@noop {} {\bibfield  {journal} {\textit {\bibinfo  {journal} {Phys.
  Rev. Lett.}}}, \bibinfo {year} {2004},\ \bibinfo {volume} {92}:\ \bibinfo
  {pages} {090402}}\BibitemShut {NoStop}%
\bibitem [{\citenamefont {Chiofalo}\ \emph {et~al.}(2002)\citenamefont
  {Chiofalo}, \citenamefont {Kokkelmans}, \citenamefont {Milstein},\ and\
  \citenamefont {Holland}}]{Chiofalo}%
  \BibitemOpen
  \bibfield  {author} {\bibinfo {author} {\bibfnamefont {M.~L.}\ \bibnamefont
  {Chiofalo}}, \bibinfo {author} {\bibfnamefont {S.~J. J. M.~F.}\ \bibnamefont
  {Kokkelmans}}, \bibinfo {author} {\bibfnamefont {J.~N.}\ \bibnamefont
  {Milstein}}, \ and\ \bibinfo {author} {\bibfnamefont {M.~J.}\ \bibnamefont
  {Holland}},\ }\bibfield  {title} {\enquote {\bibinfo {title} {Signatures of
  resonance superfluidity in a quantum {F}ermi gas},}\ }\href@noop {}
  {\bibfield  {journal} {\textit {\bibinfo  {journal} {Phys. Rev. Lett.}}},
  \bibinfo {year} {2002},\ \bibinfo {volume} {88}:\ \bibinfo {pages}
  {090402}}\BibitemShut {NoStop}%
\bibitem [{not({\natexlab{b}})}]{noteonpairdensity}%
  \BibitemOpen
  \bibinfo {note} {Note here that the definition for $n_c$ and $n_p$ differ
  from that in Ref.~\protect\cite{JS5} by a factor of 2.}\BibitemShut {Stop}%
\bibitem [{\citenamefont {Astrakharchik}\ \emph {et~al.}(2005)\citenamefont
  {Astrakharchik}, \citenamefont {Boronat}, \citenamefont {Casulleras},\ and\
  \citenamefont {Giorgini}}]{Giorgini2005}%
  \BibitemOpen
  \bibfield  {author} {\bibinfo {author} {\bibfnamefont {G.~E.}\ \bibnamefont
  {Astrakharchik}}, \bibinfo {author} {\bibfnamefont {J.}~\bibnamefont
  {Boronat}}, \bibinfo {author} {\bibfnamefont {J.}~\bibnamefont {Casulleras}},
  \ and\ \bibinfo {author} {\bibfnamefont {S.}~\bibnamefont {Giorgini}},\
  }\bibfield  {title} {\enquote {\bibinfo {title} {Momentum distribution and
  condensate fraction of a fermion gas in the {BCS-BEC} crossover},}\
  }\href@noop {} {\bibfield  {journal} {\textit {\bibinfo  {journal} {\prl}}},
  \bibinfo {year} {2005},\ \bibinfo {volume} {95}:\ \bibinfo {pages}
  {230405}},\ \bibinfo {note} {their result seems to suggest a tendency of
  decrease in the condensate fraction with an increasing particle number used
  for simulation.}\BibitemShut {Stop}%
\bibitem [{not({\natexlab{c}})}]{noteon2channel}%
  \BibitemOpen
  \bibinfo {note} {The curves in Fig.~\ref{fig:entropy} were calculated using a
  two-channel model. Nevertheless, for wide Feshbach resonances such as in
  $^6$Li and $^{40}$K, the closed-channel fraction is very small
  \protect\cite{ChenClosed,Hulet4} so that the quantitative difference in the
  entropy $s(r)$ between the two-channel and one-channel model is
  negligible.}\BibitemShut {Stop}%
\bibitem [{\citenamefont {Chen}\ \emph
  {et~al.}(2005{\natexlab{b}})\citenamefont {Chen}, \citenamefont {Stajic},\
  and\ \citenamefont {Levin}}]{ChenThermo}%
  \BibitemOpen
  \bibfield  {author} {\bibinfo {author} {\bibfnamefont {Q.~J.}\ \bibnamefont
  {Chen}}, \bibinfo {author} {\bibfnamefont {J.}~\bibnamefont {Stajic}}, \ and\
  \bibinfo {author} {\bibfnamefont {K.}~\bibnamefont {Levin}},\ }\bibfield
  {title} {\enquote {\bibinfo {title} {Thermodynamics of interacting fermions
  in atomic traps},}\ }\href@noop {} {\bibfield  {journal} {\textit {\bibinfo
  {journal} {\prl}}}, \bibinfo {year} {2005}{\natexlab{b}},\ \bibinfo {volume}
  {95}:\ \bibinfo {pages} {260405}}\BibitemShut {NoStop}%
\bibitem [{\citenamefont {Chen}\ \emph
  {et~al.}(2006{\natexlab{c}})\citenamefont {Chen}, \citenamefont {Regal},
  \citenamefont {Greiner}, \citenamefont {Jin},\ and\ \citenamefont
  {Levin}}]{Jin_us}%
  \BibitemOpen
  \bibfield  {author} {\bibinfo {author} {\bibfnamefont {Q.~J.}\ \bibnamefont
  {Chen}}, \bibinfo {author} {\bibfnamefont {C.~A.}\ \bibnamefont {Regal}},
  \bibinfo {author} {\bibfnamefont {M.}~\bibnamefont {Greiner}}, \bibinfo
  {author} {\bibfnamefont {D.~S.}\ \bibnamefont {Jin}}, \ and\ \bibinfo
  {author} {\bibfnamefont {K.}~\bibnamefont {Levin}},\ }\bibfield  {title}
  {\enquote {\bibinfo {title} {Understanding the superfluid phase diagram in
  trapped {F}ermi gases},}\ }\href@noop {} {\bibfield  {journal} {\textit
  {\bibinfo  {journal} {\pra}}}, \bibinfo {year} {2006}{\natexlab{c}},\
  \bibinfo {volume} {73}:\ \bibinfo {pages} {041601(R)}}\BibitemShut {NoStop}%
\bibitem [{not({\natexlab{d}})}]{noteoncontourplot}%
  \BibitemOpen
  \bibinfo {note} {Note that the experimental data cannot be measuring $N_c/N$
  as shown in Fig.~\ref{fig:condensatefraction}, since at $1/k_Fa=-1$, $N_c/N$
  is far below the experimental threshold of detection.}\BibitemShut {Stop}%
\bibitem [{\citenamefont {Stajic}\ \emph {et~al.}(2005)\citenamefont {Stajic},
  \citenamefont {Chen},\ and\ \citenamefont {Levin}}]{JS5}%
  \BibitemOpen
  \bibfield  {author} {\bibinfo {author} {\bibfnamefont {J.}~\bibnamefont
  {Stajic}}, \bibinfo {author} {\bibfnamefont {Q.~J.}\ \bibnamefont {Chen}}, \
  and\ \bibinfo {author} {\bibfnamefont {K.}~\bibnamefont {Levin}},\ }\bibfield
   {title} {\enquote {\bibinfo {title} {Density profiles of strongly
  interacting trapped {F}ermi gases},}\ }\href@noop {} {\bibfield  {journal}
  {\textit {\bibinfo  {journal} {Phys. Rev. Lett.}}}, \bibinfo {year} {2005},\
  \bibinfo {volume} {94}:\ \bibinfo {pages} {060401}}\BibitemShut {NoStop}%
\bibitem [{not({\natexlab{e}})}]{noteon1Ddensity}%
  \BibitemOpen
  \bibinfo {note} {While one may argue that the kink, if it exists, may be
  smoothed out by the $\int \d y\d z$ integration, we note that as of the time
  of this writing, no kink behavior has ever been reported in 3D density
  profiles obtained via an inverse Abel transformation of experimental
  data.}\BibitemShut {Stop}%
\bibitem [{\citenamefont {Chen}\ \emph
  {et~al.}(2006{\natexlab{d}})\citenamefont {Chen}, \citenamefont {Regal},
  \citenamefont {Jin},\ and\ \citenamefont {Levin}}]{Jin2_us}%
  \BibitemOpen
  \bibfield  {author} {\bibinfo {author} {\bibfnamefont {Q.~J.}\ \bibnamefont
  {Chen}}, \bibinfo {author} {\bibfnamefont {C.~A.}\ \bibnamefont {Regal}},
  \bibinfo {author} {\bibfnamefont {D.~S.}\ \bibnamefont {Jin}}, \ and\
  \bibinfo {author} {\bibfnamefont {K.}~\bibnamefont {Levin}},\ }\bibfield
  {title} {\enquote {\bibinfo {title} {Finite temperature momentum distribution
  of a trapped {F}ermi gas},}\ }\href@noop {} {\bibfield  {journal} {\textit
  {\bibinfo  {journal} {\pra}}}, \bibinfo {year} {2006}{\natexlab{d}},\
  \bibinfo {volume} {74}:\ \bibinfo {pages} {011601(R)}}\BibitemShut {NoStop}%
\bibitem [{\citenamefont {Chen}\ \emph {et~al.}(2009)\citenamefont {Chen},
  \citenamefont {He}, \citenamefont {Chien},\ and\ \citenamefont
  {Levin}}]{Chen_RPP}%
  \BibitemOpen
  \bibfield  {author} {\bibinfo {author} {\bibfnamefont {Q.~J.}\ \bibnamefont
  {Chen}}, \bibinfo {author} {\bibfnamefont {Y.}~\bibnamefont {He}}, \bibinfo
  {author} {\bibfnamefont {C.-C.}\ \bibnamefont {Chien}}, \ and\ \bibinfo
  {author} {\bibfnamefont {K.}~\bibnamefont {Levin}},\ }\bibfield  {title}
  {\enquote {\bibinfo {title} {Theory of radio frequency spectroscopy in
  ultracold {F}ermi gases and their relation to photoemission in the
  cuprates},}\ }\href@noop {} {\bibfield  {journal} {\textit {\bibinfo
  {journal} {Rep. Prog. Phys.}}}, \bibinfo {year} {2009},\ \bibinfo {volume}
  {72}:\ \bibinfo {pages} {122501}}\BibitemShut {NoStop}%
\bibitem [{\citenamefont {Schunck}\ \emph {et~al.}(2008)\citenamefont
  {Schunck}, \citenamefont {Shin}, \citenamefont {Schirotzek}, \citenamefont
  {Zwierlein},\ and\ \citenamefont {Ketterle}}]{Ketterlepairsize}%
  \BibitemOpen
  \bibfield  {author} {\bibinfo {author} {\bibfnamefont {C.~H.}\ \bibnamefont
  {Schunck}}, \bibinfo {author} {\bibfnamefont {Y.}~\bibnamefont {Shin}},
  \bibinfo {author} {\bibfnamefont {A.}~\bibnamefont {Schirotzek}}, \bibinfo
  {author} {\bibfnamefont {M.~W.}\ \bibnamefont {Zwierlein}}, \ and\ \bibinfo
  {author} {\bibfnamefont {W.}~\bibnamefont {Ketterle}},\ }\bibfield  {title}
  {\enquote {\bibinfo {title} {Determination of the fermion pair size in a
  resonantly interacting superfluid},}\ }\href@noop {} {\bibfield  {journal}
  {\textit {\bibinfo  {journal} {Nature (London)}}}, \bibinfo {year} {2008},\
  \bibinfo {volume} {454}:\ \bibinfo {pages} {739}}\BibitemShut {NoStop}%
\bibitem [{\citenamefont {Schunck}\ \emph {et~al.}(2007)\citenamefont
  {Schunck}, \citenamefont {Shin}, \citenamefont {Schirotzek}, \citenamefont
  {Zwierlein},\ and\ \citenamefont {Ketterle}}]{KetterleRF}%
  \BibitemOpen
  \bibfield  {author} {\bibinfo {author} {\bibfnamefont {C.~H.}\ \bibnamefont
  {Schunck}}, \bibinfo {author} {\bibfnamefont {Y.}~\bibnamefont {Shin}},
  \bibinfo {author} {\bibfnamefont {A.}~\bibnamefont {Schirotzek}}, \bibinfo
  {author} {\bibfnamefont {M.~W.}\ \bibnamefont {Zwierlein}}, \ and\ \bibinfo
  {author} {\bibfnamefont {W.}~\bibnamefont {Ketterle}},\ }\bibfield  {title}
  {\enquote {\bibinfo {title} {Pairing without superfluidity: {T}he ground
  state of an imbalanced {F}ermi mixture},}\ }\href@noop {} {\bibfield
  {journal} {\textit {\bibinfo  {journal} {Science}}}, \bibinfo {year} {2007},\
  \bibinfo {volume} {316}:\ \bibinfo {pages} {867}}\BibitemShut {NoStop}%
\bibitem [{\citenamefont {Yu}\ and\ \citenamefont {Baym}(2006)}]{Baym2}%
  \BibitemOpen
  \bibfield  {author} {\bibinfo {author} {\bibfnamefont {Z.~H.}\ \bibnamefont
  {Yu}}\ and\ \bibinfo {author} {\bibfnamefont {G.}~\bibnamefont {Baym}},\
  }\bibfield  {title} {\enquote {\bibinfo {title} {Spin-correlation functions
  in ultracold paired atomic-fermion systems: {Sum} rules, self-consistent
  approximations, and mean fields},}\ }\href@noop {} {\bibfield  {journal}
  {\textit {\bibinfo  {journal} {Phys. Rev. A}}}, \bibinfo {year} {2006},\
  \bibinfo {volume} {73}:\ \bibinfo {pages} {063601}}\BibitemShut {NoStop}%
\bibitem [{\citenamefont {Baym}\ \emph {et~al.}(2007)\citenamefont {Baym},
  \citenamefont {Pethick}, \citenamefont {Yu},\ and\ \citenamefont
  {Zwierlein}}]{Baym3}%
  \BibitemOpen
  \bibfield  {author} {\bibinfo {author} {\bibfnamefont {G.}~\bibnamefont
  {Baym}}, \bibinfo {author} {\bibfnamefont {C.~J.}\ \bibnamefont {Pethick}},
  \bibinfo {author} {\bibfnamefont {Z.~H.}\ \bibnamefont {Yu}}, \ and\ \bibinfo
  {author} {\bibfnamefont {M.~W.}\ \bibnamefont {Zwierlein}},\ }\bibfield
  {title} {\enquote {\bibinfo {title} {Coherence and clock shifts in ultracold
  {F}ermi gases with resonant interactions},}\ }\href@noop {} {\bibfield
  {journal} {\textit {\bibinfo  {journal} {Phys. Rev. Lett.}}}, \bibinfo {year}
  {2007},\ \bibinfo {volume} {99}:\ \bibinfo {pages} {190407}}\BibitemShut
  {NoStop}%
\bibitem [{\citenamefont {Punk}\ and\ \citenamefont {Zwerger}(2007)}]{Punk}%
  \BibitemOpen
  \bibfield  {author} {\bibinfo {author} {\bibfnamefont {M.}~\bibnamefont
  {Punk}}\ and\ \bibinfo {author} {\bibfnamefont {W.}~\bibnamefont {Zwerger}},\
  }\bibfield  {title} {\enquote {\bibinfo {title} {Theory of rf-spectroscopy of
  strongly interacting fermions},}\ }\href@noop {} {\bibfield  {journal}
  {\textit {\bibinfo  {journal} {Phys. Rev. Lett.}}}, \bibinfo {year} {2007},\
  \bibinfo {volume} {99}:\ \bibinfo {pages} {170404}}\BibitemShut {NoStop}%
\bibitem [{\citenamefont {Perali}\ \emph {et~al.}(2008)\citenamefont {Perali},
  \citenamefont {Pieri},\ and\ \citenamefont {Strinati}}]{Strinati7}%
  \BibitemOpen
  \bibfield  {author} {\bibinfo {author} {\bibfnamefont {A.}~\bibnamefont
  {Perali}}, \bibinfo {author} {\bibfnamefont {P.}~\bibnamefont {Pieri}}, \
  and\ \bibinfo {author} {\bibfnamefont {G.~C.}\ \bibnamefont {Strinati}},\
  }\bibfield  {title} {\enquote {\bibinfo {title} {Competition between
  final-state and pairing gap effects in the radio-frequency spectra of
  ultracold {Fermi} atoms},}\ }\href@noop {} {\bibfield  {journal} {\textit
  {\bibinfo  {journal} {Phys. Rev. Lett.}}}, \bibinfo {year} {2008},\ \bibinfo
  {volume} {100}:\ \bibinfo {pages} {010402}}\BibitemShut {NoStop}%
\bibitem [{\citenamefont {Basu}\ and\ \citenamefont {Mueller}(2008)}]{Basu}%
  \BibitemOpen
  \bibfield  {author} {\bibinfo {author} {\bibfnamefont {S.}~\bibnamefont
  {Basu}}\ and\ \bibinfo {author} {\bibfnamefont {E.}~\bibnamefont {Mueller}},\
  }\bibfield  {title} {\enquote {\bibinfo {title} {Final-state effects in the
  radio frequency spectrum of strongly interacting fermions},}\ }\href@noop {}
  {\bibfield  {journal} {\textit {\bibinfo  {journal} {Phys. Rev. Lett.}}},
  \bibinfo {year} {2008},\ \bibinfo {volume} {101}:\ \bibinfo {pages}
  {060405}}\BibitemShut {NoStop}%
\bibitem [{\citenamefont {He}\ \emph {et~al.}(2009)\citenamefont {He},
  \citenamefont {Chien}, \citenamefont {Chen},\ and\ \citenamefont
  {Levin}}]{ourRF3}%
  \BibitemOpen
  \bibfield  {author} {\bibinfo {author} {\bibfnamefont {Y.}~\bibnamefont
  {He}}, \bibinfo {author} {\bibfnamefont {C.~C.}\ \bibnamefont {Chien}},
  \bibinfo {author} {\bibfnamefont {Q.~J.}\ \bibnamefont {Chen}}, \ and\
  \bibinfo {author} {\bibfnamefont {K.}~\bibnamefont {Levin}},\ }\bibfield
  {title} {\enquote {\bibinfo {title} {Temperature and final state effects in
  radio frequency spectroscopy experiments on atomic {Fermi} gases},}\
  }\href@noop {} {\bibfield  {journal} {\textit {\bibinfo  {journal} {Phys.
  Rev. Lett.}}}, \bibinfo {year} {2009},\ \bibinfo {volume} {102}:\ \bibinfo
  {pages} {020402}}\BibitemShut {NoStop}%
\bibitem [{\citenamefont {Ku}\ \emph {et~al.}(2012)\citenamefont {Ku},
  \citenamefont {Sommer}, \citenamefont {Cheuk},\ and\ \citenamefont
  {Zwierlein}}]{Zwierlein2011}%
  \BibitemOpen
  \bibfield  {author} {\bibinfo {author} {\bibfnamefont {M.~J.~H.}\
  \bibnamefont {Ku}}, \bibinfo {author} {\bibfnamefont {A.~T.}\ \bibnamefont
  {Sommer}}, \bibinfo {author} {\bibfnamefont {L.~W.}\ \bibnamefont {Cheuk}}, \
  and\ \bibinfo {author} {\bibfnamefont {M.~W.}\ \bibnamefont {Zwierlein}},\
  }\bibfield  {title} {\enquote {\bibinfo {title} {Revealing the superfluid
  lambda transition in the universal thermodynamics of a unitary {Fermi}
  gas},}\ }\href@noop {} {\bibfield  {journal} {\textit {\bibinfo  {journal}
  {Science}}}, \bibinfo {year} {2012},\ \bibinfo {volume} {335}:\ \bibinfo
  {pages} {563}}\BibitemShut {NoStop}%
\bibitem [{\citenamefont {Burovski}\ \emph {et~al.}(2006)\citenamefont
  {Burovski}, \citenamefont {Prokof'ev}, \citenamefont {Svistunov},\ and\
  \citenamefont {Troyer}}]{TroyerPRL2006}%
  \BibitemOpen
  \bibfield  {author} {\bibinfo {author} {\bibfnamefont {E.}~\bibnamefont
  {Burovski}}, \bibinfo {author} {\bibfnamefont {N.}~\bibnamefont {Prokof'ev}},
  \bibinfo {author} {\bibfnamefont {B.}~\bibnamefont {Svistunov}}, \ and\
  \bibinfo {author} {\bibfnamefont {M.}~\bibnamefont {Troyer}},\ }\bibfield
  {title} {\enquote {\bibinfo {title} {Critical temperature and thermodynamics
  of attractive fermions at unitarity},}\ }\href@noop {} {\bibfield  {journal}
  {\textit {\bibinfo  {journal} {\prl}}}, \bibinfo {year} {2006},\ \bibinfo
  {volume} {96}:\ \bibinfo {pages} {160402}}\BibitemShut {NoStop}%
\bibitem [{\citenamefont {Burovski}\ \emph {et~al.}(2008)\citenamefont
  {Burovski}, \citenamefont {Kozik}, \citenamefont {Prokof'ev}, \citenamefont
  {Svistunov},\ and\ \citenamefont {Troyer}}]{TroyerPRL2008}%
  \BibitemOpen
  \bibfield  {author} {\bibinfo {author} {\bibfnamefont {E.}~\bibnamefont
  {Burovski}}, \bibinfo {author} {\bibfnamefont {E.}~\bibnamefont {Kozik}},
  \bibinfo {author} {\bibfnamefont {N.}~\bibnamefont {Prokof'ev}}, \bibinfo
  {author} {\bibfnamefont {B.}~\bibnamefont {Svistunov}}, \ and\ \bibinfo
  {author} {\bibfnamefont {M.}~\bibnamefont {Troyer}},\ }\bibfield  {title}
  {\enquote {\bibinfo {title} {Critical temperature curve in {BEC-BCS}
  crossover},}\ }\href@noop {} {\bibfield  {journal} {\textit {\bibinfo
  {journal} {\prl}}}, \bibinfo {year} {2008},\ \bibinfo {volume} {101}:\
  \bibinfo {pages} {090402}}\BibitemShut {NoStop}%
\bibitem [{\citenamefont {Goulko}\ and\ \citenamefont
  {Wingate}(2010)}]{Wingate}%
  \BibitemOpen
  \bibfield  {author} {\bibinfo {author} {\bibfnamefont {O.}~\bibnamefont
  {Goulko}}\ and\ \bibinfo {author} {\bibfnamefont {M.}~\bibnamefont
  {Wingate}},\ }\bibfield  {title} {\enquote {\bibinfo {title} {Thermodynamics
  of balanced and slightly spin-imbalanced {F}ermi gases at unitarity},}\
  }\href@noop {} {\bibfield  {journal} {\textit {\bibinfo  {journal} {Phys.
  Rev. A}}}, \bibinfo {year} {2010},\ \bibinfo {volume} {82}:\ \bibinfo {pages}
  {053621}}\BibitemShut {NoStop}%
\bibitem [{\citenamefont {Kinnunen}\ \emph {et~al.}(2004)\citenamefont
  {Kinnunen}, \citenamefont {Rodriguez},\ and\ \citenamefont
  {T\"orm\"a}}]{Torma2}%
  \BibitemOpen
  \bibfield  {author} {\bibinfo {author} {\bibfnamefont {J.}~\bibnamefont
  {Kinnunen}}, \bibinfo {author} {\bibfnamefont {M.}~\bibnamefont {Rodriguez}},
  \ and\ \bibinfo {author} {\bibfnamefont {P.}~\bibnamefont {T\"orm\"a}},\
  }\bibfield  {title} {\enquote {\bibinfo {title} {Pairing gap and in-gap
  excitations in trapped fermionic superfluids},}\ }\href@noop {} {\bibfield
  {journal} {\textit {\bibinfo  {journal} {Science}}}, \bibinfo {year} {2004},\
  \bibinfo {volume} {305}:\ \bibinfo {pages} {1131}}\BibitemShut {NoStop}%
\bibitem [{\citenamefont {He}\ \emph {et~al.}(2005)\citenamefont {He},
  \citenamefont {Chen},\ and\ \citenamefont {Levin}}]{heyan}%
  \BibitemOpen
  \bibfield  {author} {\bibinfo {author} {\bibfnamefont {Y.}~\bibnamefont
  {He}}, \bibinfo {author} {\bibfnamefont {Q.~J.}\ \bibnamefont {Chen}}, \ and\
  \bibinfo {author} {\bibfnamefont {K.}~\bibnamefont {Levin}},\ }\bibfield
  {title} {\enquote {\bibinfo {title} {Radio-frequency spectroscopy and the
  pairing gap in trapped {Fermi} gases},}\ }\href@noop {} {\bibfield  {journal}
  {\textit {\bibinfo  {journal} {Phys. Rev. A}}}, \bibinfo {year} {2005},\
  \bibinfo {volume} {72}:\ \bibinfo {pages} {011602(R)}}\BibitemShut {NoStop}%
\bibitem [{\citenamefont {Massignan}\ \emph {et~al.}(2008)\citenamefont
  {Massignan}, \citenamefont {Bruun},\ and\ \citenamefont {Stoof}}]{Stoof3}%
  \BibitemOpen
  \bibfield  {author} {\bibinfo {author} {\bibfnamefont {P.}~\bibnamefont
  {Massignan}}, \bibinfo {author} {\bibfnamefont {G.~M.}\ \bibnamefont
  {Bruun}}, \ and\ \bibinfo {author} {\bibfnamefont {H.~T.~C.}\ \bibnamefont
  {Stoof}},\ }\bibfield  {title} {\enquote {\bibinfo {title} {Twin peaks in rf
  spectra of {F}ermi gases at unitarity},}\ }\href@noop {} {\bibfield
  {journal} {\textit {\bibinfo  {journal} {\pra}}}, \bibinfo {year} {2008},\
  \bibinfo {volume} {77}:\ \bibinfo {pages} {031601(R)}}\BibitemShut {NoStop}%
\bibitem [{\citenamefont {Stewart}\ \emph {et~al.}(2008)\citenamefont
  {Stewart}, \citenamefont {Gaebler},\ and\ \citenamefont {Jin}}]{Jin6}%
  \BibitemOpen
  \bibfield  {author} {\bibinfo {author} {\bibfnamefont {J.~T.}\ \bibnamefont
  {Stewart}}, \bibinfo {author} {\bibfnamefont {J.~P.}\ \bibnamefont
  {Gaebler}}, \ and\ \bibinfo {author} {\bibfnamefont {D.~S.}\ \bibnamefont
  {Jin}},\ }\bibfield  {title} {\enquote {\bibinfo {title} {Using photoemission
  spectroscopy to probe a strongly interacting {Fermi} gas},}\ }\href@noop {}
  {\bibfield  {journal} {\textit {\bibinfo  {journal} {Nature (London)}}},
  \bibinfo {year} {2008},\ \bibinfo {volume} {454}:\ \bibinfo {pages}
  {744}}\BibitemShut {NoStop}%
\bibitem [{\citenamefont {Chen}\ and\ \citenamefont {Levin}(2009)}]{Chen_MRRF}%
  \BibitemOpen
  \bibfield  {author} {\bibinfo {author} {\bibfnamefont {Q.~J.}\ \bibnamefont
  {Chen}}\ and\ \bibinfo {author} {\bibfnamefont {K.}~\bibnamefont {Levin}},\
  }\bibfield  {title} {\enquote {\bibinfo {title} {Momentum resolved radio
  frequency spectroscopy in trapped {F}ermi gases},}\ }\href@noop {} {\bibfield
   {journal} {\textit {\bibinfo  {journal} {\prl}}}, \bibinfo {year} {2009},\
  \bibinfo {volume} {102}:\ \bibinfo {pages} {190402}}\BibitemShut {NoStop}%
\bibitem [{Jin()}]{JinPrivate}%
  \BibitemOpen
  \bibinfo {note} {D.S. Jin, private communications; D.S. Jin, American
  Physical Society March Meeting Talk B8.00002, 2009, abstract available at
  http://meetings.aps.org/link/BAPS.2009.MAR.B8.2}\BibitemShut {NoStop}%
\bibitem [{\citenamefont {Gaebler}\ \emph {et~al.}(2010)\citenamefont
  {Gaebler}, \citenamefont {Stewart}, \citenamefont {Drake}, \citenamefont
  {Jin}, \citenamefont {Perali}, \citenamefont {Pieri},\ and\ \citenamefont
  {Strinati}}]{JinStrinati_nphys}%
  \BibitemOpen
  \bibfield  {author} {\bibinfo {author} {\bibfnamefont {J.~P.}\ \bibnamefont
  {Gaebler}}, \bibinfo {author} {\bibfnamefont {J.~T.}\ \bibnamefont
  {Stewart}}, \bibinfo {author} {\bibfnamefont {T.~E.}\ \bibnamefont {Drake}},
  \bibinfo {author} {\bibfnamefont {D.~S.}\ \bibnamefont {Jin}}, \bibinfo
  {author} {\bibfnamefont {A.}~\bibnamefont {Perali}}, \bibinfo {author}
  {\bibfnamefont {P.}~\bibnamefont {Pieri}}, \ and\ \bibinfo {author}
  {\bibfnamefont {G.~C.}\ \bibnamefont {Strinati}},\ }\bibfield  {title}
  {\enquote {\bibinfo {title} {Observation of pseudogap behaviour in a strongly
  interacting {F}ermi gas},}\ }\href@noop {} {\bibfield  {journal} {\textit
  {\bibinfo  {journal} {Nat. Phys.}}}, \bibinfo {year} {2010},\ \bibinfo
  {volume} {6}:\ \bibinfo {pages} {569}}\BibitemShut {NoStop}%
\bibitem [{\citenamefont {Perali}\ \emph {et~al.}(2011)\citenamefont {Perali},
  \citenamefont {Palestini}, \citenamefont {Pieri}, \citenamefont {Strinati},
  \citenamefont {Stewart}, \citenamefont {Gaebler}, \citenamefont {Drake},\
  and\ \citenamefont {Jin}}]{StrinatiJin}%
  \BibitemOpen
  \bibfield  {author} {\bibinfo {author} {\bibfnamefont {A.}~\bibnamefont
  {Perali}}, \bibinfo {author} {\bibfnamefont {F.}~\bibnamefont {Palestini}},
  \bibinfo {author} {\bibfnamefont {P.}~\bibnamefont {Pieri}}, \bibinfo
  {author} {\bibfnamefont {G.~C.}\ \bibnamefont {Strinati}}, \bibinfo {author}
  {\bibfnamefont {J.~T.}\ \bibnamefont {Stewart}}, \bibinfo {author}
  {\bibfnamefont {J.~P.}\ \bibnamefont {Gaebler}}, \bibinfo {author}
  {\bibfnamefont {T.~E.}\ \bibnamefont {Drake}}, \ and\ \bibinfo {author}
  {\bibfnamefont {D.~S.}\ \bibnamefont {Jin}},\ }\bibfield  {title} {\enquote
  {\bibinfo {title} {Evolution of the normal state of a strongly interacting
  {Fermi} gas from a pseudogap phase to a molecular {Bose} gas},}\ }\href@noop
  {} {\bibfield  {journal} {\textit {\bibinfo  {journal} {Phys. Rev. Lett.}}},
  \bibinfo {year} {2011},\ \bibinfo {volume} {106}:\ \bibinfo {pages}
  {060402}}\BibitemShut {NoStop}%
\bibitem [{\citenamefont {Perali}\ \emph {et~al.}(2002)\citenamefont {Perali},
  \citenamefont {Pieri}, \citenamefont {Strinati},\ and\ \citenamefont
  {Castellani}}]{Strinaticuprates}%
  \BibitemOpen
  \bibfield  {author} {\bibinfo {author} {\bibfnamefont {A.}~\bibnamefont
  {Perali}}, \bibinfo {author} {\bibfnamefont {P.}~\bibnamefont {Pieri}},
  \bibinfo {author} {\bibfnamefont {G.~C.}\ \bibnamefont {Strinati}}, \ and\
  \bibinfo {author} {\bibfnamefont {C.}~\bibnamefont {Castellani}},\ }\bibfield
   {title} {\enquote {\bibinfo {title} {Pseudogap and spectral function from
  superconducting fluctuations to the bosonic limit},}\ }\href@noop {}
  {\bibfield  {journal} {\textit {\bibinfo  {journal} {\prb}}}, \bibinfo {year}
  {2002},\ \bibinfo {volume} {66}:\ \bibinfo {pages} {024510}}\BibitemShut
  {NoStop}%
\bibitem [{\citenamefont {Pieri}\ \emph
  {et~al.}(2004{\natexlab{b}})\citenamefont {Pieri}, \citenamefont {Pisani},\
  and\ \citenamefont {Strinati}}]{Strinati2}%
  \BibitemOpen
  \bibfield  {author} {\bibinfo {author} {\bibfnamefont {P.}~\bibnamefont
  {Pieri}}, \bibinfo {author} {\bibfnamefont {L.}~\bibnamefont {Pisani}}, \
  and\ \bibinfo {author} {\bibfnamefont {G.~C.}\ \bibnamefont {Strinati}},\
  }\bibfield  {title} {\enquote {\bibinfo {title} {Pairing fluctuation effects
  on the single-particle spectra for the superconducting state},}\ }\href@noop
  {} {\bibfield  {journal} {\textit {\bibinfo  {journal} {Phys. Rev. Lett.}}},
  \bibinfo {year} {2004}{\natexlab{b}},\ \bibinfo {volume} {92}:\ \bibinfo
  {pages} {110401}}\BibitemShut {NoStop}%
\bibitem [{\citenamefont {Shin}\ \emph {et~al.}(2006)\citenamefont {Shin},
  \citenamefont {Zwierlein}, \citenamefont {Schunck}, \citenamefont
  {Schirotzek},\ and\ \citenamefont {Ketterle}}]{MITPRL06}%
  \BibitemOpen
  \bibfield  {author} {\bibinfo {author} {\bibfnamefont {Y.}~\bibnamefont
  {Shin}}, \bibinfo {author} {\bibfnamefont {M.~W.}\ \bibnamefont {Zwierlein}},
  \bibinfo {author} {\bibfnamefont {C.~H.}\ \bibnamefont {Schunck}}, \bibinfo
  {author} {\bibfnamefont {A.}~\bibnamefont {Schirotzek}}, \ and\ \bibinfo
  {author} {\bibfnamefont {W.}~\bibnamefont {Ketterle}},\ }\bibfield  {title}
  {\enquote {\bibinfo {title} {Observation of phase separation in a strongly
  interacting imbalanced {F}ermi gas},}\ }\href@noop {} {\bibfield  {journal}
  {\textit {\bibinfo  {journal} {Phys. Rev. Lett.}}}, \bibinfo {year} {2006},\
  \bibinfo {volume} {97}:\ \bibinfo {pages} {030401}}\BibitemShut {NoStop}%
\bibitem [{\citenamefont {Nascimb\`ene}\ \emph {et~al.}(2010)\citenamefont
  {Nascimb\`ene}, \citenamefont {Navon}, \citenamefont {Jiang}, \citenamefont
  {Chevy},\ and\ \citenamefont {Salomon}}]{Salomon2010Nature}%
  \BibitemOpen
  \bibfield  {author} {\bibinfo {author} {\bibfnamefont {S.}~\bibnamefont
  {Nascimb\`ene}}, \bibinfo {author} {\bibfnamefont {N.}~\bibnamefont {Navon}},
  \bibinfo {author} {\bibfnamefont {K.}~\bibnamefont {Jiang}}, \bibinfo
  {author} {\bibfnamefont {F.}~\bibnamefont {Chevy}}, \ and\ \bibinfo {author}
  {\bibfnamefont {C.}~\bibnamefont {Salomon}},\ }\bibfield  {title} {\enquote
  {\bibinfo {title} {Exploring the thermodynamics of a universal {F}ermi
  gas},}\ }\href@noop {} {\bibfield  {journal} {\textit {\bibinfo  {journal}
  {Nature (London)}}}, \bibinfo {year} {2010},\ \bibinfo {volume} {463}:\
  \bibinfo {pages} {1057}}\BibitemShut {NoStop}%
\bibitem [{\citenamefont {Nascimb\`ene}\ \emph {et~al.}(2011)\citenamefont
  {Nascimb\`ene}, \citenamefont {Navon}, \citenamefont {Pilati}, \citenamefont
  {Chevy}, \citenamefont {Giorgini}, \citenamefont {Georges},\ and\
  \citenamefont {Salomon}}]{SalomonPRL106}%
  \BibitemOpen
  \bibfield  {author} {\bibinfo {author} {\bibfnamefont {S.}~\bibnamefont
  {Nascimb\`ene}}, \bibinfo {author} {\bibfnamefont {N.}~\bibnamefont {Navon}},
  \bibinfo {author} {\bibfnamefont {S.}~\bibnamefont {Pilati}}, \bibinfo
  {author} {\bibfnamefont {F.}~\bibnamefont {Chevy}}, \bibinfo {author}
  {\bibfnamefont {S.}~\bibnamefont {Giorgini}}, \bibinfo {author}
  {\bibfnamefont {A.}~\bibnamefont {Georges}}, \ and\ \bibinfo {author}
  {\bibfnamefont {C.}~\bibnamefont {Salomon}},\ }\bibfield  {title} {\enquote
  {\bibinfo {title} {Fermi-liquid behavior of the normal phase of a strongly
  interacting gas of cold atoms},}\ }\href@noop {} {\bibfield  {journal}
  {\textit {\bibinfo  {journal} {Phys. Rev. Lett.}}}, \bibinfo {year} {2011},\
  \bibinfo {volume} {106}:\ \bibinfo {pages} {215303}}\BibitemShut {NoStop}%
\bibitem [{\citenamefont {Gorkov}\ and\ \citenamefont
  {Melik-Barkhudarov}(1961)}]{GMB}%
  \BibitemOpen
  \bibfield  {author} {\bibinfo {author} {\bibfnamefont {L.~P.}\ \bibnamefont
  {Gorkov}}\ and\ \bibinfo {author} {\bibfnamefont {T.~K.}\ \bibnamefont
  {Melik-Barkhudarov}},\ }\bibfield  {title} {\enquote {\bibinfo {title}
  {Contribution to the theory of superfluidity in an imperfect {F}ermi gas,},}\
  }\href@noop {} {\bibfield  {journal} {\textit {\bibinfo  {journal} {Sov.
  Phys. JETP}}}, \bibinfo {year} {1961},\ \bibinfo {volume} {13}:\ \bibinfo
  {pages} {1018}}\BibitemShut {NoStop}%
\bibitem [{\citenamefont {Heiselberg}\ \emph {et~al.}(2000)\citenamefont
  {Heiselberg}, \citenamefont {Pethick}, \citenamefont {Smith},\ and\
  \citenamefont {Viverit}}]{Heiselberg2000}%
  \BibitemOpen
  \bibfield  {author} {\bibinfo {author} {\bibfnamefont {H.}~\bibnamefont
  {Heiselberg}}, \bibinfo {author} {\bibfnamefont {C.~J.}\ \bibnamefont
  {Pethick}}, \bibinfo {author} {\bibfnamefont {H.}~\bibnamefont {Smith}}, \
  and\ \bibinfo {author} {\bibfnamefont {L.}~\bibnamefont {Viverit}},\
  }\bibfield  {title} {\enquote {\bibinfo {title} {Influence of induced
  interactions on the superfluid transition in dilute {F}ermi gases},}\
  }\href@noop {} {\bibfield  {journal} {\textit {\bibinfo  {journal} {\prl}}},
  \bibinfo {year} {2000},\ \bibinfo {volume} {85}:\ \bibinfo {pages}
  {2418}}\BibitemShut {NoStop}%
\bibitem [{\citenamefont {Kim}\ \emph {et~al.}(2009)\citenamefont {Kim},
  \citenamefont {Torma},\ and\ \citenamefont {Martikainen}}]{KimTorma2009}%
  \BibitemOpen
  \bibfield  {author} {\bibinfo {author} {\bibfnamefont {D.-H.}\ \bibnamefont
  {Kim}}, \bibinfo {author} {\bibfnamefont {P.}~\bibnamefont {Torma}}, \ and\
  \bibinfo {author} {\bibfnamefont {J.-P.}\ \bibnamefont {Martikainen}},\
  }\bibfield  {title} {\enquote {\bibinfo {title} {Induced interactions for
  ultracold {F}ermi gases in optical lattices},}\ }\href@noop {} {\bibfield
  {journal} {\textit {\bibinfo  {journal} {\prl}}}, \bibinfo {year} {2009},\
  \bibinfo {volume} {102}:\ \bibinfo {pages} {245301}}\BibitemShut {NoStop}%
\bibitem [{\citenamefont {Martikainen}\ \emph {et~al.}(2009)\citenamefont
  {Martikainen}, \citenamefont {Kinnunen}, \citenamefont {Torma},\ and\
  \citenamefont {Pethick}}]{TormaPethick2009}%
  \BibitemOpen
  \bibfield  {author} {\bibinfo {author} {\bibfnamefont {J.-P.}\ \bibnamefont
  {Martikainen}}, \bibinfo {author} {\bibfnamefont {J.~J.}\ \bibnamefont
  {Kinnunen}}, \bibinfo {author} {\bibfnamefont {P.}~\bibnamefont {Torma}}, \
  and\ \bibinfo {author} {\bibfnamefont {C.~J.}\ \bibnamefont {Pethick}},\
  }\bibfield  {title} {\enquote {\bibinfo {title} {Induced interactions and the
  superfluid transition temperature in a three-component {F}ermi gas},}\
  }\href@noop {} {\bibfield  {journal} {\textit {\bibinfo  {journal} {\prl}}},
  \bibinfo {year} {2009},\ \bibinfo {volume} {103}:\ \bibinfo {pages}
  {260403}}\BibitemShut {NoStop}%
\bibitem [{\citenamefont {Yu}\ \emph {et~al.}(2009)\citenamefont {Yu},
  \citenamefont {Huang},\ and\ \citenamefont {Yin}}]{Yin2009}%
  \BibitemOpen
  \bibfield  {author} {\bibinfo {author} {\bibfnamefont {Z.-Q.}\ \bibnamefont
  {Yu}}, \bibinfo {author} {\bibfnamefont {K.}~\bibnamefont {Huang}}, \ and\
  \bibinfo {author} {\bibfnamefont {L.}~\bibnamefont {Yin}},\ }\bibfield
  {title} {\enquote {\bibinfo {title} {Induced interaction in a {Fermi} gas
  with a {BEC-BCS} crossover},}\ }\href@noop {} {\bibfield  {journal} {\textit
  {\bibinfo  {journal} {\pra}}}, \bibinfo {year} {2009},\ \bibinfo {volume}
  {79}:\ \bibinfo {pages} {053636}}\BibitemShut {NoStop}%
\bibitem [{\citenamefont {Chen}()}]{ParticleHoleChannel}%
  \BibitemOpen
  \bibfield  {author} {\bibinfo {author} {\bibfnamefont {Q.~J.}\ \bibnamefont
  {Chen}},\ }\href@noop {} {\enquote {\bibinfo {title} {Effect of the
  particle-hole channel on {BCS--Bose-Einstein} condensation crossover in
  atomic {Fermi} gases},}\ }\bibinfo {note} {E-print
  arXiv:1109.2307}\BibitemShut {NoStop}%
\bibitem [{\citenamefont {Chen}\ and\ \citenamefont
  {Levin}(2005)}]{ChenClosed}%
  \BibitemOpen
  \bibfield  {author} {\bibinfo {author} {\bibfnamefont {Q.~J.}\ \bibnamefont
  {Chen}}\ and\ \bibinfo {author} {\bibfnamefont {K.}~\bibnamefont {Levin}},\
  }\bibfield  {title} {\enquote {\bibinfo {title} {Population of closed-channel
  molecules in trapped {F}ermi gases with broad {F}eshbach resonances},}\
  }\href@noop {} {\bibfield  {journal} {\textit {\bibinfo  {journal} {\prl}}},
  \bibinfo {year} {2005},\ \bibinfo {volume} {95}:\ \bibinfo {pages}
  {260406}}\BibitemShut {NoStop}%
\bibitem [{\citenamefont {Partridge}\ \emph {et~al.}(2005)\citenamefont
  {Partridge}, \citenamefont {Strecker}, \citenamefont {Kamar}, \citenamefont
  {Jack},\ and\ \citenamefont {Hulet}}]{Hulet4}%
  \BibitemOpen
  \bibfield  {author} {\bibinfo {author} {\bibfnamefont {G.~B.}\ \bibnamefont
  {Partridge}}, \bibinfo {author} {\bibfnamefont {K.~E.}\ \bibnamefont
  {Strecker}}, \bibinfo {author} {\bibfnamefont {R.~I.}\ \bibnamefont {Kamar}},
  \bibinfo {author} {\bibfnamefont {M.~W.}\ \bibnamefont {Jack}}, \ and\
  \bibinfo {author} {\bibfnamefont {R.~G.}\ \bibnamefont {Hulet}},\ }\bibfield
  {title} {\enquote {\bibinfo {title} {Molecular probe of pairing in the
  {BEC-BCS} crossover},}\ }\href@noop {} {\bibfield  {journal} {\textit
  {\bibinfo  {journal} {Phys. Rev. Lett.}}}, \bibinfo {year} {2005},\ \bibinfo
  {volume} {95}:\ \bibinfo {pages} {020404}}\BibitemShut {NoStop}%
\bibitem [{\citenamefont {Guo}\ \emph {et~al.}(2010)\citenamefont {Guo},
  \citenamefont {Chien},\ and\ \citenamefont {Levin}}]{GuoPRL105}%
  \BibitemOpen
  \bibfield  {author} {\bibinfo {author} {\bibfnamefont {H.}~\bibnamefont
  {Guo}}, \bibinfo {author} {\bibfnamefont {C.-C.}\ \bibnamefont {Chien}}, \
  and\ \bibinfo {author} {\bibfnamefont {K.}~\bibnamefont {Levin}},\ }\bibfield
   {title} {\enquote {\bibinfo {title} {Establishing the presence of coherence
  in atomic {F}ermi superfluids: {S}pin-flip and spin-preserving {Bragg}
  scattering at finite temperatures},}\ }\href@noop {} {\bibfield  {journal}
  {\textit {\bibinfo  {journal} {Phys. Rev. Lett.}}}, \bibinfo {year} {2010},\
  \bibinfo {volume} {105}:\ \bibinfo {pages} {120401}}\BibitemShut {NoStop}%
\bibitem [{\citenamefont {Lingham}\ \emph {et~al.}(2014)\citenamefont
  {Lingham}, \citenamefont {Fenech}, \citenamefont {Hoinka},\ and\
  \citenamefont {Vale}}]{ValePRL112}%
  \BibitemOpen
  \bibfield  {author} {\bibinfo {author} {\bibfnamefont {M.~G.}\ \bibnamefont
  {Lingham}}, \bibinfo {author} {\bibfnamefont {K.}~\bibnamefont {Fenech}},
  \bibinfo {author} {\bibfnamefont {S.}~\bibnamefont {Hoinka}}, \ and\ \bibinfo
  {author} {\bibfnamefont {C.~J.}\ \bibnamefont {Vale}},\ }\bibfield  {title}
  {\enquote {\bibinfo {title} {Local observation of pair condensation in a
  {F}ermi gas at unitarity},}\ }\href@noop {} {\bibfield  {journal} {\textit
  {\bibinfo  {journal} {\prl}}}, \bibinfo {year} {2014},\ \bibinfo {volume}
  {112}:\ \bibinfo {pages} {100404}}\BibitemShut {NoStop}%
\end{thebibliography}
%merlin.mbs apsrev4-1.bst 2010-07-25 4.21a (PWD, AO, DPC) hacked
%Control: key (0)
%Control: author (72) initials jnrlst
%Control: editor formatted (1) identically to author
%Control: production of article title (1) required
%Control: page (0) single
%Control: year (1) truncated
%Control: production of eprint (0) enabled
%

\end{document}